\documentclass{aa}
\usepackage{graphicx}
\usepackage{bm}
\usepackage{times}
\usepackage{amsmath}
\usepackage{amssymb}
\usepackage{natbib}
\usepackage{color}
\usepackage{hyperref}
\usepackage{verbatim}
\bibpunct{(}{)}{;}{a}{}{,}
\graphicspath{{./png/}}
\newcommand{\pd}{\partial}
\newcommand{\gggg}{\mbox{\boldmath $g$} {}}
\newcommand{\OO}{\bm{\Omega}}
\newcommand{\nab}{\mbox{\boldmath $\nabla$} {}}
\newcommand{\chiSGS}{\chi_{\rm SGS}}
\newcommand{\Pra}{{\rm Pr}}
\newcommand{\PraSGS}{{\rm Pr}_{\rm SGS}}
\newcommand{\Pm}{{\rm Pr}_{\rm M}}
\newcommand{\Ta}{{\rm Ta}}
\newcommand{\Ra}{{\rm Ra}}
\newcommand{\Rey}{{\rm Re}}
\newcommand{\Pe}{{\rm Pe}}
\newcommand{\Co}{{\rm Co}}
\newcommand{\chiSGSm}{\chi^{\rm m}_{\rm SGS}}
\newcommand{\urms}{u_{\rm rms}}
\newcommand{\kf}{k_{\rm f}}
\newcommand{\ReM}{{\rm Re}_{\rm M}}
\newcommand{\BBB}{{\bm B}}
\newcommand{\meanv}[1]{\overline{\bm #1}}
\newcommand{\brac}[1]{\langle #1 \rangle}
\newcommand{\s}{\,{\rm s}}
\newcommand{\kg}{\,{\rm kg}}
\newcommand{\m}{\,{\rm m}}
\newcommand{\Table}[1]{Table~\ref{#1}}
\newcommand{\Fig}[1]{Fig.~\ref{#1}}
\newcommand{\Eq}[1]{Eq.~(\ref{#1})}
\def\onehalf{{\textstyle{1\over2}}}
\newcommand{\uuu}{{\bm u}}
\newcommand{\mfi}[2]{\overline{#1}_{#2}}
\newcommand{\RHK}{R'_{\rm HK}}
\newcommand{\mean}[1]{\overline{#1}}
\newcommand{\Ro}{{\rm Ro}}

\begin{document}

\authorrunning{Viviani et al.}
\titlerunning{Axi- to nonaxisymmetric dynamo mode transition in
  solar-like stars}

  \title{Transition from axi- to nonaxisymmetric
     dynamo modes in spherical convection models of solar-like stars}

 \author{M. Viviani \inst{1}
          \and
           J. Warnecke \inst{1,2}
           \and
          M. J. K\"apyl\"a \inst{1,2}
          \and
          P. J. K\"apyl\"a \inst{3,2,1,4}
         \and
             N. Olspert \inst{2}
           \and
          E. M. Cole-Kodikara\inst{5}
         \and\\
         J. J. Lehtinen \inst{1,2}
         \and 
          A. Brandenburg \inst{4,6,7,8}
        }
 
   \institute{Max Planck Institute for Solar System Research,
              Justus-von-Liebig-Weg 3, D-37077 G\"ottingen, Germany\\
              \email{viviani@mps.mpg.de} 
         \and
	      ReSoLVE Centre of Excellence, Department of Computer Science, 
	      Aalto University, PO Box 15400, FI-00076 Aalto, Finland
         \and
	       Leibniz Institute for Astrophysics Potsdam, An der Sternwarte 16, 
	       D-14482 Potsdam, Germany
             \and NORDITA, KTH Royal Institute of Technology and Stockholm University, 
              Roslagstullsbacken 23, SE-10691 Stockholm, Sweden
     \and
               Department of Physics, Gustaf H\"allstr\"omin katu 2a
              (PO Box 64), FI-00014 University of Helsinki, Finland
         \and Department of Astronomy, AlbaNova University Center,
              Stockholm University, SE-10691 Stockholm, Sweden
         \and JILA and Department of Astrophysical and Planetary Sciences,
              Box 440, University of Colorado, Boulder, CO 80303, USA
         \and Laboratory for Atmospheric and Space Physics,
              3665 Discovery Drive, Boulder, CO 80303, USA
}

   \date{\today, $ $Revision: 1.741 $ $}

   \abstract{
     Both dynamo theory and observations of stellar large-scale
     magnetic fields suggest a change from nearly axisymmetric configurations
     at solar rotation rates to nonaxisymmetric configurations
     for rapid rotation.
   }
   {
     We seek to understand this transition using numerical simulations.
   }
   {
     We use three-dimensional simulations of turbulent convection and
    consider rotation rates between one and thirty times the solar value.
   }
   {
     We find a transition from axi- to nonaxisymmetric solutions at
     around 1.8 times the solar rotation rate.
     This transition coincides with a
     change in the rotation profile from antisolar- to solar-like
     differential rotation with a faster equator and slow poles.
     In the solar-like rotation regime, the field configuration consists of an axisymmetric oscillatory field
     accompanied by an $m=1$ azimuthal mode (two active longitudes), which also shows temporal
     variability. At slow (rapid) rotation, the axisymmetric (nonaxisymmetric) mode dominates.
     The axisymmetric mode produces latitudinal dynamo waves with
     polarity reversals, while the nonaxisymmetric
     mode often exhibits a slow drift in the rotating reference frame and 
     the strength of the active longitudes changes cyclically
     over time between the different hemispheres. 
     In the majority of cases we find retrograde waves,
     while prograde ones are more often found from observations.
     Most of the obtained dynamo solutions exhibit cyclic
     variability either caused by latitudinal or azimuthal dynamo
     waves.
     In an activity--period diagram, the cycle lengths normalized by the 
     rotation period form two different populations
     as a function of rotation rate or magnetic activity level. 
    The slowly rotating axisymmetric population lies close to what
    is called the inactive branch in observations, where the stars are however
    believed to have solar-like differential rotation, while the rapidly rotating
    models are close to the superactive branch with a declining cycle
    to rotation frequency ratio with increasing rotation rate.
   }
   {
We can successfully reproduce the transition from axi- to
nonaxisymmetric dynamo solutions for high rotation rates, but
high-resolution simulations are required to limit the effect of rotational
quenching of convection at rotation rates above twenty times the solar
value.
  }


  \maketitle


\section{Introduction}

Large-scale magnetic fields in late-type stars are thought to be
maintained by a dynamo process within or just below the convection
zone \citep[e.g.][]{O03,2013SAAS...39..187C}.
In the relatively slowly rotating and magnetically inactive Sun,
the dynamo process is often described by a
classical $\alpha\Omega$ dynamo, where shearing due to differential
rotation produces toroidal magnetic field from a poloidal one
($\Omega$ effect), and
cyclonic convection ($\alpha$ effect) is responsible for
regenerating the poloidal
field \citep{Pa55a}.
Younger late-type stars rotate much faster than the Sun
and they also exhibit more vigorous magnetic activity.
Theoretical models have long indicated that the differential
    rotation stays roughly constant as a function of rotation
    \citep[e.g.][]{KR99}. The interpretation of observational data is
    much more challenging. Recent studies show
either a mild decrease \citep[e.g.][]{Lehtinen16},
or a mild increase \citep{RRB13,RG15,Distefano2017} of the relative latitudinal
differential rotation,
indicating a broad agreement with the theoretical expectation.
Therefore, the main effect of increased rotation is a relative dominance of the $\alpha$
effect compared with differential rotation in maintaining the
toroidal field.
Hence, in view of dynamo theory \cite[e.g.][]{KR80}, dynamos in rapidly
rotating stars operate in a regime where
dynamo action is nearly fully maintained by cyclonic convection
($\alpha^2$ dynamo).
Since the early theoretical work it has been known that in the rapid
rotation regime the $\alpha$ effect becomes increasingly anisotropic
\citep{Rue78}. 
An indication of this has been seen
in wedges at moderate rotation rates \citep{WRKKB16}.
Such an anisotropic $\alpha$ effect can promote nonaxisymmetric large-scale
magnetic field configurations
\citep[e.g.][]{RWBMT90,MB95,MBBT95,2017MNRAS.466.3007P}.

Solar and stellar dynamos tend to manifest themselves very differently in
observations. The solar magnetic field exhibits cyclic behavior,
where the activity indicators vary over an approximate 11 year cycle,
and during each activity cycle the polarity of the field reverses, resulting in
a magnetic cycle of roughly 22 years. During one activity cycle, the
location in which sunspots appear migrates from mid-latitudes
towards the equator. This is commonly thought to trace the latitudinal
dynamo wave, that is, a predominantly toroidal component of the large-scale magnetic
field that
migrates towards the equator.
In the Sun, the longitudinal distribution of
sunspots indicates that the solar large-scale magnetic field is mostly
axisymmetric \citep[e.g.][]{PBKT06}.
In late-type stars with rapid rotation, by contrast,
much larger spots located at high latitudes or even polar regions
have been observed using Doppler imaging (DI), Zeeman Doppler imaging (ZDI),
and interferometry \citep[e.g.][]{Jarvinen2008, Hackman2016, Roettenbacher2016}.
Many studies have reported highly nonaxisymmetric configurations
of the spots \citep[e.g.][]{Jetsu1996, BT98}, referred to as active longitudes,
while especially the indirect imaging of the surface magnetic field using ZDI
tends to yield more axisymmetric configurations
\citep{Rosen2016, 2016MNRAS.462.4442S}.

Especially interesting are the recent results by \citet{Lehtinen16} regarding
a sample of solar-like stars, obtained by analyzing photometric light curves, 
that show a rather sharp transition from
stars with magnetic cycles and no active longitudes to stars
with both cycles and active longitudes as the activity level or rotation rate
of the stars increases.
This result can be interpreted in terms of rapid rotators hosting
nonaxisymmetric dynamos and moderate rotators axisymmetric ones.
Furthermore, some studies have reported cyclic behavior related to the
active longitudes,
in the form of the activity periodically switching from one longitude
to the other on the same hemisphere \citep{Berdyugina2002} in an abrupt flip-flop event
\citep{JPT93}.
Other studies report irregular polarity changes between the two
longitudes, not necessarily connected to the overall cyclic
variability of the star \citep{Hackman2013, Olspert15}.

The stellar cycles remain poorly characterized, however.
Nevertheless,
it is clear that many late-type stars exhibit time variability that appears
cyclic.
This is especially well manifested by studies of stellar samples,
such as the intensively investigated Mount Wilson chromospheric activity data base
\citep{Baliunas1995, Olah2016, BMJRCMPW18,OLKPG17}.
Even if the range of periods that can be studied is severely limited by the nature of
the data,
it being too short to study long cycles while the rotational and seasonal
time scales limit the periods at the short end,
it is clear that stellar cycles
are common, and even multiple superimposed ones can occur in one and the same object
\citep{Olah2009, Lehtinen16}.
There are also indications that the stars tend to cluster into distinct
activity branches in a diagram where
the ratio of the cycle period over rotation period is plotted against the rotation
rate or activity level \citep{SB99, Lehtinen16, Brandenburg2017},
but the existence of these branches continues to raise debate
\citep{Reinhold2017, Distefano2017,BMJRCMPW18,OLKPG17}.

The steadily increasing computational resources have enabled the
large-scale use of self-consistent three-dimensional convection
simulations to study the mechanisms that drive dynamo action in stars.
Recent three-dimensional numerical simulations have been
successful in reproducing many aspects of the solar dynamo, such as
cyclic magnetic activity and equatorward migration
\citep[e.g.][]{GCS10,KMB12,ABMT15}, the existence of multiple dynamo
modes \citep{KKOBWKP16,BSCC16}, and irregular behavior \citep{ABMT15,KKOBWKP16,KKOWB16}.
There are also studies that investigate the dependence of the
dynamo solutions on rotation rate, but these have so far
either been limited to wedges with limited longitudinal extent
\citep{KMCWB13, KKOWB16,WKKB16,W17} or the
range of
rotation rates investigated have been restricted to narrow regions in the vicinity
of the solar rotation rate \citep{SBCBN17} or three times the solar rotation
rate \citep{NBBMT13}.
The first indications of stellar dynamos changing from axisymmetric
to nonaxisymmetric were reported by \cite{KMCWB13}, \cite{CKMB14}, and
\cite{YGCR15}, occurring
in the regime of moderate rotation. However, the parameter ranges were rather
limited in these studies.

Planetary dynamo simulations \citep[e.g.][]{IK00,SPD12,GDW12}, which typically
have much lower density stratification than their stellar counterparts, show
that a transition from multipolar to dipolar magnetic field configurations
exists at sufficiently rapid rotation. 
Dipolar solutions have also been found in models with
high density stratification and low Prandtl number
\citep{Jones2014,YCMGRPW15,2016arXiv161202870D}.
Simulations of fully convective stars also favored the occurrence of
dipolar solutions \citep{DSB06}, but with a transition to multipolar
solutions at slower rotation \citep{2008ApJ...676.1262B}.

Intriguingly, some ZDI studies suggest that both weak multipolar and strong
dipolar magnetic field configurations can occur with very similar stellar parameters
in rapidly rotating low-mass (spectral type M) 
stars \citep[e.g.,][]{Morin2010, SHCWIJJMMR11}.
These observations challenge the simple picture of the M-star
dynamos being of classical $\alpha^2$ type.
Namely, in this case the theoretical expectation is
that, because of the Coriolis number being large due to long convective turnover
times, the $\alpha$ effect becomes strongly anisotropic and results in nonaxisymmetric 
(multipolar) fields \citep[e.g.][]{RWBMT90,MB95,MBBT95,2017MNRAS.466.3007P}.
Numerical simulations
have revealed bistable dynamo solutions in the rapid rotation regime
where both configurations can be found with the same system parameters but
different initial conditions \citep[e.g.][]{SPD12,GDW12}. However, the dipolar
solution is typically realized only with a strong initial field.

The goal of the present paper is to carry out a systematic survey of
convective dynamo simulations trying to understand the change of magnetic field
generation from a young rapidly rotating Sun to its present rotation
rate.
We are specifically studying the transition of the dynamo solutions
from axisymmetric to nonaxisymmetric ones.

\section{Model and setup} \label{sect:model}

We use spherical polar coordinates ($r$,$\theta$,$\phi$) to model the
magnetohydrodynamics (MHD) in convective envelopes of solar-like stars.
The general model and setup are detailed in
\cite{KMCWB13}.
For most of the runs we use the full azimuthal extent ($0 \leq \phi
\leq 2 \pi $). However, for some runs we consider only a quarter of the full range
($0 \leq \phi \leq \pi/2$), which we call $\pi/2$ wedges for short.
We omit the poles and thus model the star between $\pm75^\circ$ latitude
($\theta_0 \leq \theta \leq \pi-\theta_0$, with $\theta_0=15^\circ$),
and model only the convection zone of the star in radius
($0.7 R \leq r \leq R$, where $R$ is the radius of the star).

\subsection{Basic equations}

We solve the compressible MHD equations
\begin{equation}
\frac{\pd \bm A}{\pd t} = {\bm u}\times{\bm B} - \mu_0\eta {\bm J},
\end{equation}
\begin{equation}
\frac{D \ln \rho}{Dt} = -\bm\nabla\cdot\bm{u},
\end{equation}
\begin{equation}
\frac{D\bm{u}}{Dt} = \bm{g} -2\bm\Omega_0\times\bm{u}+\frac{1}{\rho}
\left(\bm{J}\times\bm{B}-\bm\nabla p
+\bm\nabla \cdot 2\nu\rho\bm{\mathsf{S}}\right),
\end{equation}
\begin{equation}
T\frac{D s}{Dt} = \frac{1}{\rho}\left[-\bm\nabla \cdot
\left({\bm F^{\rm rad}}+ {\bm F^{\rm SGS}}\right) +
\mu_0 \eta {\bm J}^2\right] +2\nu \bm{\mathsf{S}}^2,
\label{equ:ss}
\end{equation}
where ${\bm A}$ is the magnetic vector potential, $\bm{u}$ is the velocity,
$D/Dt = \pd/\pd t + \bm{u} \cdot \bm\nabla$ is the
Lagrangian time derivative, ${\bm B} =\bm\nabla\times{\bm A}$ is the magnetic field,
${\bm J} =\mu_0^{-1}\bm\nabla\times{\bm B}$ is the current density,
$\mu_0$ and  $\rho$ are the vacuum
permeability and the plasma density, respectively,
$\nu$ and $\eta$ are the constant kinematic viscosity and magnetic
diffusivity, respectively,
$\gggg=-GM{\bm r}/r^ 3$ is the gravitational acceleration where $G$ is
the gravitational constant and $M$ is the mass of the star,
$\OO_0=\Omega_0(\cos\theta,-\sin\theta,0)$ is the rotation vector, where
$\Omega_0$ is the rotation rate of the frame of reference,
$\bm{\mathsf{S}}$ is the rate-of-strain tensor,
$s$ is the specific entropy and equations above are solved together with an equation of state for
the pressure $p$, assuming an ideal gas $p=(\gamma -1)\rho e$, where
$e=c_VT$ is the internal energy, $T$ is the temperature, and
$\gamma=c_P/c_V$ is the ratio of specific heats at constant
pressure and volume, respectively.
The radiative and sub-grid-scale (SGS) heat fluxes are given by $\bm F^{\rm rad}=-K\nab T$
and $\bm F^{\rm SGS}=-\chiSGS\rho T\nab s$, respectively,
where $K$ is the radiative heat conductivity and $\chiSGS$ is the SGS
heat diffusivity.

\begin{table*}[t!]
\centering
\caption[]{Summary of the runs.}
       \label{tab:runs}
      $$
          \begin{array}{p{0.025\linewidth}crrcrcccrcrrrcccc}
            \hline
            \hline
Run & \rm Grid &\tilde\Omega & \Pra & \PraSGS & \Pm & \Ta  & \Ra & \Rey & \Pe & \rm Re_M & \Co & \Gamma_\rho &\mbox{$\Delta t$} & \Delta_\Omega^{(r)} & \Delta_\Omega^{(\theta)} & |\Delta \Omega_r | & { |\Delta \Omega_{\theta} |} \\
\hline
A1& 144 \times 288 \times 576 &  1.0 & 58 &  2.50 &  1.00 &  6.32 \cdot 10^{6} & 6.54 \cdot 10^{7}  &  40 & 100 &  40 &  1.6 &  22 & 22 & -0.26 & -0.37 & 0.26 & 0.37 \\ 
A2 & 144 \times 288 \times 576 &  1.0 & 69 &  0.30 &  1.00 &  4.39 \cdot 10^{6} & 8.00 \cdot 10^{5}  &  36 &  10 &  36 &  1.4 & 21 & 23 & -0.22 & -0.24 & 0.22 & 0.24 \\ 
B & 144 \times 288 \times 576  &  1.5 & 58 &  2.50 &  1.00 &  1.42 \cdot 10^{7} &  6.54\cdot 10^{7} &  40 & 100 &  40 &  2.4 &  22 & 32 & -0.11 & -0.17 & 0.22 & 0.24 \\ 
C1& 144 \times 288 \times 576   &  1.8 & 58 &  2.50 &  1.00 &  2.03 \cdot 10^{7} & 6.54\cdot 10^{7}  &  41 & 102 &  41 &  2.8 &  22 & 26 & -0.08 & -0.11& 0.14 & 0.20 \\ 
C2 & 144 \times 288 \times 576  &  1.8 & 58 &  1.00 &  1.00 &  2.03 \cdot 10^{7} & 1.29\cdot 10^{7}  &  43 &  43 &  43 &  2.6 &  22 & 45 & 0.78 & -0.35 & 0.13 & 0.17 \\ 
\hline
C3 & 144 \times 288 \times 576  &  1.8 & 77 &  0.33 &  1.00 &  1.14 \cdot 10^{7} & 7.00\cdot 10^{5}  &  28 &   9 &  28 &  3.0 &  20 & 88 &  0.07 &  0.17 & 0.12 & 0.30 \\ 
D & 128 \times 256 \times 512 &  2.1 & 67 &  3.00 &  1.00 &  2.03 \cdot 10^{7} & 4.55\cdot 10^{7} &  32 &  98 &  32 &  3.5 &  26 & 29 & 0.003 &  0.007 & 0.008 & 0.01 \\ 
E  & 128 \times 256 \times 512 &  2.9 & 78 &  3.50 &  1.00 &  2.64 \cdot 10^{7} & 3.11\cdot 10^{7}  &  25 &  90 &  25 &  5.0 &  24 & 87 & 0.06 &  0.06 & 0.18 & 0.17 \\ 
F1& 128 \times 256 \times 512 &  4.3 & 66 &  3.00 &  1.00 &  8.10 \cdot 10^{7} & 3.31\cdot 10^{7} &  28 &  86 &  28 &  7.9 &  23 & 33 & 0.01 &  0.04 & 0.06 & 0.15 \\ 
F2  & 144 \times 288 \times 576  & 4.3 & 57 &  1.00 &  1.00 &  1.17 \cdot 10^{8} &  1.29\cdot 10^{7} &  33 &  33 &  33 &  8.3 &  19 & 37 & 0.03 &  0.06 & 0.13 & 0.24 \\ 
F3& 144 \times 288 \times 576 &  4.3 & 58 &  0.25 &  1.00 &  1.17 \cdot 10^{8} & 9.00\cdot 10^{5}  &  27 &   6 &  27 &  9.8 &  18 & 49 & 0.02 &  0.07 & 0.10 & 0.28 \\ 
G$^{a}$ & 256 \times 512 \times 1024  &  4.9 & 43 &  1.20 &  1.00 &  3.47\cdot 10^8 & 4.55\cdot 10^7  &  50 &  61 &  50 &  9.3 &  21 & 37 & 0.03 &  0.04 & 0.03 & 0.10 \\ 
G$^{W}$ & 180 \times 256 \times 128   & 4.8 & 67 &  2.00 &  1.00 &  1.25\cdot 10^8 & 4.00 \cdot 10^7  &  34 &  68 &  34 &  8.3 &  31 & 22 & 0.05 & 0.07 & 0.21 & 0.29 \\ 
H & 128 \times 256 \times 512   &  7.1 & 69 &  3.00 &  1.00 &  2.25 \cdot 10^{8} &  2.04\cdot 10^{7} &  24 &  72 &  24 & 15.6 &  21 & 200 & 0.01 &  0.03 & 0.10 & 0.20 \\ 
H$^{a}$ & 256 \times 512 \times 1024  &  7.8 & 51 &  1.40 &  1.00 &  6.61\cdot 10^8 & 5.21 \cdot 10^7  &  40 &  56 &  40 & 16.1 &  18 & 36 & 0.004  &  0.014 & 0.03 & 0.10 \\ 
I & 128 \times 256 \times 512   &  9.6 & 71 &  2.08 &  1.04 &  4.63\cdot 10^8 & 3.93\cdot 10^{7}  &  26 &  55 &  27 & 20.4 &  28 & 52 & 0.01 &  0.03 & 0.11 & 0.23 \\ 
I$^{W}$ & 128 \times 256 \times 128  & 9.6 & 71 &  2.08 &  1.04 &  4.63\cdot 10^8 & 3.83\cdot 10^7 &  27 &  56 &  28 & 19.9 & 28 & 20 & 0.01 &  0.02 & 0.11 & 0.20 \\ 
J & 128 \times 256 \times 512  & 14.5 & 62 &  2.50 &  1.00 &  1.30 \cdot 10^9 & 1.12 \cdot 10^7  &  25 &  63 &  25 & 36.1 &  18 & 62 & -0.001 &  0.01 & 0.01 & 0.14 \\ 
J$^{W}$ & 180 \times 256 \times 128  & 15.5 & 69 &  2.00 &  1.00 &  1.30\cdot 10^9 &3.93\cdot 10^7  &  21 &  43 &  21 & 41.7 &  26 & 53 & 0.004  &  0.009 & 0.05 & 0.13 \\ 
K1 & 128 \times 256 \times 512  &21.4 & 74 &  3.00 &  1.00 &  2.03\cdot 10^{9} &  1.00\cdot 10^7 &  16 &  50 &  16 & 67.5 & 13 & 18 & -0.001 &   0.007 & 0.03 & 0.15 \\
K2 & 128 \times 256 \times 512  & 21.4 & 55 &  2.25 &  1.00 &  3.60\cdot 10^{9} & 1.56\cdot 10^7 &  21 &  48 &  21 & 71.2 &  13 & 18 & -0.001 &  0.005 & 0.03 & 0.11 \\
L$^{a}$& 256 \times 512 \times 1024  & 23.3 &60 &1.60 &1.00 &4.6 \cdot 10^9 & 4.58 \cdot 10^7 & 21 &32 &21 &83.4 &15 & 51 & 1\cdot 10^{-4}  & 0.002 & 0.003 & 0.04 \\
L$^{W}$ &180 \times 256 \times 128 & 23.3 & 70 &  2.00 &  1.00 &  2.92 \cdot 10^9 &  4.00\cdot10^7 &  16 &  33 &  16 & 82.4 &  24 & 53 & -1\cdot 10^{-4}  &  0.003 & 0.002 & 0.07 \\ 
M & 128 \times 256 \times 512 & 28.5 & 61 &  2.50 &  1.00 &  5.18\cdot 10^9 & 6.00\cdot10^6  &  18 &  46 &  18 & 98.7 & 9 & 24 & -0.001  & 0.003 & 0.02 & 0.10 \\ 
M$^{a}$ & 256 \times 512 \times 1024 & 28.5 & 31 &  2.50 &  1.00 &  2.07\cdot 10^{10} & 1.48 \cdot 10^5  &  33 &  82 &  33 &109.9 &  15 & 33 & -7\cdot 10^{-5} &   9\cdot 10^{-4} & 0.002 & 0.03 \\
M$^{W}$ & 180 \times 256 \times 128  & 31.0 & 71 &  2.00 &  1.00 &  5.18\cdot 10^9 &  1.03\cdot 10^8 & 14 & 28 & 14 &126.5 & 21 & 49  &  -1\cdot 10^{-4}  &  0.002 & 0.003 & 0.07 \\ 
\hline
        \end{array}
          $$
\tablefoot{
The quantities in the second to eighth column are input
  parameters of the runs whereas the quantities from the ninth to the
  eighteenth
column are outcomes of the simulations. 
  Superscripts ${a}$ denote high-resolution runs
  and superscripts ${W}$ denote $\pi/2$ wedges.
  The horizontal line denotes the transition from axisymmetric (antisolar) magnetic field (rotation profile)
  to a nonaxisymmetric (solar) one. 
$\Delta t$ indicates the time span of the saturated stage in years.
$\Delta_\Omega^{(r)}$ and $\Delta_\Omega^{(\theta)}$ indicate the
relative radial and latitudinal differential rotation, see \Eq{eq:rel_diff}, whereas 
$|\Delta \Omega_r |$ and $ {|\Delta \Omega_{\theta} |}$ are the
absolute radial and latitudinal differential rotation;
see \Eq{eq:abs_diff}.
}
\end{table*}

\subsection{Setup characteristics}
\label{sec:initcond}
The initial stratification is isentropic, where the hydrostatic
temperature gradient is defined via an adiabatic polytropic index of $n_{\rm
  ad}=1.5$.
We initialize the magnetic field with a weak white-noise Gaussian
seed field.
More details about our initial setup can be found in \cite{KMCWB13}.

Most of our runs use a grid covering the full azimuthal extent, but we perform
some comparison runs, labelled with superscript ``W'' for $\pi/2$ wedges
with reduced longitudinal extent.
In all cases, we assume periodicity
in the azimuthal direction for all quantities.
For the magnetic field, we apply perfect conductor boundary conditions
at the bottom and both latitudinal boundaries, and at the top boundary we
use a radial field condition.
Stress-free, impenetrable boundaries are used for the velocity on all radial and latitudinal
boundaries.
The boundary condition of entropy is set by assuming a constant
radiative heat flux at the bottom of the computational domain.
The thermodynamic quantities have zero first derivatives on both latitudinal
boundaries, leading to zero energy fluxes there.
At the top boundary, the temperature follows a black body condition.
The exact equations for these conditions are described in \cite{KMCWB13,KKOWB16}.

Our simulations are defined by the following non-dimensional
parameters.
As input parameters we quote the Taylor number
\begin{equation}
\Ta=[2\Omega_0 (0.3R)^2/\nu]^2,
\end{equation}
the fluid, SGS, and magnetic Prandtl numbers
\begin{equation}
\Pra=\frac{\nu}{\chi_{\rm m}},\quad \PraSGS=\frac{\nu}{\chiSGSm},\quad \Pm=\frac{\nu}{\eta},
\end{equation}
where $\chi_{\rm m}=K(r_{\rm m})/c_P\rho(r_{\rm m})$ and
$\chiSGSm=\chiSGS(r_{\rm m})$ are evaluated at $r_{\rm m}=0.85R$.
The Rayleigh number is obtained from the hydrostatic stratification, 
evolving a 1D model, and is given by
\begin{equation}
\Ra\!=\!\frac{GM(0.3R)^4}{\nu \chiSGSm R^2}
  \bigg(-\frac{1}{c_{\rm P}}\frac{{\rm d}s_{\rm hs}}{{\rm d}r}
  \bigg)_{(r=0.85R)},
\label{equ:Co}
\end{equation}
where $s_{\rm hs}$ is the hydrostatic entropy.

Useful diagnostic parameters are the density contrast
\begin{equation}
\Gamma_\rho\equiv\rho(r=0.7R)/\rho(R),
\end{equation}
fluid and magnetic Reynolds numbers
and the P\'eclet number,
\begin{equation}
\Rey=\frac{\urms}{\nu \kf},\quad \ReM=\frac{\urms}{\eta \kf},\quad
\Pe=\frac{\urms}{\chiSGSm \kf},
\end{equation}
where $\kf=2\pi/0.3R\approx21/R$ is an estimate of the wavenumber of
the largest eddies.
The Coriolis number is defined as
\begin{equation}
\Co=\frac{2\Omega_0}{\urms \kf},
\label{CoDef}
\end{equation}
where $\urms=\sqrt{(3/2)\brac{u_r^2+u_\theta^2}_{r\theta\phi t}}$ is
the rms velocity and the subscripts indicate averaging over $r$,
$\theta$,
$\phi$, and a time interval during which the run is thermally
relaxed and which typically covers at least 
one magnetic diffusion time.

We define mean quantities as averages over the $\phi$-coordinate and denote
them by an overbar, for example $\brac{\BBB}_\phi=\meanv{B}$. The difference
between the total and the mean, for example $\BBB^\prime=\BBB-\meanv{B}$, are
the fluctuations.
Furthermore, we indicate volume averages using $\brac{\cdot}_V$.

For the purpose of this paper, it is convenient to normalize the
rotation rate by the solar value, so we define
\begin{equation}
\tilde\Omega\equiv\Omega_0/\Omega_\odot,
\label{OmegaTildeDef}
\end{equation}
where $\Omega_\odot$ is the solar rotation rate.
Moreover, we use $\Omega_\odot=2.7\times10^{-6} \s^{-1}$, the solar
radius $R=7\times10^{8} \m$, $\rho(0.7R)=200 \kg/\m^3$, and
$\mu_0=4\pi\cdot10^{-7}$~H~m$^{-1}$ to normalize
our quantities to physical units.

The simulations were performed 
  using the {\sc Pencil
  Code}\footnote{\url{https://github.com/pencil-code/}}.
The code employs a
high-order finite difference method for solving the compressible
equations of magnetohydrodynamics.

\begin{figure*}[t]
\centering
\includegraphics[width=\columnwidth]{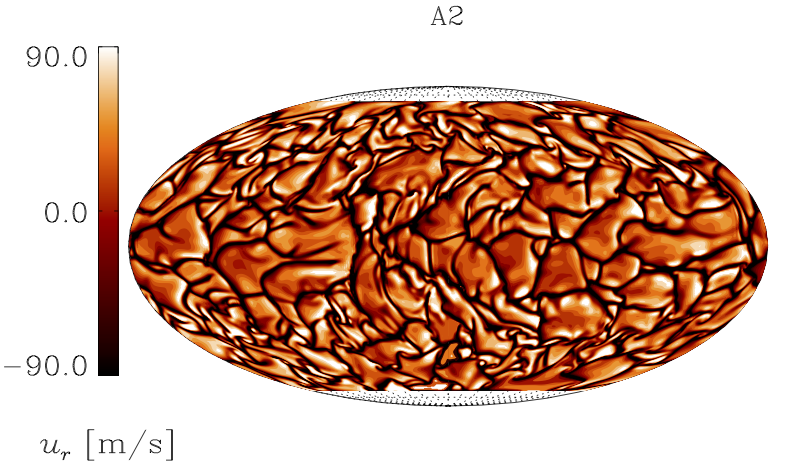} 
\includegraphics[width=\columnwidth]{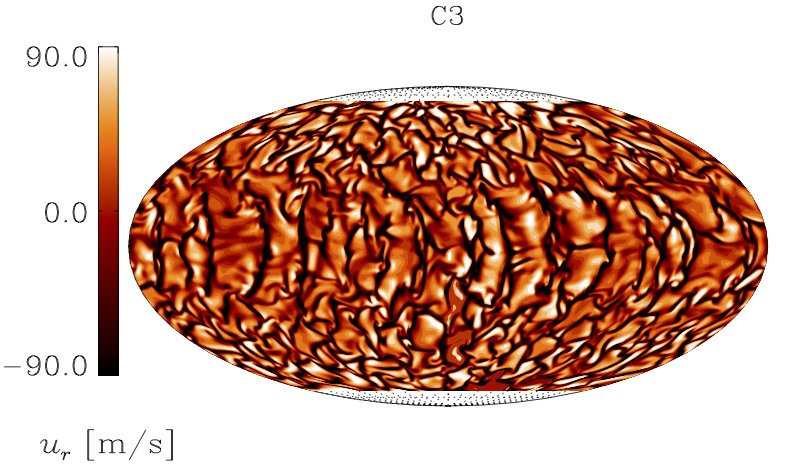}  
\includegraphics[width=\columnwidth]{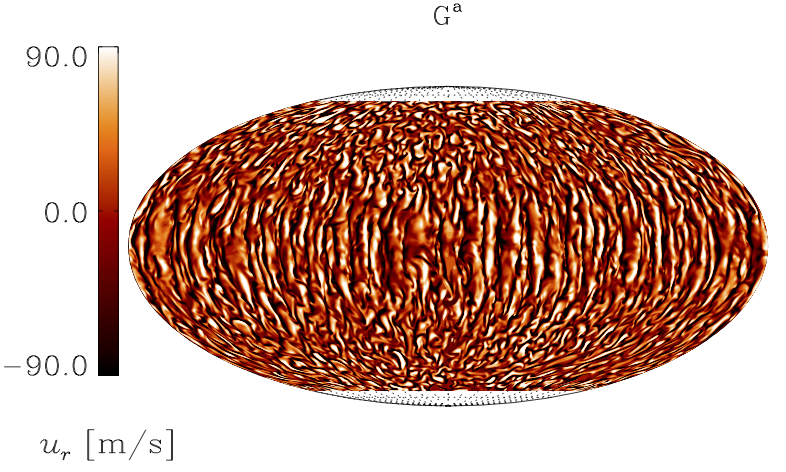} 
\includegraphics[width=\columnwidth]{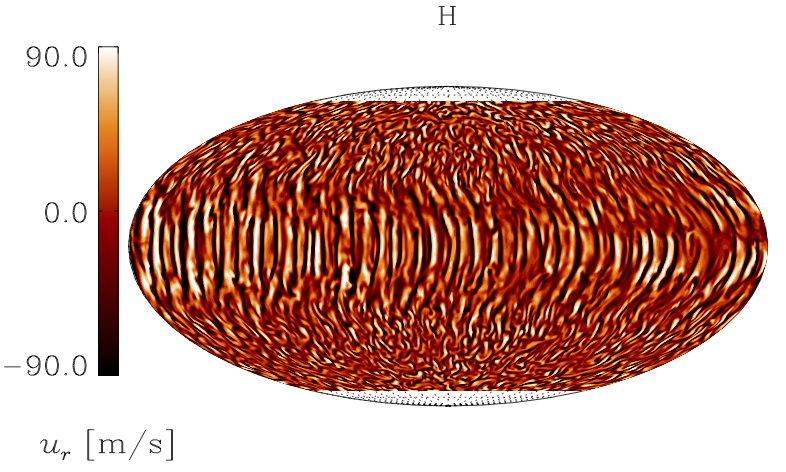} 
\includegraphics[width=\columnwidth]{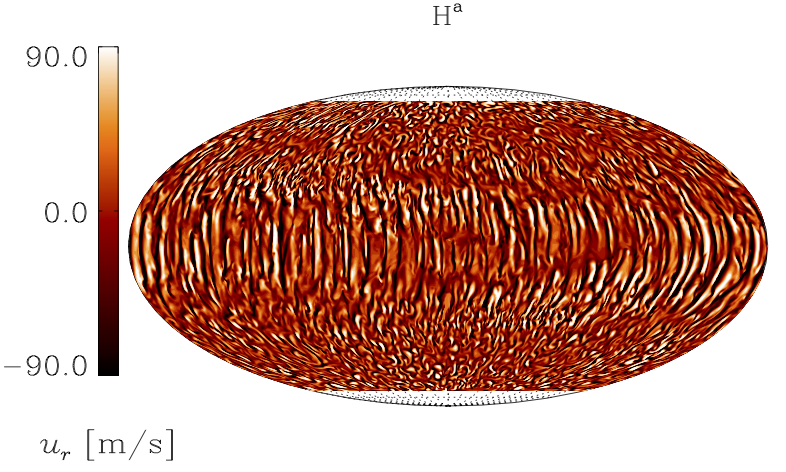} 
\includegraphics[width=\columnwidth]{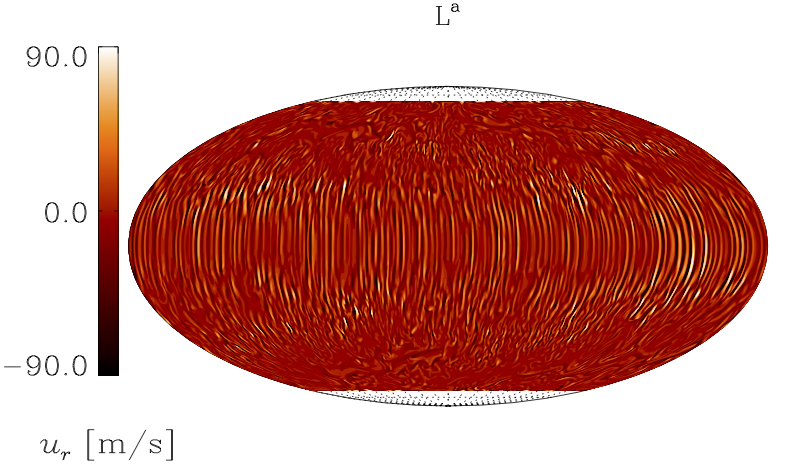} 
\caption{
Mollweide projection of radial velocity $u_r$ at $r=0.98R$ for Runs~A2, C3, G$^a$,
H, H$^a$, and L$^a$.
}\label{fig:pmoll_uu1}
\end{figure*}

\section{Results}
\label{sec:results}

We consider a number of runs that probe the rotational dependence
in the range $\tilde\Omega$=$1$--$31$, corresponding to
$\Co$=$1.6$--$126.5$; see
\Table{tab:runs}. 
The range in $\Co$ is larger than that in $\tilde\Omega$, because
faster rotation leads to lower supercriticality of convection,
resulting in a decreased $\urms$ and increased $\Co$; see \Eq{CoDef}.
For some rotation rates we consider different values of the SGS
Prandtl number, resulting in different Rayleigh and P\'eclet numbers
and different levels of supercriticality.
Runs~E, F1, and H are direct continuations of Runs~A, B, and C of
\cite{CKMB14} and Run~F1 was already discussed as Run~E4
\citep{KMCWB13}.
Run~G$^{\rm W}$ has been analyzed as Run~I in
\cite{WKKB14}, as Run~A1 in \cite{WKKB16}, as Run~D3 in
\cite{KKOBWKP16} and in \cite{WRKKB16}.
Furthermore, Runs~A1 and A2 correspond to $2\pi$ extensions of the
$\pi/2$ wedges of Set~F in \cite{KKOWB16}, whereas Run~E corresponds to Set~E in
\cite{KKOWB16} and Run~B1 in \cite{WKKB16}.
We also include a selection of models (Runs~A2, C3, F3) with a lower
$\PraSGS$ to compare with other studies where such parameter regimes
are explored \cite[e.g.][]{BBBMT10,NBBMT13,FF14,HRY16}.
The numerical studies were carried out over an extended period
of time, during which the setups have been continuously refined.
This, and the aim to compare to other studies, 
explains the heterogeneity in the choice of parameters.
The physical run time of the saturated stage is denoted by $\Delta t$.

\subsection{Overview of convective states}

All our models have a density stratification that is much smaller than in
the Sun.
Therefore, the effects of small-scale convection near the surface and
the
resulting low local Coriolis numbers in those layers are not captured.
This can be achieved only at very high
resolution \cite[e.g.,][]{HRY14} and is not feasible for parameter
studies such as those done here.
Thus, the effects of rotation are more strongly
imprinted in the velocity field near the surfaces of our models
than what is expected in actual stars.
This is manifested in \Fig{fig:pmoll_uu1} where the radial velocity $u_r$
is shown for several runs with increasing rotation rate.
The size of the convection cells at high latitudes decreases as the
rotation rate is increased.
Also, we observe the appearance of elongated in latitude columnar
structures near the equator at about twice the solar rotation rate.
These structures, often referred to as banana cells, persist for
all higher rotation rates investigated, their azimuthal and radial
extents reducing as function of rotation, while the latitudinal
extent remains roughly constant.
The reason for their emergence is the strong rotational influence on the flow
and the geometry of the system. Strong rotation tries to force convection into
Taylor-Proudman balance resulting in columnar cells which are aligned with the
rotation vector. Such cells are connected over the equator only outside the
tangent cylinder in a spherical shell, manifesting themselves as elongated
structures at low latitudes. Such convective modes can also lead to
equatorial acceleration as observed in the simulations and in the Sun
\citep{Busse1970}.
In the Sun, the small-scale granulation near the surface masks direct
observation of larger-scale convective modes. However, also helioseismic
results suggest that large-scale convective structures exceeding the
supergranular scale of 20--30 Mm are weak \citep[e.g,][]{HDS12}.

\begin{figure}[t]
\centering
\includegraphics[width=\columnwidth]{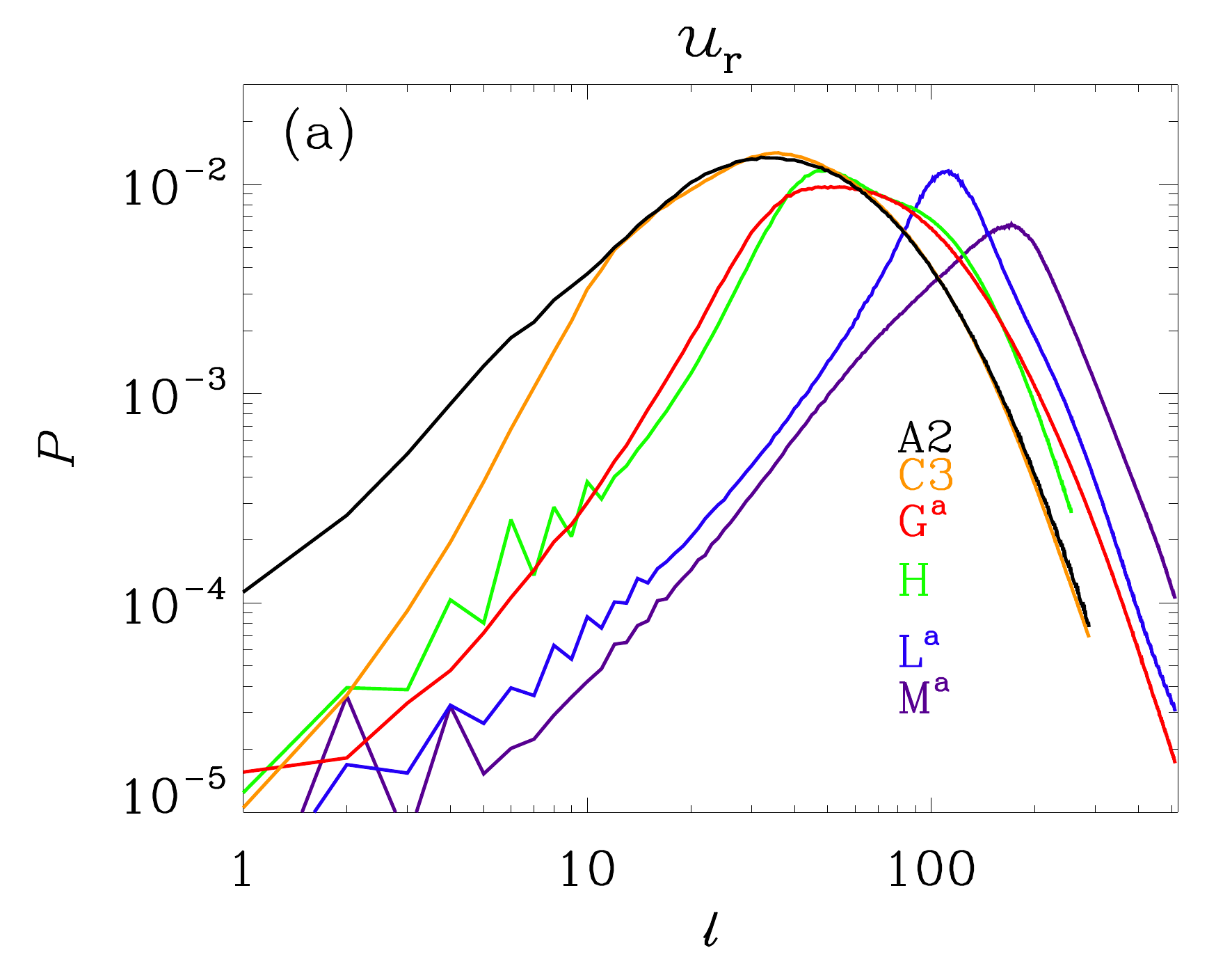} \\
\includegraphics[width=\columnwidth]{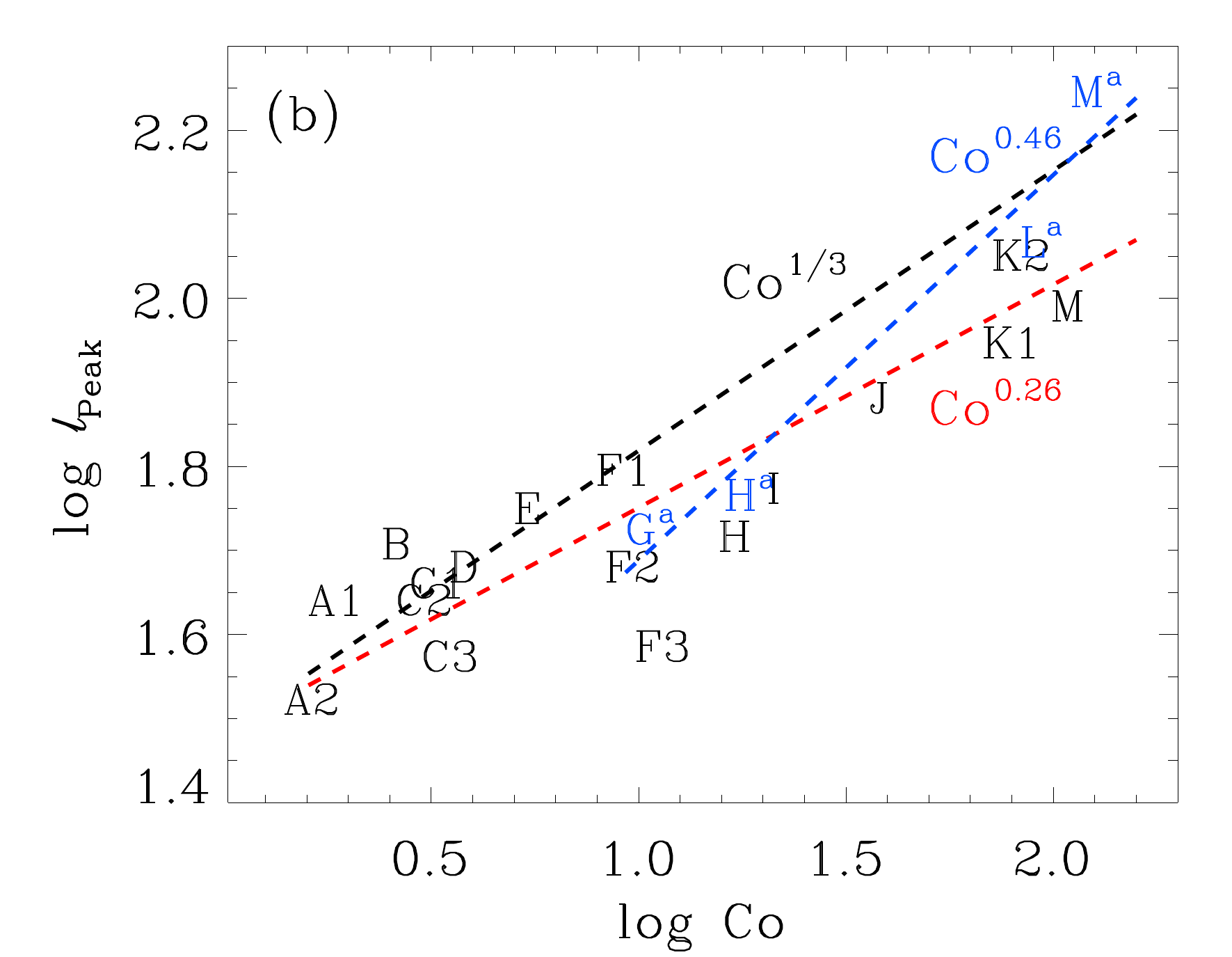} 
\caption{
(a) Normalized power spectra $P$ of the radial velocity as function of degree $\ell$
for Runs~A2, C3, G$^a$, H, L$^{a}$, and M$^{a}$ with increasing rotational influence.
(b) Degree of peak power $\ell_{\rm peak}$ estimated from the
power spectra plotted over Coriolis number $\Co$. The runs are
indicated with their run names.
The red dashed line is a power law fit
including all the runs, the blue dashed line 
is the fit for the high resolution runs, 
while the black dashed line is the expected
slope from theoretical estimates \citep{Ch61}.}
\label{fig:spectra}
\end{figure}

To quantify the size of convective structures as a function of
rotation we compute the power spectra of the radial velocity near the
surface, see \Fig{fig:spectra}.
We use a spherical harmonics decomposition to calculate the coefficients
$\hat{u}_{r}^{\ell m}$,
where $\ell,m$ are the order of the spherical harmonics and the azimuthal number, respectively.
The details on the decomposition can be found in
Appendix \ref{sec:appendixA}.
The power at each $\ell$ is
\begin{equation}
P=\frac{E_{\rm kin}^{(\ell)}}{\sum \limits_{\ell} E_{\rm kin}^{(\ell)}}, \quad
E_{\rm kin}^{(\ell)}= \sum \limits_{m=0}^\ell C_m |\hat{u}_{r}^{\ell m}|^2,
\end{equation}
where $C_m=2-\delta_{m0}$.
We find that for more rapid rotation the radial kinetic energy peaks
at smaller scales (higher $\ell$, close to $\ell=100$ for Run~L$^a$) and the kinetic energy at
large scales (lower $\ell$) becomes smaller; see \Fig{fig:spectra}a.
The increasing rotational influence is clearly seen in \Fig{fig:spectra}b, 
where we plot the value of $\ell$
at the maximum of the radial velocity spectra
as a function of the Coriolis number for all runs.
The dependence is consistent with a power law with $\rm Co^{0.26}$,
which is relatively close to the theoretically expected 1/3
scaling for rotating hydrodynamic convection near onset \citep{Ch61}.
This is shallower than the slope of about 1/2 found for the horizontal
velocity spectra in the simulations of \cite{FH16b}.
When we only consider the high-resolution runs (blue line in
\Fig{fig:spectra}b), we observe a steeper trend ($\rm Co^{0.46}$).
Especially at rapid rotation, the high-resolution
runs start deviating significantly from their low-resolution counterparts,
the scale of convection being reduced much more strongly in the former
class of runs.

To look at the energy of the radial velocity field at different values of $m$, 
we decompose it at the surface, as described in Appendix \ref{sec:appendixA}.
In \Fig{fig:naxikin} we plot the kinetic energy for $0 \leq \ell \leq 10$.
The total kinetic energy at the surface is decreasing with rotation (panel a), 
and most of the kinetic energy is contained in the small-scales (panel b, orange line). 
While the fifth nonaxisymmetric mode is mostly constant with increasing rotation (red line), 
the axisymmetric mode ($m=0$) is varying strongly, having sometimes comparable 
or even higher energy than $m=5$.

Nonaxisymmetric structures in the velocity field are also visible in \Fig{fig:pmoll_uu1} around the equator,
in particular for Run~L$^a$. This is in agreement with previous studies \citep[e.g.][]{BBBMT08},
which reported the presence of clear nonaxisymmetric large-scale flows
for hydrodynamic simulations
in parameter regimes near the onset of convection.
These localized nonaxisymmetric structures are similar to the ``relaxation oscillations,"
first seen in planetary simulations \citep[][]{Bu01}.
Those are explained by realizing that,
at intermediate Rayleigh numbers, differential rotation tends to suppress
the convective 
cells and, as a result, they localize in groups 
across longitude, leaving the rest of the azimuthal domain dominated by
the axisymmetric differential rotation.

\begin{figure}[t]
\centering
\includegraphics[width=\columnwidth]{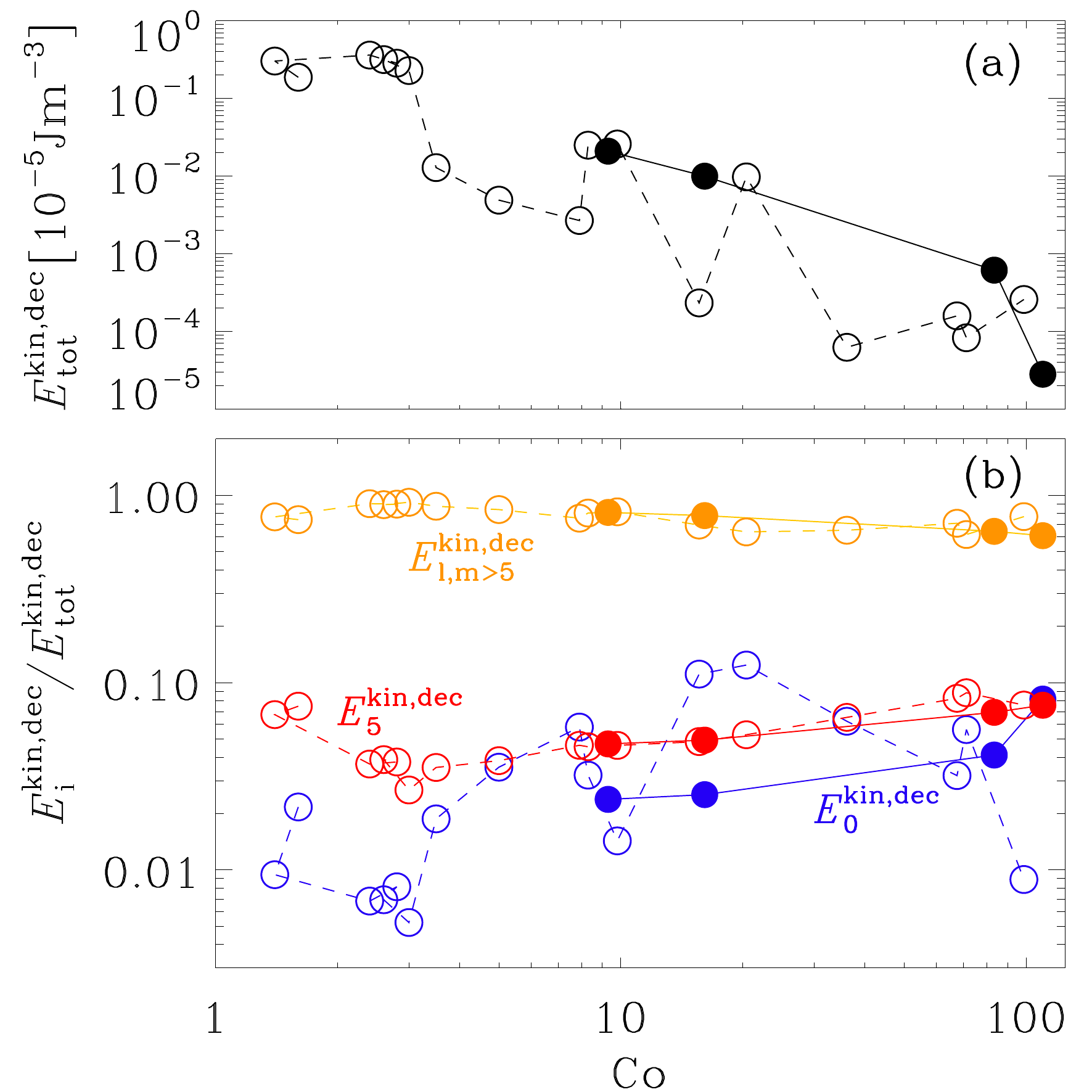} 
\caption{Kinetic energy of the decomposition as a function of Coriolis number $\Co$ for all
  $2\pi$ runs showing the total energy (a),
  axisymmetric ($m=0$, blue), fifth nonaxisymmetric mode ($m=5$, red) 
  and small-scale ($l,m > 5$, orange) contribution (b).
  All the energies in panel b are normalized to the total energy (panel a).
Filled circles connected by a continuous line indicate high resolution
runs.}
\label{fig:naxikin}
\end{figure}

\subsection{Mean flows}
\begin{table}[t!]
\centering
\caption[]{Volume averaged kinetic and magnetic energy densities in
  units of $10^5$J~m$^{-3}$.}
       \label{tab:ene}
      $$
          \begin{array}{p{0.05\linewidth}cccccccc}
            \hline
            \hline
            \noalign{\smallskip}
Run & E_{\rm kin} & E_{\rm kin}^{\rm DR} & E_{\rm kin}^{\rm MC} & E_{\rm kin}^{\rm fluc} & E_{\rm mag} & E_{\rm mag}^{\rm tor} & E_{\rm mag}^{\rm pol} & E_{\rm mag}^{\rm fluc} \\
            \hline
            A1 &  4.428 &  1.152 &  0.015 &  3.261 &  0.876 &  0.050 &  0.055 &  0.771 \\ 
            A2 &  5.055 &  0.858 &  0.015 &  4.182 &  0.995 &  0.047 &  0.055 &  0.893 \\ 
            B &  3.263 &  0.358 &  0.005 &  2.901 &  0.715 &  0.055 &  0.037 &  0.623 \\ 
            C1 &  3.153 &  0.164 &  0.003 &  2.986 &  0.504 &  0.035 &  0.026 &  0.442 \\ 
            C2 &  3.631 &  0.128 &  0.003 &  3.500 &  0.488 &  0.028 &  0.023 &  0.438 \\ 
            \hline
            C3 &  6.572 &  3.941 &  0.003 &  2.628 &  0.891 &  0.177 &  0.023 &  0.692 \\ 
            D &  3.181 &  0.873 &  0.003 &  2.305 &  0.671 &  0.042 &  0.012 &  0.617 \\ 
            E &  4.189 &  2.317 &  0.001 &  1.871 &  0.579 &  0.073 &  0.023 &  0.483 \\ 
            F1 &  2.485 &  0.842 &  0.002 &  1.642 &  1.363 &  0.166 &  0.017 &  1.181 \\ 
            F2 &  2.898 &  1.101 &  0.002 &  1.794 &  1.082 &  0.088 &  0.023 &  0.971 \\ 
            F3 &  2.700 &  1.263 &  0.001 &  1.437 &  0.767 &  0.208 &  0.018 &  0.541 \\ 
            G$^{a}$ & 2.748 & 0.820 & 0.001 & 1.926 & 0.754 & 0.076 & 0.014 & 0.664 \\ 
            G$^{W}$ &  3.506 &  1.653 &  0.003 &  1.851 &  0.986 &  0.193 &  0.132 &  0.661 \\ 
            H &  2.153 &  0.845 &  0.001 &  1.306 &  1.049 &  0.058 &  0.028 &  0.963 \\ 
            H$^{a}$ &  1.704 &  0.354 &  0.001 &  1.349 &  1.449 &  0.111 &  0.029 &  1.309  \\ 
            I & 1.706 &  0.570 &  0.001 &  1.135 &  1.361 &  0.065 &  0.036 &  1.260  \\ 
            I$^{W}$ & 1.625 &  0.483 &  0.001 &  1.141 &  1.197 &  0.247 &  0.230 &  0.720 \\
            J & 0.580 &  0.346 &  0.000 &  0.234 &  0.113 &  0.024 &  0.006 &  0.083 \\ 
            J$^{W}$ &  0.786 &  0.101 &  0.000 &  0.685 &  0.900 &  0.102 &  0.230 &  0.568  \\ 
            K1 &  2.325 &  1.624 &  0.000 &  0.701 &  0.426 &  0.216 &  0.025 &  0.185 \\ 
            K2 &  1.549 &  0.934 &  0.000 &  0.615 &  1.029 &  0.358 &  0.153 &  0.518 \\ 
            L$^{a}$ &0.708 &  0.155 &  0.000 &  0.552 &  1.928 &  0.031 &  0.018 &  1.878 \\ 
            L$^{W}$ &  0.415 &  0.023 &  0.000 &  0.391 &  1.102 &  0.129 &  0.393 &  0.580  \\ 
            M & 2.053 &  1.433 &  0.000 &  0.620 &  0.967 &  0.337 &  0.152 &  0.477 \\ 
            M$^{a}$ & 0.393 &  0.008 &  0.000 &  0.385 &  2.793 &  0.057 &  0.062 &  2.674 \\ 
            M$^{W}$ &  0.328 &  0.025 &  0.000 &  0.303 &  1.024 &  0.138 &  0.407 &  0.479  \\ 
\hline
          \end{array}
          $$
\end{table}

To estimate the rotational influence on the convection we also
calculated the volume averaged total kinetic energy density and its
contributions; see \Table{tab:ene}.
The total kinetic energy density is given by
\begin{equation}
E_{\rm kin}=\brac{\onehalf \rho \uuu^2}_V,
\end{equation}
and the contributions contained in differential rotation and
meridional circulation are, respectively:
\begin{equation}
E_{\rm kin}^{\rm DR}=\brac{\onehalf \rho \mfi{u}{\phi}^2}_V, \quad
E_{\rm kin}^{\rm MC}=\brac{\onehalf \rho(\mfi{u}{r}^2+\mfi{u}{\theta}^2)}_V.
\end{equation}
The contribution from the nonaxisymmetric flows
\begin{equation}
E_{\rm kin}^{\rm fluc}
= E_{\rm kin}-(E_{\rm kin}^{\rm DR}+E_{\rm kin}^{\rm MC}).
\end{equation}
The total kinetic energy decreases nearly monotonically as a function of
rotation.
This clearly shows the rotational quenching of convection,
which is related to an increasing critical Rayleigh number
in rapidly rotating systems.
As a result, the flow becomes less supercritical for convection the
higher the rotation rate, which is also reflected in the monotonous
decrease of the nonaxisymmetric energy that also contains the fluctuations
due to convective turbulence. The energy contained in differential
rotation and meridional circulation shows a decreasing overall trend as
function of rotation. In general, the capability of the flow to extract
energy from thermal energy is decreased by rotation.
Comparison to $\pi/2$ wedge simulations indicates some differences in the dynamics
of the flow, but it is hard to discern any systematic behavior. For
a moderate rotation Run~G, the $\pi/2$ wedge (Run~G$^{
\rm W}$) has an excess of every
type of
kinetic energy, while in the rapid rotation regime (Runs~I, J, L, M) the
flow energies have a tendency to be lower than in the corresponding runs
with full azimuthal extent.

\begin{figure}[t]
\centering
\includegraphics[width=.49\columnwidth]{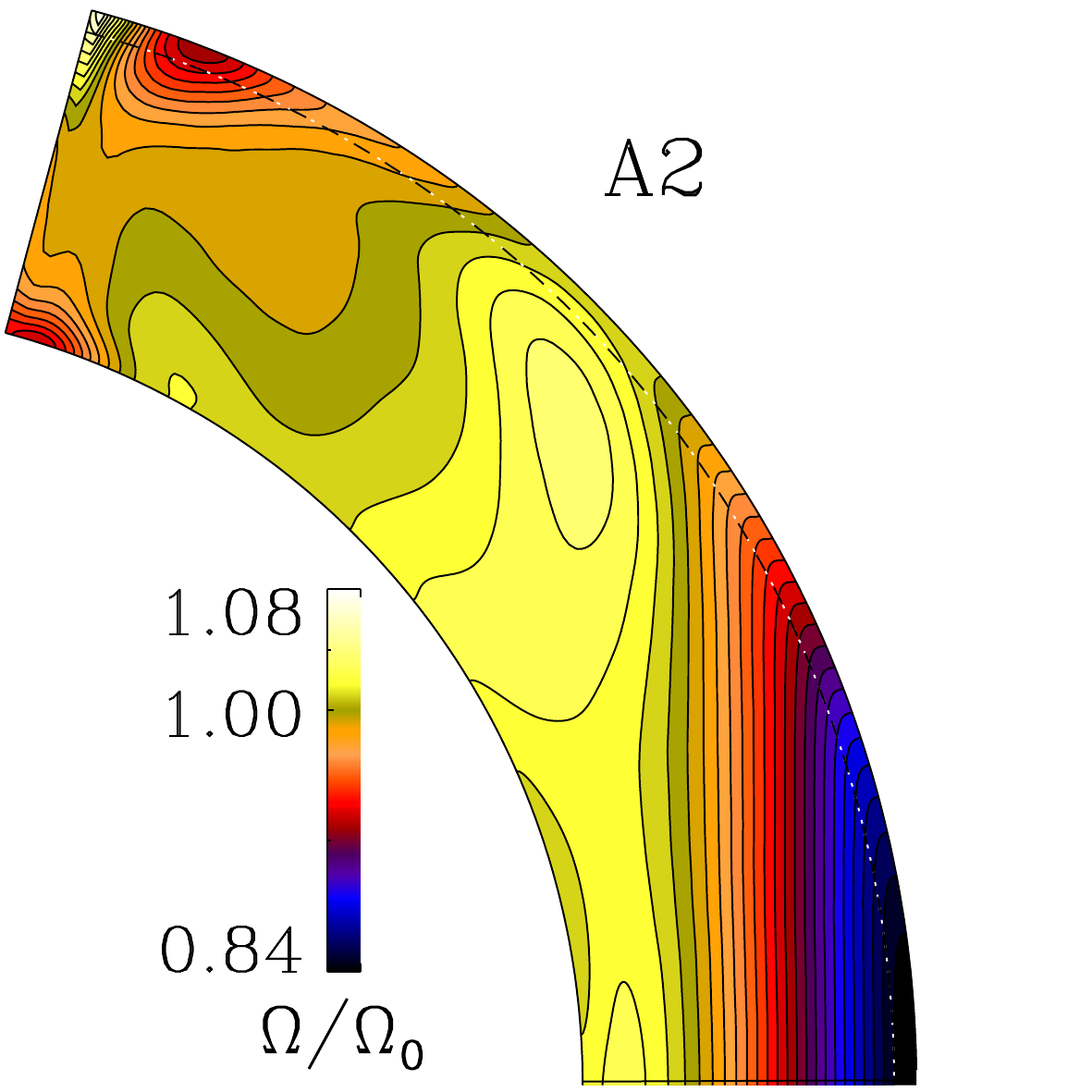}  
\includegraphics[width=.49\columnwidth]{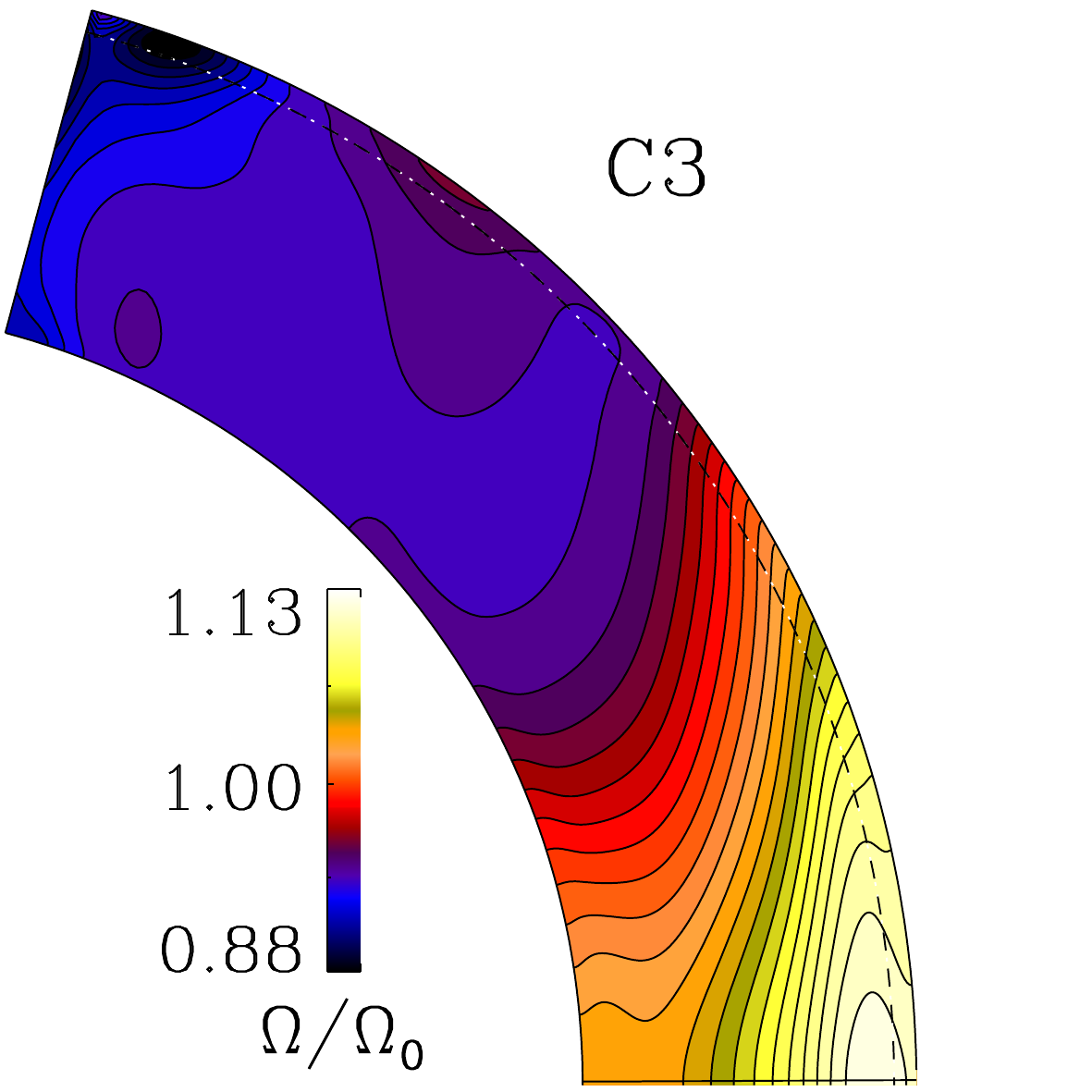}
\includegraphics[width=.49\columnwidth]{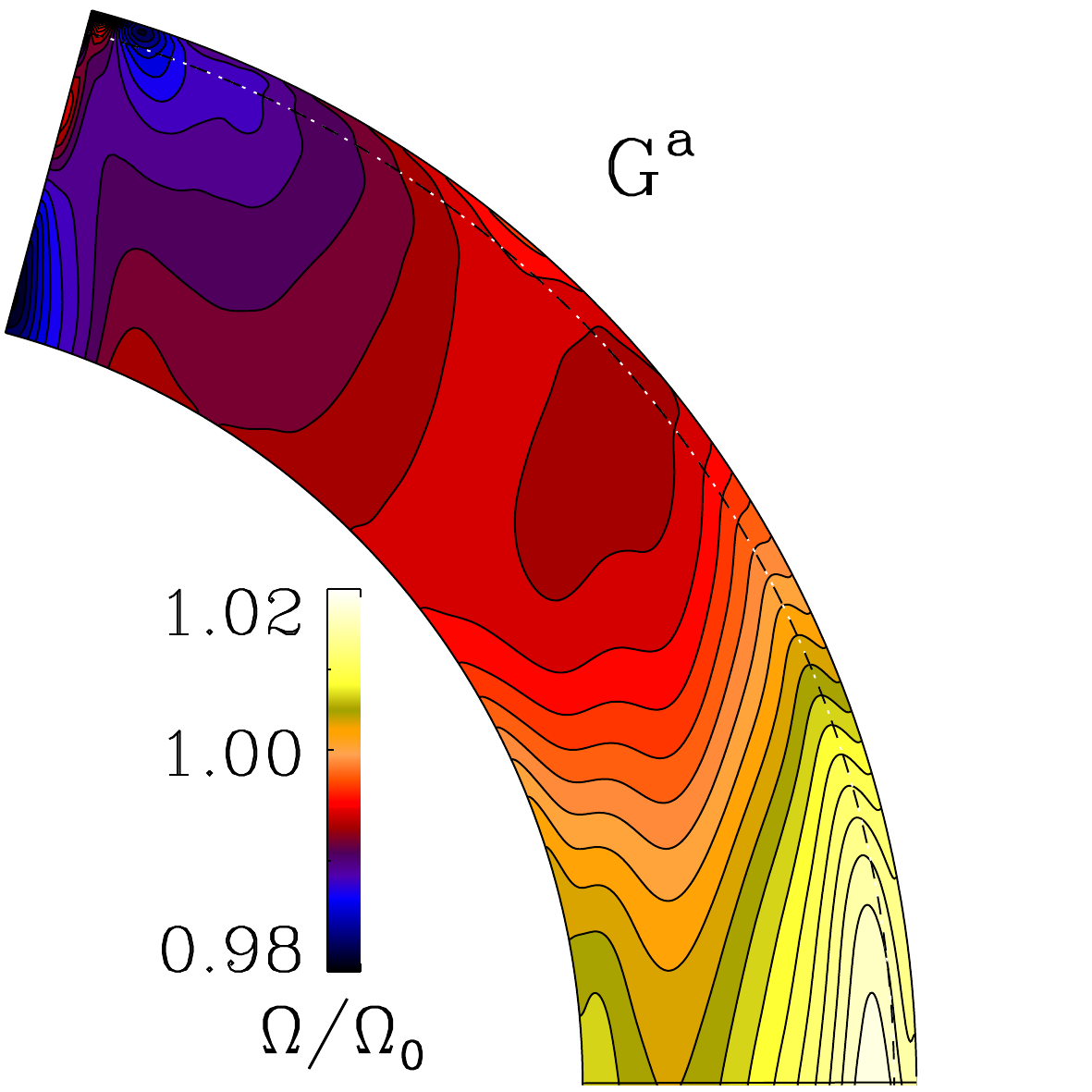}  
\includegraphics[width=.49\columnwidth]{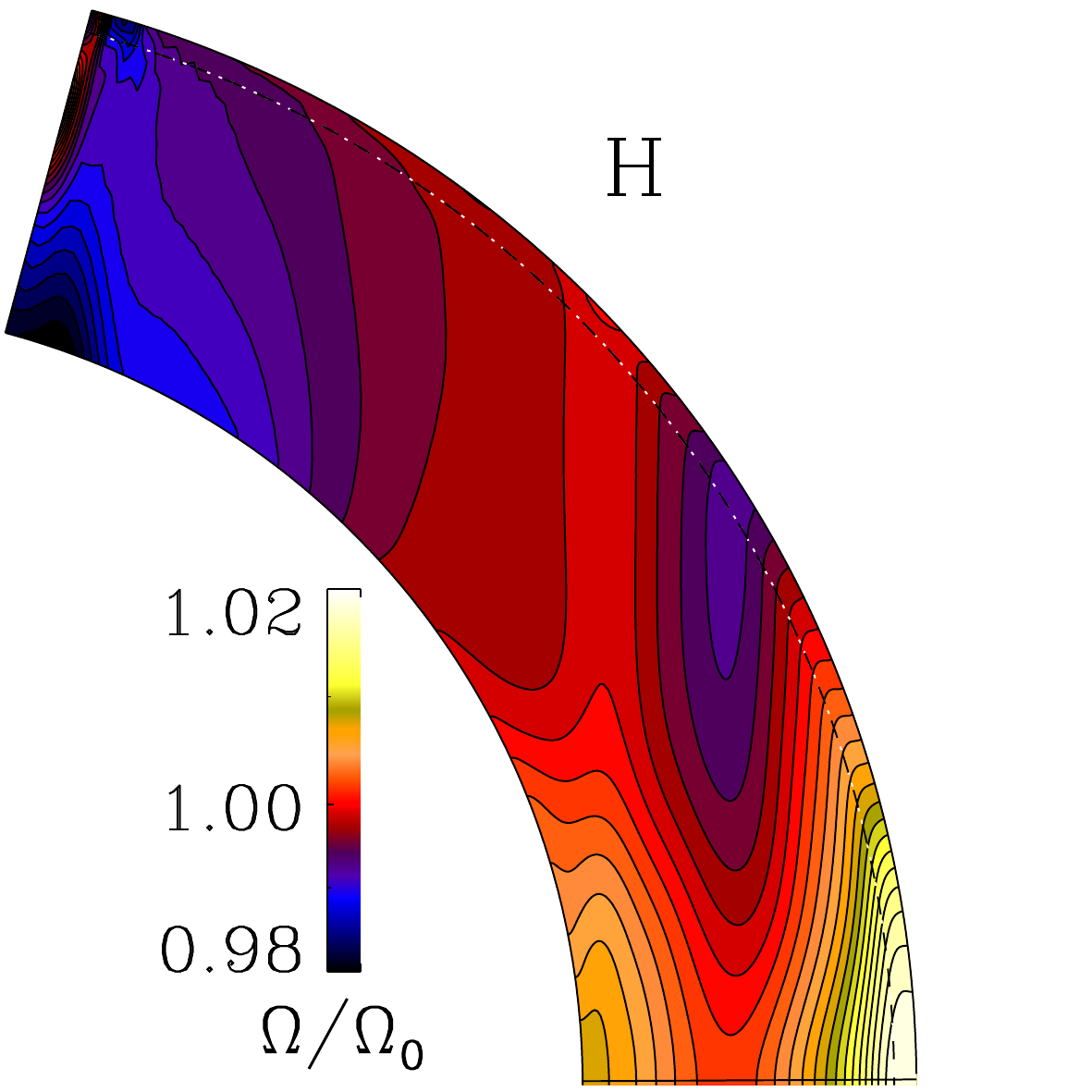}
\includegraphics[width=.49\columnwidth]{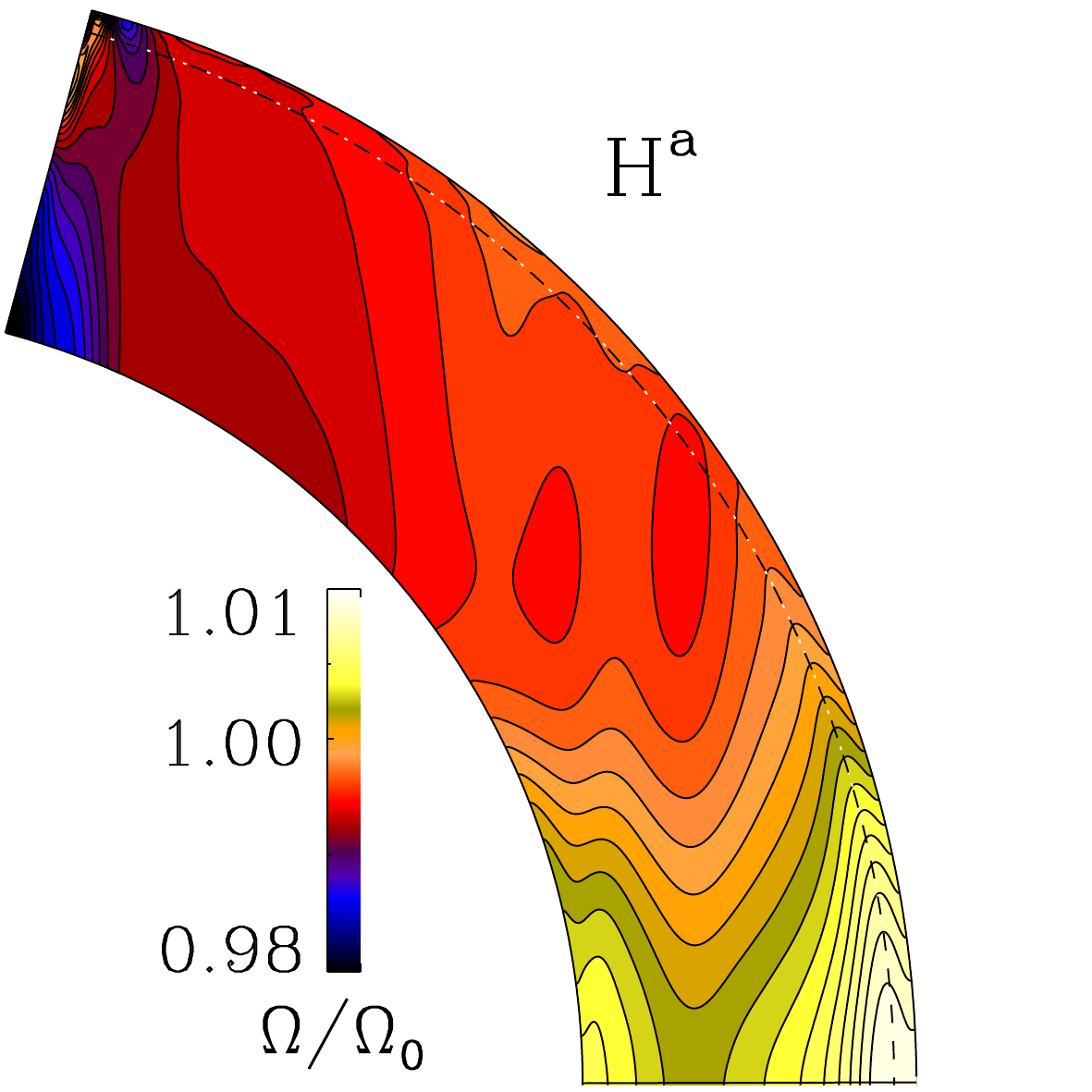} 
\includegraphics[width=.49\columnwidth]{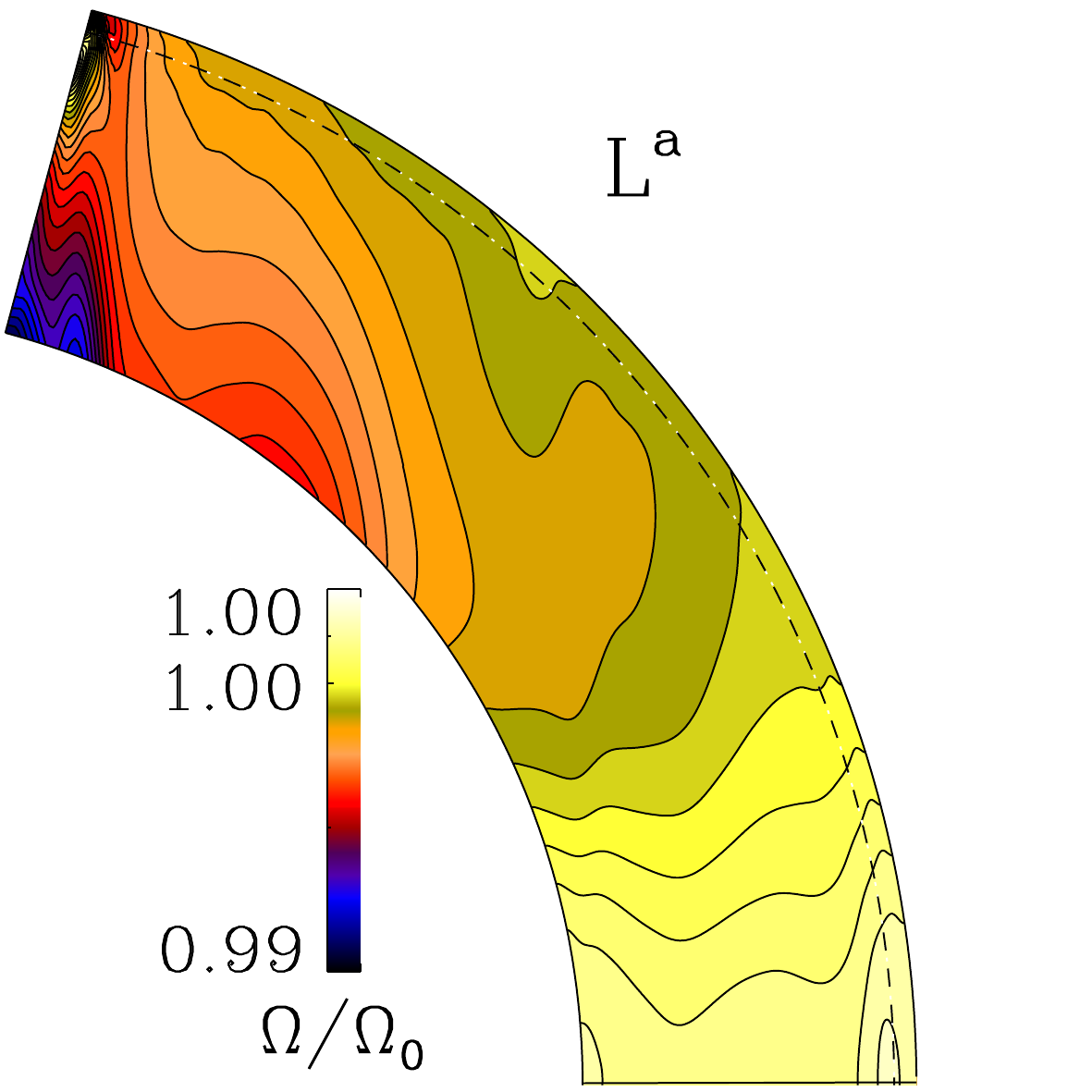}
\caption{Normalized angular velocity $\Omega(r,\theta)$ of Runs~A2,
  C3, G$^a$, H, H$^a$, and L$^a$.
The dashed lines denote the radius $r=0.98R$, which is used for the
further analysis.}
\label{fig:pOm}
\end{figure}

\subsection{Differential rotation}

The rotation also influences the generation of mean flows as for
example the differential rotation.
To illustrate this, we plot the profiles of angular velocity,
$\Omega(r,\theta)=\overline{u}_\phi(r,\theta)/r\sin\theta+\Omega_0$,
for six representative runs (Runs~A2, C3, G$^{\rm a}$, H, H$^{\rm a}$ and
L$^{\rm a}$) in \Fig{fig:pOm}.
We find antisolar differential rotation for the solar
rotation rate 
(Runs~A1 and A2), which is consistent with previous numerical studies
\citep[e.g.][]{GYMRW14,KKB14}. 
This might be due to the too high
overall convective velocities or too high concentration of power at large
spatial scales \citep{FH16} realized in the simulations in comparison to the
Sun.
The antisolar rotation switches to solar-like at slightly more rapid
rotation corresponding to $\Co=3.0$.
For higher rotation rates the differential rotation develops a minimum at
mid-latitudes.
Such a configuration has been shown to be important in producing equatorward
migrating magnetic activity \citep{WKKB14}.
We find such minima also at moderate rotation, up to roughly seven times
solar rotation rate (Run~H).
At higher rotation rates very little differential rotation is
generated overall,
and the mid-latitude minimum becomes progressively weaker.

 \begin{figure}[t]
 \centering
 \includegraphics[width=\columnwidth]{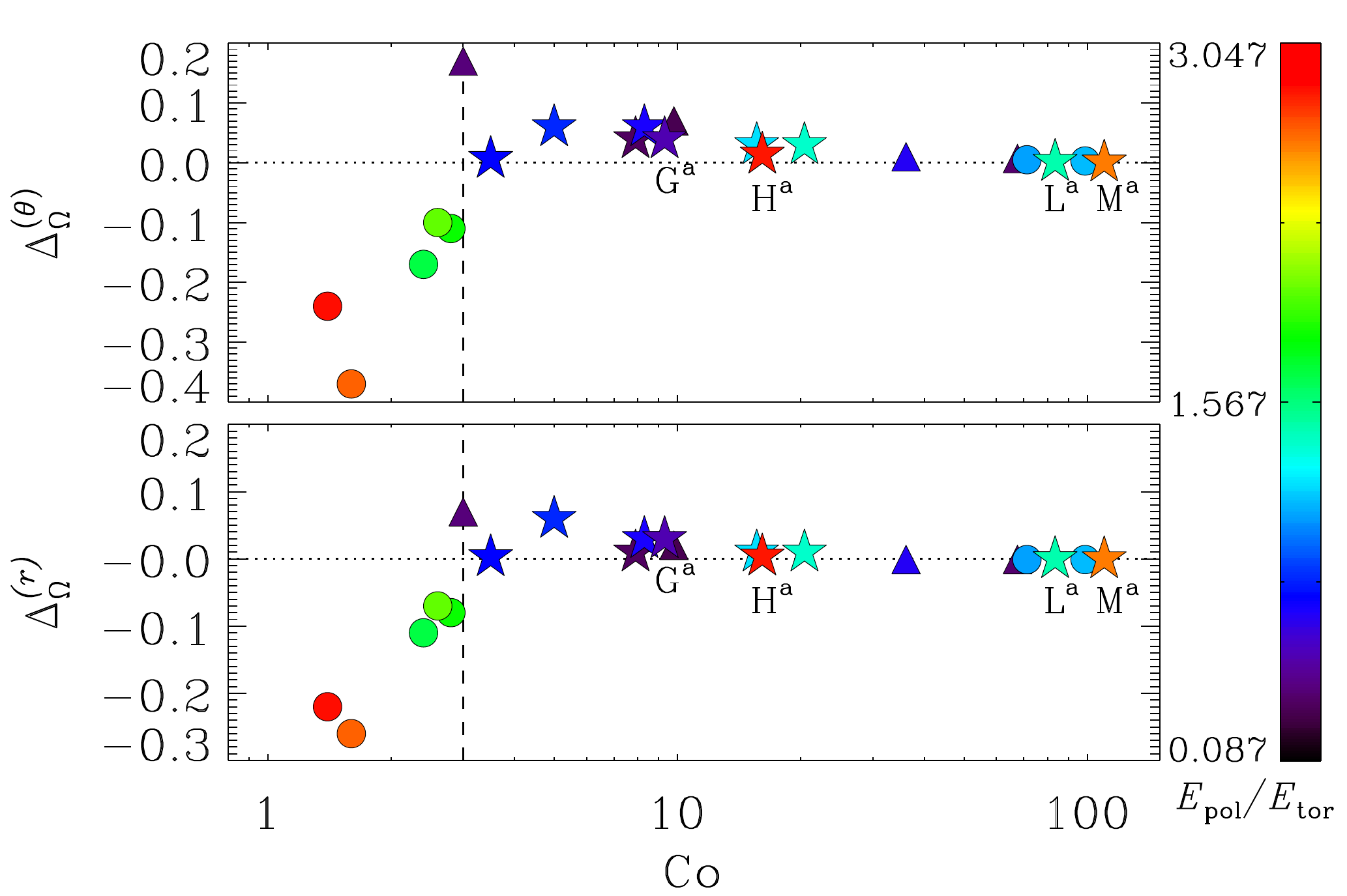}
\caption{
  Relative latitudinal differential rotation $\Delta_\Omega^{(\theta)}$ (top panel) 
  and relative radial differential rotation $\Delta_\Omega^{(r)}$ (bottom panel) 
  for all $2\pi$ runs. 
  The shape
 of the plotted symbols indicates the degree of nonaxisymmetry of the magnetic field
 (sphere -- axisymmetric; triangle -- mixed; 
star -- nonaxisymmetric) while the color indicates the ratio of
poloidal energy  $E_{\rm mag}^{\rm pol}$ to toroidal energy  $E_{\rm
  mag}^{\rm tor}$, see \Table{tab:ene}.
The dashed line ($\Co=3$) indicates the transition from antisolar to
solar-like latitudinal differential rotation and the dotted lines
indicate the zero.
The high resolution runs G$^a$, H$^a$, L$^a$, and M$^a$ are marked for better visibility.}
\label{fig:Deltas}
\end{figure}

\begin{figure}[h!]
\centering
\includegraphics[width=\columnwidth]{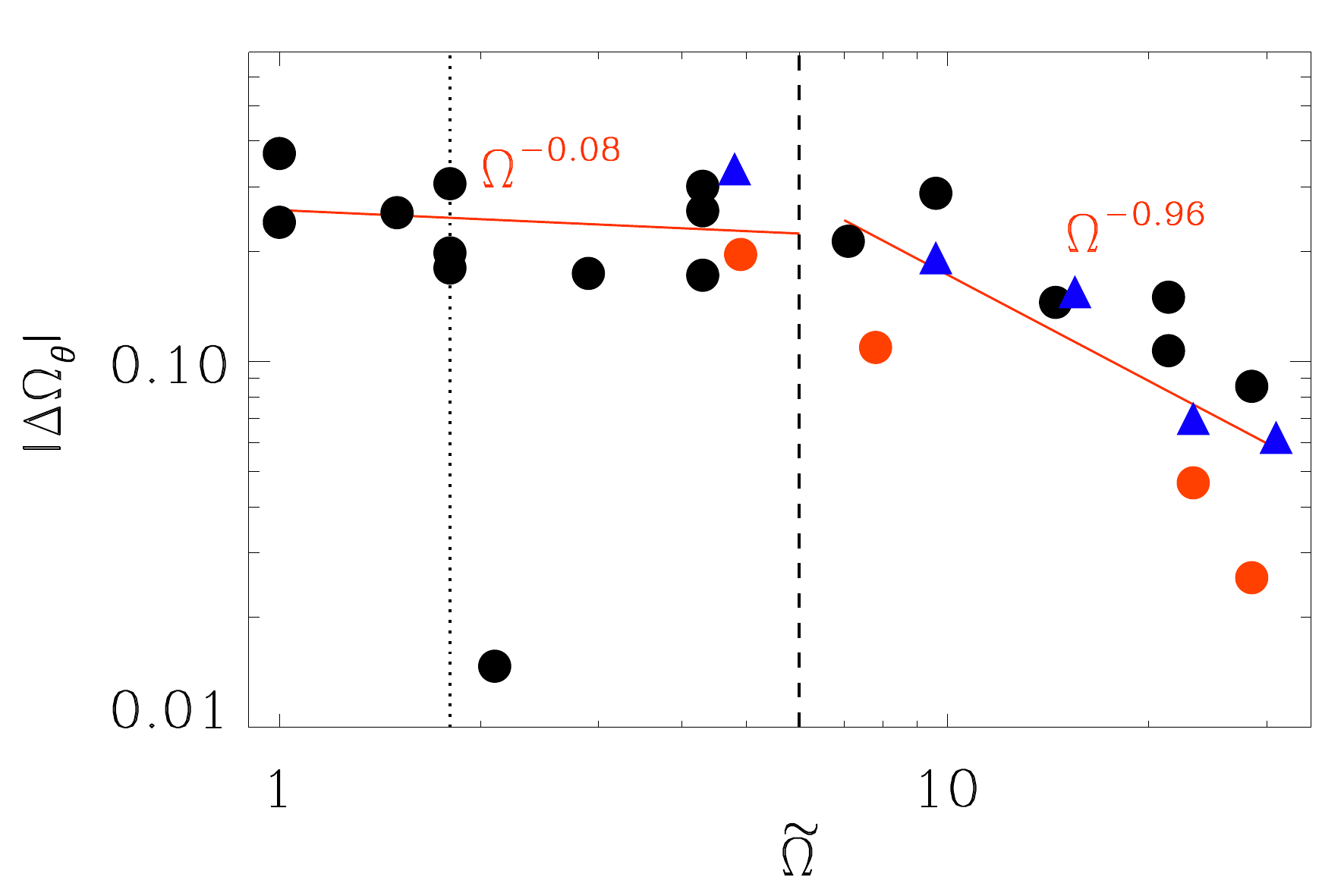}
\caption{
  The modulus of the absolute latitudinal differential rotation,
 $\Delta\Omega_{\theta}=\Delta_\Omega^{(\theta)}\Omega_0$, normalized by the
  solar rotation rate,
  as function of rotation rate. 
  The red lines result from fitting.
  The black dashed line indicates the break in the slope.
  Red and black circles stand for high- and low-resolution $2\pi$
  runs, respectively, while blue triangles show the $\pi/2$ wedges.
  We note that for the fit for moderate rotation, we do not
  take into account Run~D with very low values.
  The dotted line indicates the transition from antisolar to
  solar-like latitudinal differential rotation.
}
\label{fig:pkom}
\end{figure}

We quantify the relative radial and latitudinal differential rotation
using
\begin{eqnarray}
\Delta_\Omega^{(r)}=\frac{\Omega_{\rm eq}-\Omega_{\rm
  bot}}{\Omega_{\rm eq}} \; \mbox{and} \;
  \Delta_\Omega^{(\theta)}=\frac{\Omega_{\rm eq}-\Omega_{\rm
  pole}}{\Omega_{\rm eq}},
\label{eq:rel_diff}
\end{eqnarray}
where $\Omega_{\rm eq}=\Omega(R,\pi/2)$ and $\Omega_{\rm
  bot}=\Omega(0.7R,\pi/2)$ are the angular velocities at the top and
bottom of the convection zone at the equator, respectively, and $\Omega_{\rm
  pole}=[\Omega(R,\theta_0) + \Omega(R,\pi-\theta_0)]/2$ is the
angular velocity at the latitudinal boundaries.
Negative/positive values of $\Delta_\Omega^{(\theta)}$ indicate
antisolar (fast poles, slow equator)/solar-like (fast equator, slow poles)
differential rotation.
In \Table{tab:runs} we list these numbers from our simulations, and
notice
that a transition from strong antisolar to significantly weaker solar-like
differential rotation occurs at about
$\tilde\Omega\approx1.8$ ($\rm Co \approx 3$; Run C3).
We also plot $\Delta_\Omega^{(r)}$ and $\Delta_\Omega^{(\theta)}$
as functions of $\Co$ for all the $2\pi$ runs in \Fig{fig:Deltas}.
There, we indicate the transition point with a vertical dashed line. 
As we will later
discuss in detail, this point also marks the change of the dynamo modes
from axisymmetric to nonaxisymmetric ones. From this plot it is evident that,
as the rotation increases, both relative differential
rotation measures approach zero.
From Tables~\ref{tab:runs} and \ref{tab:ene} we also see that near the transition, the
rotation profile is sensitive to changes in the convective efficiency,
as indicated by the Rayleigh number.
In Run~C3 with a low $\Pr_{\rm SGS}$ and lower Rayleigh
and Reynolds numbers
than in the more turbulent
Runs~C1 and C2, the rotation profile is solar-like, while in the others
it is antisolar.
This transition and its sensitivity to the efficiency of convection
has been studied in detail by, e.g., \cite{GYMRW14} and \cite{KKB14}.

Note that $\Delta_\Omega^{(r)}$ and $\Delta_\Omega^{(\theta)}$ measure only the
difference between certain points and neglect the actual latitudinal
variation, which can be more complicated.
In the case of wedge geometry the flows near the
latitudinal boundaries may not be representative of what takes place
at high latitudes in real stars. This can lead to
unrepresentative results, in particular for the latitudinal
differential rotation in cases where the latitudinal profile is
non-monotonic \citep[cf.][]{KKKBOP15}. 

The antisolar regime typically shows strong negative radial and
latitudinal shear \citep{GYMRW14}, whereas magnetic fields tend to
quench the differential rotation \citep[e.g.][]{FF14,KKKBOP15}. Our results
are in agreement with those aforementioned studies. 
Another important aspect is the dependence of absolute
differential rotation, defined as
\begin{equation}
\Delta \Omega_r = \Delta_\Omega^{(r)} \tilde\Omega ,\quad \Delta
\Omega_{\theta} = \Delta_\Omega^{(\theta)} \tilde\Omega ,
\label{eq:abs_diff}
\end{equation}
on the rotation rate itself.
The broad range of probed rotation rates allows us to search for a power-law behavior 
of the form
\begin{equation}
|\Delta \Omega_{\theta} | \propto \Omega^q.
\label{eq:q}
\end{equation}
In \Fig{fig:pkom}, 
we do not find, however, a single power-law that would describe
the behavior at all rotation rates. 
For slow and moderate rotation, up to $\tilde\Omega\approx 5$, we fit a slope 
of $q\approx-0.08$, while for the highest rotation rates investigated,
$\tilde\Omega \approx 5-31$,
we find a steeper power law with $q\approx-0.96$.

In \Table{tab:DROmega}, we compare our results with those of some observational
studies \citep[][]{RG15,Lehtinen16} and a mean-field model \citep[][]{KR99}.
Our results for the low to intermediate rotation rates agree with these studies,
but the power law we find for the rapid rotation regime is much steeper and 
therefore in disagreement with them.
This disagreement cannot be
explained by the lack of supercriticality as the high-resolution runs
show even weaker latitudinal differential rotation than their low-resolution
counterparts.
However, the magnetic fields in the rapidly rotating high-resolution
runs (H$^a$, L$^a$, and especially in M$^a$) are generally stronger than in the lower
resolution ones, possibly also contributing to the reduced differential
rotation \citep[cf.][]{KKOWB16}.

\begin{table}[t!]
\centering
\caption[]{
  Scaling of absolute differential rotation with rotation of
  some recent observational studies, models and our work using the
  exponent $q$; see \Eq{eq:q}.}
\label{tab:DROmega}
\begin{tabular}{ll}
\hline
\hline
$q$ & Reference \\
\hline
$-0.08$ & This work (slow rotation) \\
$-0.96$ & This work (rapid rotation) \\
$-0.36$ & \cite{Lehtinen16} \\
$+0.29$ & \cite{RG15} \\
$-0.15$ (G2, mean) & \cite{KR99} \\
$-0.04$ (K5, mean) & \cite{KR99} \\
\hline
\end{tabular}
\end{table}

\subsection{Overview of magnetic states}

All the runs discussed in this work produce large-scale magnetic fields.
Similar runs were recently analyzed by \citet{WRKKB16} using the test-field method,
who measured significant turbulent effects contributing to the magnetic field generation.
Therefore, we attribute the magnetic fields seen in the current runs to the turbulent dynamo mechanism.
To describe the magnetic solutions, we first look at the
volume-averaged magnetic energy densities.
We define them analogously to their kinetic
counterparts. We use 
\begin{equation}
E_{\rm mag}=\brac{\BBB^2}_V/2\mu_0
\end{equation}
for the total magnetic energy density,
\begin{equation}
E_{\rm mag}^{\rm tor}=\brac{\mfi{B}{\phi}^2}_V/2\mu_0, \quad E_{\rm mag}^{\rm pol}=\brac{\mfi{B}{r}^2+\mfi{B}{\theta}^2}_V/2\mu_0
\end{equation}
for the contribution of mean toroidal and mean poloidal fields,
and
\begin{equation}
E_{\rm mag}^{\rm fluc}= E_{\rm mag}-(E_{\rm mag}^{\rm tor}
+E_{\rm mag}^{\rm pol})
\end{equation}
for the contribution of fluctuating magnetic fields.
These quantities are listed in \Table{tab:ene}.
We find that for all the runs, the contributions from fluctuating
magnetic fields dominate the magnetic energy.
The axisymmetric contributions
contain on the order of 5 to 10 per cent of the total
magnetic energy in the majority of the runs, and exceeds 15 per cent
only in Runs~C3, F3, K1, and K2. These runs are characterized either by a
low $\PraSGS$ (C3 and F3) or rapid rotation (K1 and K2), both leading to
reduced supercriticality of convection.

\begin{figure}[h!]
\centering
\includegraphics[width=1.0\columnwidth]{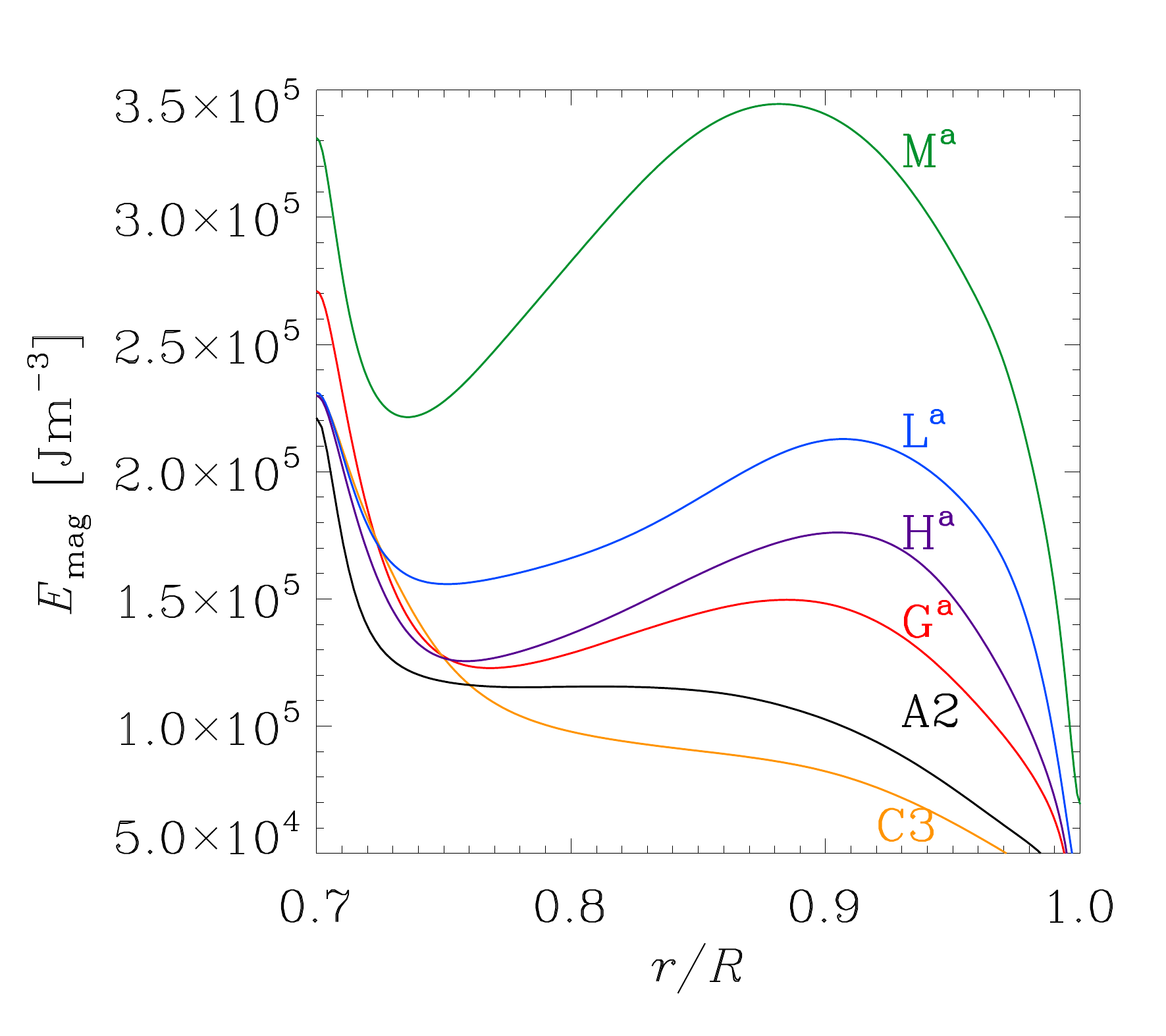}
\caption{
Radial profiles of the total magnetic energy density $E_{\rm mag}$ averaged over time, latitude, and azimuth
for Runs~A2, C3, G$^a$, H$^a$, L$^a$, and M$^a$. \label{fig:Emagr}
}\end{figure}

\begin{figure}[h!]
\centering
\includegraphics[width=1.0\columnwidth]{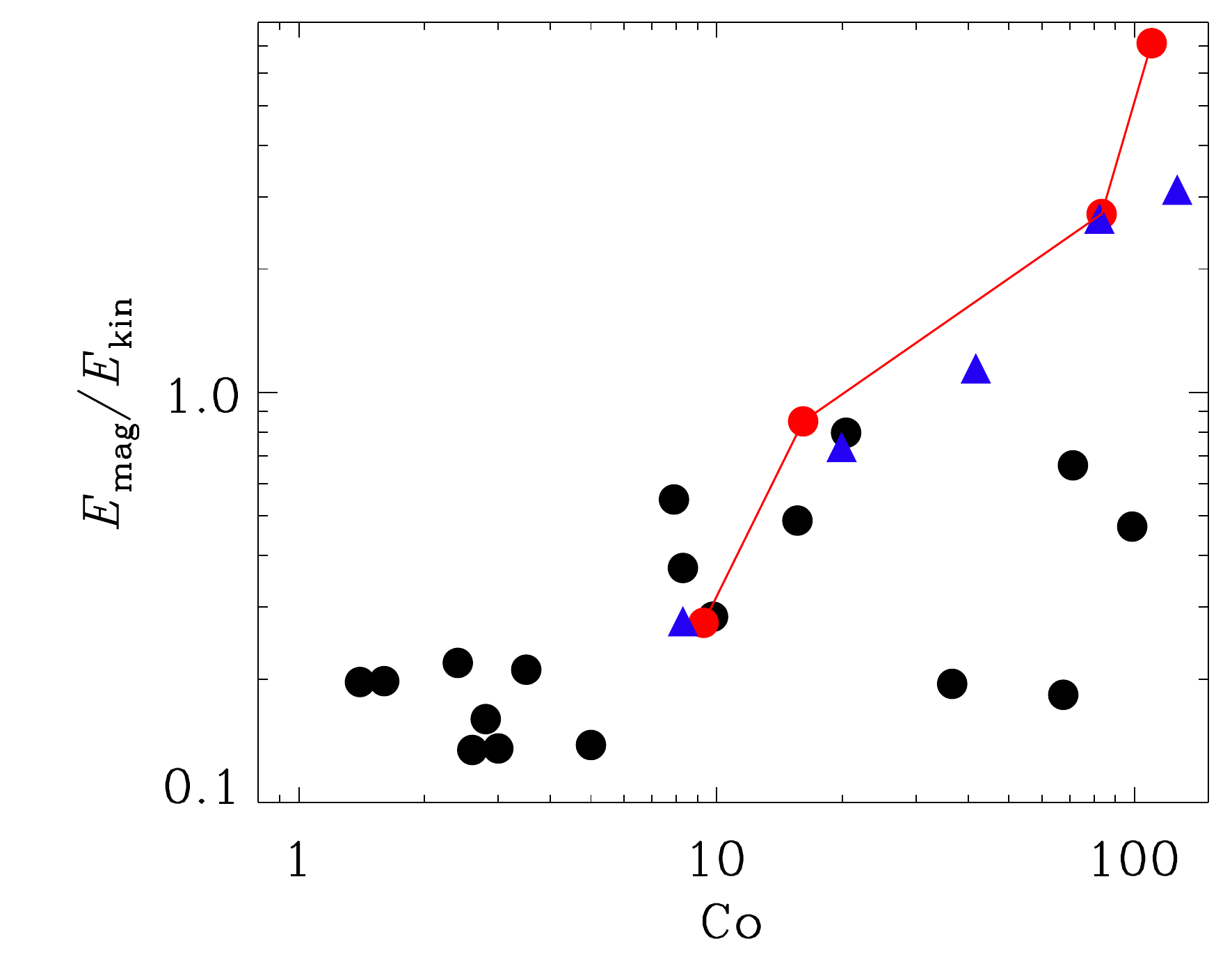}
\caption{
Ratio of total magnetic to kinetic energies $E_{\rm mag}/E_{\rm kin}$ as a function of Coriolis number $\Co$.
The red filled symbols (connected by a line) denotes high resolution runs.
Blue triangles refer to $\pi/2$ wedge runs.
}\label{fig:EkinEmag}
\end{figure}

In contrast to the kinetic energy, we do not find a clear trend
for magnetic energies as a function of rotation rate.
In the rapid rotation regime, the high-resolution runs L$^a$ and M$^a$ exhibit
magnetic fields with an energy that significantly exceeds the kinetic one by factors
of roughly 3 and 8, respectively.
If we look at the radial profile of the magnetic energy
density for a selection of runs (\Fig{fig:Emagr}), we find that the
magnetic field in the upper half of the convection zone increases with
rotation. 
As discussed earlier, we observe a simultaneous, nearly monotonic,
decrease of the kinetic energy as a function
of rotation rate.
Therefore, the ratio between the magnetic and kinetic energies,
which is a measure of
the dynamo efficiency, is actually steeply increasing as a
function of rotation, as can be seen from \Fig{fig:EkinEmag}.
We find that in the low-resolution cases the dynamo is clearly less
efficient in the
rapid rotation regime in comparison to the high-resolution cases. We also observe that the
$\pi/2$ wedge runs produce a far more
efficient dynamo in the rapid rotation regime than the
corresponding low-resolution runs with full azimuthal extent. 
This is possibly explained by the somewhat higher stratification and Rayleigh
numbers in the $\pi/2$ wedge runs in comparison to those
in the low resolution full $2\pi$ models.
We can conclude that the convective efficiency directly influences the
dynamo efficiency and therefore the magnetic energy production.

In \Fig{fig:Deltas}
we have studied the dependence of the overall magnetic topology on the amount of differential
rotation generated in the system. The ratio of poloidal to toroidal magnetic energies,
shown with the color of the symbols, changes systematically from mostly poloidal
field configurations at very low rotation rates to toroidal ones at
moderate and rapid rotation.
The energy ratio gradually decreases, and with rotation rates exceeding the
antisolar to solar transition point, dominantly toroidal fields are seen.
The strongest toroidal fields are generated for moderate rotation.
At the highest rotation rates, the ratio of toroidal and poloidal
becomes again lower in the high-resolution runs, while the low-resolution
counterparts fail to show systematic behavior.
In the run with the highest rotation rate, M$^{a}$, the poloidal component is again
dominating.
By inspecting \Table{tab:ene}, we notice that the models with reduced $\phi$ extent tend to produce a larger
poloidal to toroidal energy ratio than the corresponding runs covering the full azimuthal
extent.

\begin{figure*}[t]
\centering
\includegraphics[width=\columnwidth]{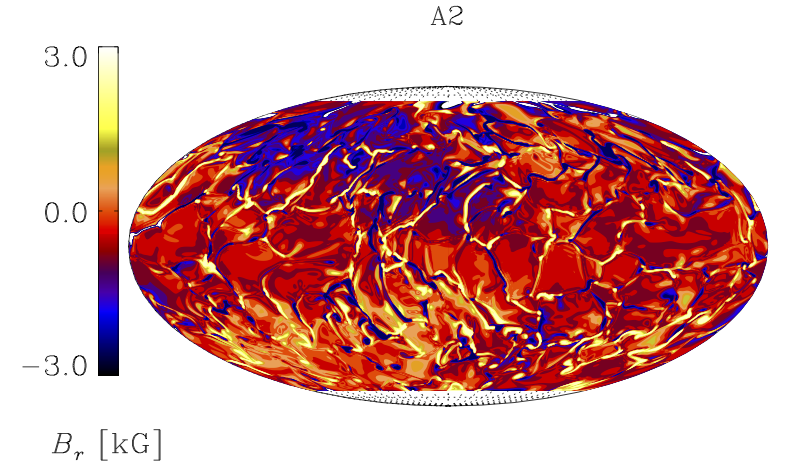}
\includegraphics[width=\columnwidth]{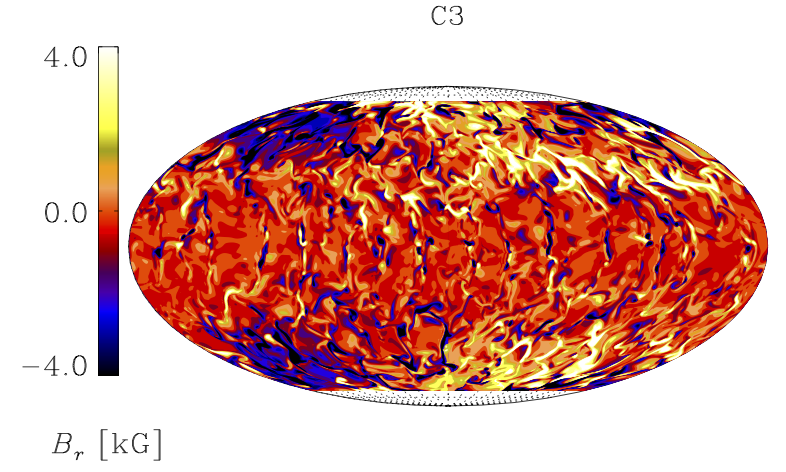}
\includegraphics[width=\columnwidth]{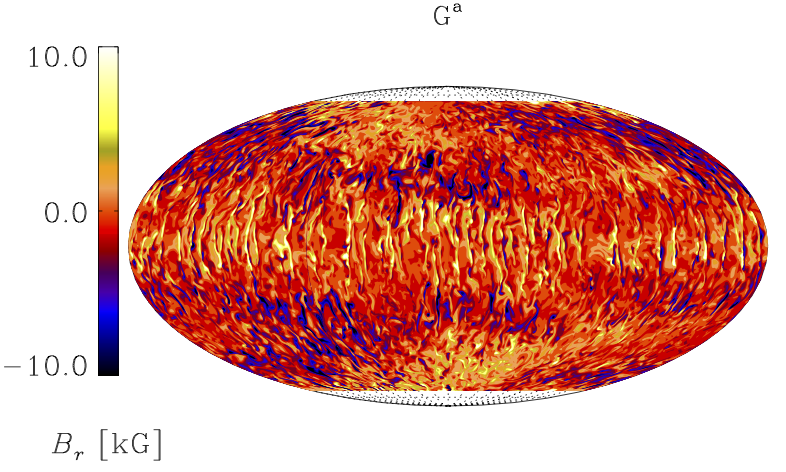}
\includegraphics[width=\columnwidth]{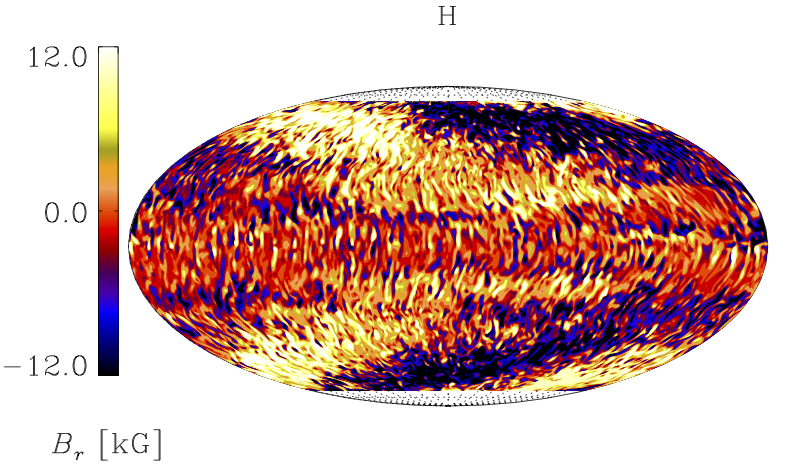}
\includegraphics[width=\columnwidth]{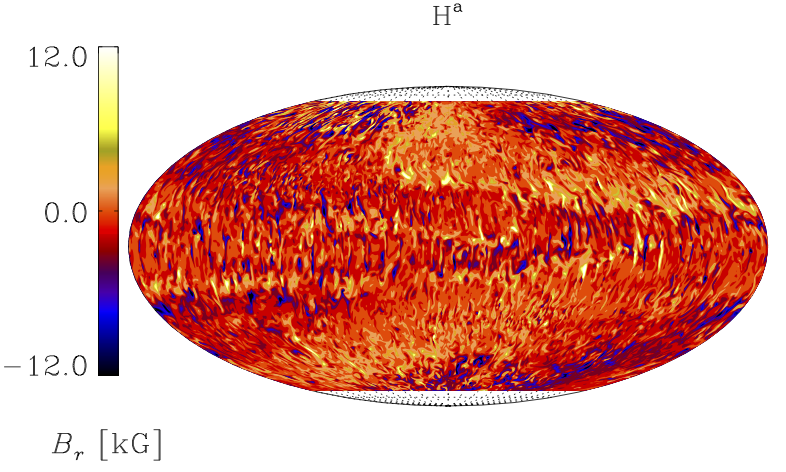}
\includegraphics[width=\columnwidth]{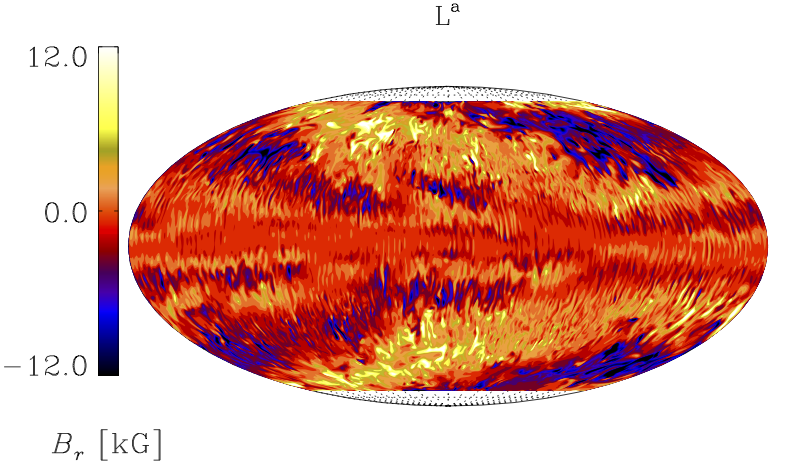}
\caption{
  Radial magnetic field $B_r$ at $r=0.98R_\odot$ from the same
  runs as in \Fig{fig:pmoll_uu1}.
}\label{fig:pmoll_bb1}
\end{figure*}

To investigate the spatial structure of the magnetic field, we show
in \Fig{fig:pmoll_bb1} snapshots of $B_r$ for six representative runs.
At low rotation rates, most of the
magnetic field is concentrated in the downflows between the convective cells,
while at high rotation rates,
the scale of convection, still clearly affecting the magnetic field,
thereby leaving a small-scale imprint on it,
is significantly reduced. Nevertheless, global-scale
magnetic field configurations clearly emerge in the high-latitude
regions.
It is immediately apparent that
a nonaxisymmetric large-scale pattern is visible in all cases.
In the slowly rotating cases, the nonaxisymmetric component is sub-dominant and
the equatorial symmetry of the field is clearly dipolar (antisymmetric with respect to
the equator). In all the runs with solar-like differential rotation, however, the
field configuration is observed to be symmetric (or quadrupolar) with respect
to the equator, 
even though a more detailed analysis revealed that the parity of the solutions is not pure.
A weaker antisymmetric (dipolar) component is present at all times, and the global parity
undergoes some fluctuations.
The quadrupolar component remains most significant at all times, however. 
This result is in agreement with some ZDI measurements of solar-like stars 
\citep[e.g.][]{Hackman2016,Rosen2016}.
However, we should point out that our results can be influenced by the wedge assumption
in latitude and need to be verified in full spherical geometry.

We also depict the overall nonaxisymmetry of the large-scale magnetic
field solutions with the shape of the symbol in \Fig{fig:Deltas}.
Again, on the lower rotation side of the break-point identified, the magnetic fields are
mostly axisymmetric (circular symbol). On the rapid rotation side, the fields exhibit
a significant nonaxisymmetric component (triangles) and finally turn into completely
nonaxisymmetric ones (stars). The resolution plays also a significant role for the
nonaxisymmetry measure: the higher resolution runs show preferentially nonaxisymmetric
configurations, while the lower-resolution runs turn back to axisymmetry at the highest
rotation rates investigated.

\begin{figure}[t]
\centering
\includegraphics[width=1.0\columnwidth]{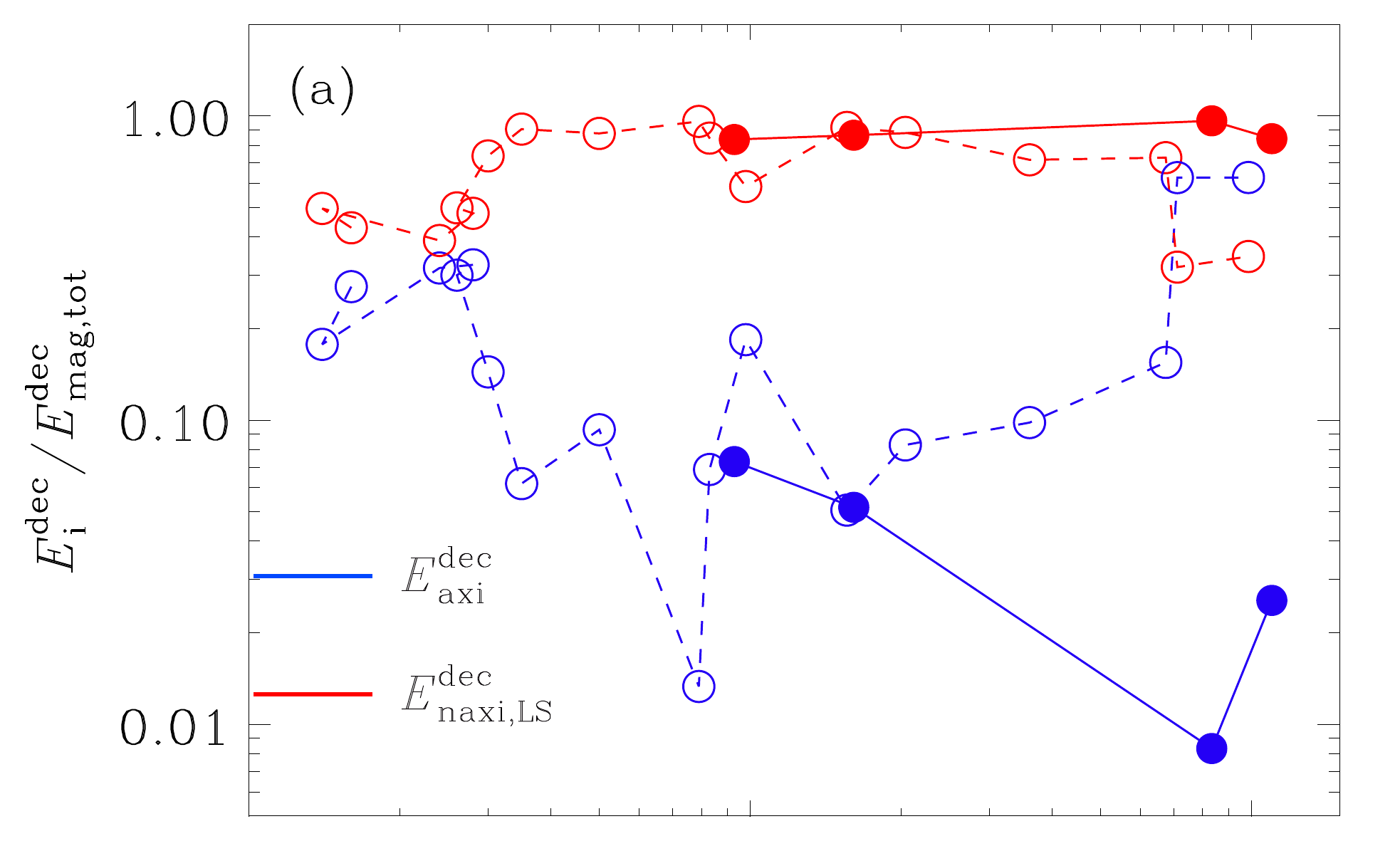} \\
\includegraphics[width=\columnwidth]{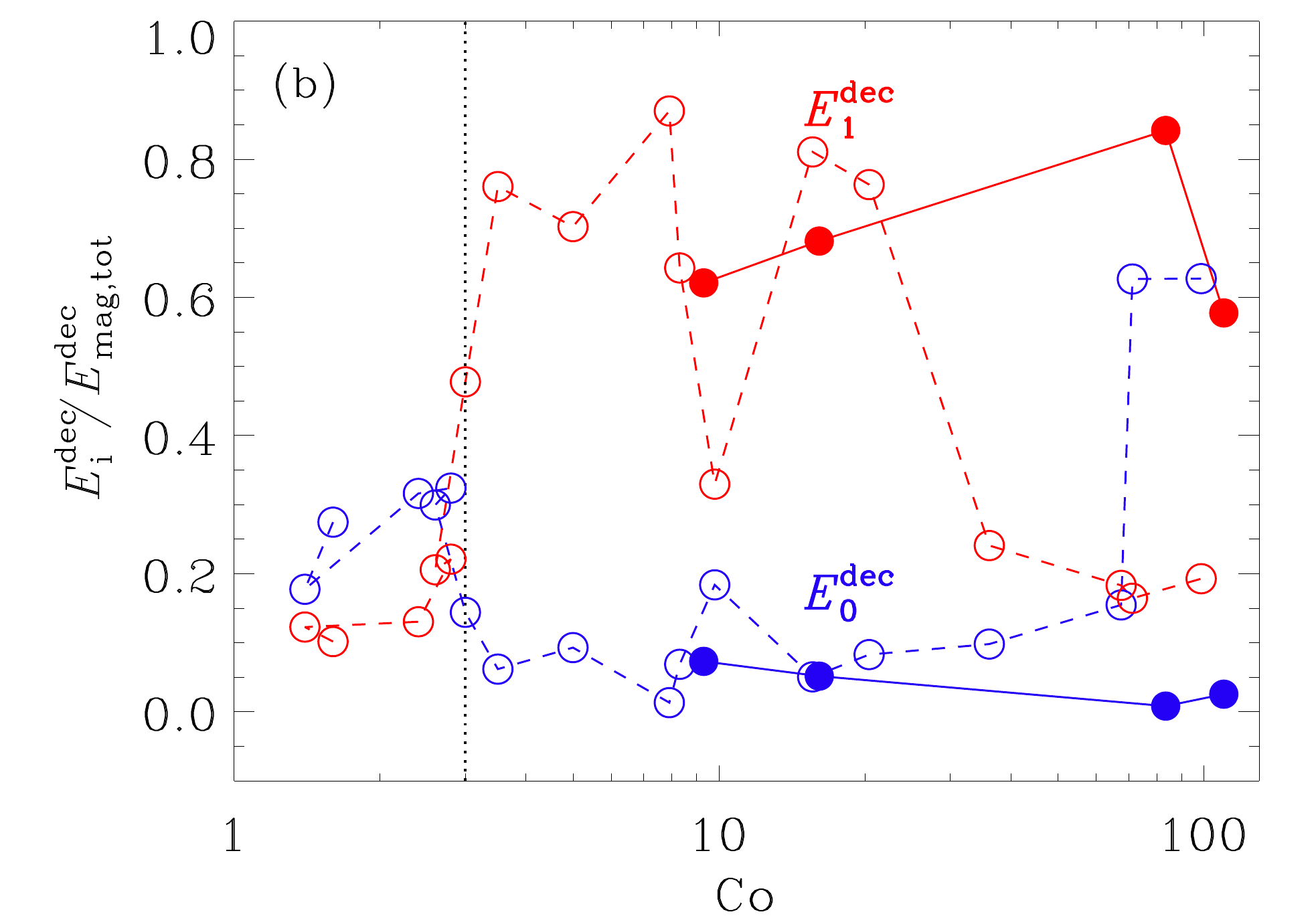} 
\caption{
Axisymmetric mean ($m=0$, blue) energy versus nonaxisymmetric
large-scale ($m=1,5$; red) energy fraction at the surface (a).
Axisymmetric mean ($m=0$; blue) energy fraction
versus the first nonaxisymmetric mode ($m=1$; red) energy fraction (b). 
The dotted black line denotes the axisymmetric to nonaxisymmetric transition at
region $\Co \approx 3$.
In both plots, the dashed red and blue lines connect $2\pi$ runs; 
filled symbols, connected with solid lines, denote the high resolution runs.}\label{fig:m0m1}
\end{figure}

\subsection{Degree of nonaxisymmetry}

Large-scale nonaxisymmetric magnetic fields, as seen in
\Fig{fig:pmoll_bb1}, are included in the definition of $E^{\rm
  fluc}_{\rm mag}$ in \Table{tab:ene}, as this quantity is 
the difference between total and azimuthally averaged (mean) magnetic energies.
This term contains, therefore, both small-scale fluctuations and
the large-scale nonaxisymmetric contributions. Thus the diagnostics
introduced so
far describe the large-scale fields in the system only roughly.

To obtain a more complete picture,
we perform a spherical harmonics decomposition
of the radial components of the vector fields at
$r=0.98R$
with the method described in Appendix \ref{sec:appendixA}.
The $m=0$ mode contains the axisymmetric (mean) part of the radial
magnetic field, the $m=1$ mode is the first nonaxisymmetric 
one, $m=2$ is the second one, and so on.
For the $\pi/2$ wedges, the first nonaxisymmetric mode is $m=4$.
The energies of the modes resulting from the decomposition are listed in \Table{tab:dec}.
Depending on the dominant large-scale component, we call the magnetic
fields nonaxisymmetric or axisymmetric -- even though their small-scale
contributions, which are always nonaxisymmetric, might be more energetic.

\begin{table*}[t!]
\centering
\caption[]{Energy densities of the radial magnetic field and dynamo
  cycle properties.}
       \label{tab:dec}
      $$
          \begin{array}{p{0.025\linewidth}lllllllllllllll}
            \hline
            \hline
            \noalign{\smallskip}
Run & E_{\rm mag}^{\rm surf} & E_{\rm mag,tot}^{\rm dec} & E_{0}^{\rm dec} & E_{1}^{\rm dec} & E_2^{\rm dec} & E_3^{\rm dec} & E_4^{\rm dec} & E_5^{\rm dec} & E_{l, m > 5}^{\rm dec} & P_{\rm ADW} &P_{\rm ADW} & P_{\rm ADW} & D &\tau_{\rm cyc}\\
&&&&&&&&&&[{\rm yr}]&[P_{\rm 0}]&[P_{\rm DR}] &&[{\rm yr}]\\
\hline
            \hline
           A1 & 0.211 & 2.1(-2) & 5.8(-3) & 2.1(-3) & 1.9(-3) & 1.7(-3) & 1.7(-3) & 1.5(-3) & 6.1(-3) &  &  &  & & 3.72_{(m0)} \\
           A2 &  0.188 & 2.4(-2) & 4.3(-3) & 3.0(-3) & 2.5(-3) & 2.4(-3) & 2.2(-3) & 2.0(-3) & 7.9(-3) &  &  &  & & 4.13_{(m0)} \\
           B & 0.183 & 2.3(-2) & 7.3(-3) & 3.0(-3) & 1.6(-3) & 1.6(-3) & 1.5(-3) & 1.3(-3) & 6.8(-3) &  &  &  & & 2.45_{(m0)} \\
           \underline{C1} & 0.137 & 1.7(-2) & 5.7(-3) & 3.9(-3) & 1.3(-3) & 1.2(-3) & 1.1(-3) & 9.2(-4) & 3.5(-3) &  &  & & & 3.53_{(m0)} \\ 
           \underline{C2} & 0.128 & 1.6(-2) & 4.7(-3) & 3.2(-3) & 1.5(-3) & 1.2(-3) & 1.0(-3) & 8.4(-4) & 3.2(-3) &  &  &  & & 4.37_{(m0)} \\ 
           \hline   
           \underline{C3} & 0.142 & 2.4(-2) & 3.5(-3) & 1.2(-2) & 2.3(-3) & 1.7(-3) & 1.3(-3) & 1.0(-3) & 2.8(-3) & 19.53 & 474 & 44.83 & \mbox{R} & 3.13_{(m0)} \\ 
           D & 0.180 & 5.1(-2) & 3.1(-3) & 3.9(-2) & 3.4(-3) & 1.9(-3) & 1.2(-3) & 8.3(-4) & 1.7(-3)  & 14.14 & 410 & 24.67 & \mbox{R} & 18.25_{(m0)} \\ 
           \underline{E} & 0.147 & 3.1(-2) & 2.9(-3) & 2.1(-2) & 2.4(-3) & 1.4(-3) & 9.4(-4) & 6.0(-4) & 9.9(-4) &  39.87 & 1542 & 82.53 & \mbox{R} & 10.31_{(m1)} \\ 
           F1 & 0.290 & 0.111 & 1.5(-3) & 9.7(-2) & 4.8(-3) & 2.4(-3) & 1.5(-3) & 1.1(-3) & 3.1(-3) &  4.22 & 245 & 5.92 & \mbox{R} & 6.68_{(m0)} \\ 
           \underline{F2} & 0.220 & 4.9(-2) & 3.3(-3) & 3.1(-2) & 4.3(-3) & 2.5(-3) & 1.7(-3) & 1.4(-3) & 4.2(-3) & 5.94 & 346 & 6.55 & \mbox{R} & 8.05_{(m1)}\\ 
           \underline{F3} & 0.086 & 1.6(-2) & 2.9(-3) & 5.2(-3) & 1.5(-3) & 1.1(-3) & 8.3(-4) & 6.6(-4) & 3.6(-3) & 10.49 & 611 & 8.20 & \mbox{R} & 5.74_{(m0)} \\ 
           \underline{ G$^{a}$} & 0.254 & 5.5(-2) & 4.0(-3) & 3.4(-2) & 5.4(-3) & 2.9(-3) & 2.0(-3) & 1.4(-3) & 5.0(-3) & 8.93 & 583 & 8.60 &  \mbox{R} & 7.43_{(m1)} \\
           \underline{G}$^{W}$ & 0.286 & 3.9(-2) & 2.9(-2) & 0.0 & 0.0 & 0.0 & 6.9(-3) & 0.0 & 2.7(-3) &  &  &  & & 2.37_{(m0)}\\
           \underline{ H} & 0.053 & 2.2(-2) & 1.1(-3) & 1.8(-2) & 1.1(-3) & 5.5(-4) & 3.9(-4) & 2.9(-4) & 7.2(-4) & 42.69 & 1.9(+3) & 12.15 & \mbox{R} & 27.34_{(m1)} \\ 
           \underline{ H$^{a}$} & 0.274 & 7.7(-2) & 4.0(-3) & 5.3(-2) & 6.6(-3) & 3.4(-3) & 2.4(-3) & 1.7(-3) & 6.5(-3) & 24.36 & 2.6(+3) & 15.84 & \mbox{R} & 7.17_{(m1)} \\ 
           \underline{I} & 0.274 & 0.107 & 8.9(-3) & 8.2(-2) & 6.0(-3) & 3.2(-3) & 2.1(-3) & 1.5(-3) & 3.7(-3) & 11.66 & 1.5(+3) & 7.07 & \mbox{R} & 7.75_{(m1)}  \\ 
           \underline{I}$^{W}$ & 0.220 & 4.0(-2) & 3.0(-2) & 0.0 & 0.0 & 0.0 & 7.6(-3) & 0.0 & 2.6(-3) &  &  &  & & 4.44_{(m0)} \\ 
           \underline{J} & 0.014 & 2.0(-3) & 2.0(-4) & 4.9(-4) & 4.3(-4) & 2.7(-4) & 1.6(-4) & 1.1(-4) & 3.8(-4) & 6.0(+3) & 5.4(+5) & 382.59 & \mbox{SW, P} & 8.25_{(m0)} \\ 
            \underline{J}$^{W}$ & 0.421 & 0.129 & 7.2(-2) & 0.0 & 0.0 & 0.0 & 4.6(-2) & 0.0 & 1.1(-2) &  &  &  & & 4.05_{(m0)} \\ 
            \underline{K1} & 0.025 & 3.7(-3) & 5.7(-4) & 6.8(-4) & 1 .0(-3) & 6.2(-4) & 2.2(-4) & 1.3(-4) & 4.3(-4) & 4.89 & 1.4(+3) & 0.01 & \mbox{P} & 1.24_{(m0)}\\ 
            \underline{K2} & 0.193 & 5.9(-2) & 3.7(-2) & 9.8(-3) & 3.7(-3) & 2.(-3) & 1.7(-3) & 1.5(-3) & 3.3(-3) &  &  &  & & 5.10_{(m0)} \\
            \underline{L$^{a}$} & 0.475 & 0.292 & 2.4(-3) & 0.246 & 1.3(-2) & 1.2(-2) & 5.7(-3) & 4.5(-3) & 8.3(-3) & 56.53 & 1.78(+4) & 14.66 & \mbox{SW, R} & 3.13_{(m1)}\\ 
            \underline{L}$^{W}$ & 0.509 & 0.218 & 0.123 & 0.0 & 0.0 & 0.0 & 8.1(-2) & 0.0 & 1.3(-2) &  &  &  &  & 5.68_{(m0)}\\ 
            \underline{M} & 0.133 & 4.9(-2) & 3.0(-2) & 9.3(-3) & 3.5(-3) & 1.9(-3) & 1.1(-3) & 8.0(-4) & 1.3(-3) &  &  &  &  & 6.64_{(m0)}\\
            M${^a}$ & 0.907 & 0.514 & 1.3(-2) & 0.297 & 4.5(-2) & 4.0(-2) & 2.9(-2) & 2.1(-2) & 6.8(-2) & 151.41 & 2.9(+4) & 12.3 & \mbox{SW, P} & 16.45_{(m1)} \\
            \underline{M}$^{W}$ & 0.462 & 0.197 & 0.135 & 0.0 & 0.0 & 0.0 & 5.1(-2) & 0.0 & 1.1(-2) &  &  &  & & 4.10_{(m0)}\\
\hline
        \end{array}
        $$
\tablefoot{The data for the energy densities is quoted near the surface ($r=0.98R$)
in units of $10^5$J~m$^{-3}$. Here $E_{\rm mag}^{\rm surf}$ is the total energy density,
$E_{\rm mag,tot}^{\rm dec}=\brac{(\BBB_{\rm tot}^{\rm dec})^{2}}_{\theta \phi t} /2\mu_0$
is the magnetic energy density obtained
from the decomposition over the first 10 harmonics, while $E_{\rm m}^{\rm dec}$
denote the magnetic energy densities for the corresponding azimuthal wavenumbers with $m=0,...,5$,
and $E_{\rm l,m>5}^{\rm dec}$ the magnetic energy density in scales that are considered 
to be small-scale ($m >5$).
The rotation period $P_{\rm ADW}$ of the azimuthal dynamo wave (ADW)
is computed as the latitudinal and temporal average 
of the derivative of the maximum phase of the dynamo mode
($P_{\rm ADW}=2\pi / <{\rm d x_{\rm max, m1}}/{\rm dt}>_{t,\theta}$).
The column $P_{\rm ADW} [P_0]$ indicates the average period of the ADW
compared to the bulk rotation ($P_0=2\pi/\Omega_0$).
The column $P_{\rm ADW} [P_{\rm DR}]$ indicates the average period of the ADW
compared to the period of the differential rotation.
$D$ indicates if the ADW is moving in the retrograde (R) or prograde (P) direction.
SW indicates a standing wave.
Furthermore, $\tau_{\rm cyc}$ is the characteristic timescale of the
change of the dynamo solution. That coincides with the time evolution
of the dominating dynamo mode, 
indicated in the parenthesis.
If the solution exhibits oscillatory behavior, the run label is
underlined.
The numbers in parentheses indicate the exponent of 10.
}\end{table*}

The distribution of the radial magnetic energy density
near the surface of the star is presented in
\Fig{fig:m0m1}a as a function of $\Co$.
Here we show the axisymmetric 
and the magnetic energy in
the large-scale nonaxisymmetric field ($1 \leq \ell \leq 5$), normalized
by the total magnetic energy. 
We find an inversion between the energies in the axisymmetric 
and nonaxisymmetric components that coincides also with the transition from
antisolar- to solar-like differential rotation at $\rm Co \approx 3$.
The runs show a nonaxisymmetric magnetic field until $\rm Co \approx 70$,
but at higher $\Co$ the high resolution runs remain nonaxisymmetric, 
while the low resolution runs return to an axisymmetric configuration,
indicating that high resolution is needed at such high rotation rates to capture
the small scales.
This could explain the lack of nonaxisymmetric solutions in the study of
\cite{BBBMT10}.
This conjecture is supported by the fact that in the higher resolution
simulations of \cite{NBBMT13} significantly clearer nonaxisymmetric
features are seen (their Figs.\ 4--6), although they are confined to
low latitudes. Those simulations were made wit
$\tilde\Omega=3$, albeit with a lower thermal Prandtl
number, as well as different viscosity and diffusivity profiles than in
the current simulations \citep[cf.\ Appendix~A of][for a comparison
of different setups]{KKOWB16}.
Our Runs~C3 and F3 also produce strong nonaxisymmetric
large-scale fields at high latitudes despite their lower values of
$\PraSGS$. This could be an indication of the influence of the
latitudinal boundaries in the current simulations.

The simulations of \cite{FF14} and \cite{HRY16}, on the other hand,
used the solar rotation rate and a further decreased thermal Prandtl
number resulting in a laminar heat transport to force a solar-like
rotation profile. The large-scale
magnetic fields in those simulations are characterized by dominant
low-latitude axisymmetric fields which show apparently random polarity
reversals. The results of these studies are most closely related to
our slowly rotating Runs~A1, A2, and B which also produce
predominantly axisymmetric large-scale fields, although with
antisolar differential rotation.
This seems to suggest that
axisymmetric fields are preferred at slow rotation irrespective of the
differential rotation profile.

From \Table{tab:dec} we notice that $m=1$ is the first large-scale
nonaxisymmetric mode excited as the rotation increases.
Some higher $m$ modes get excited, too, but they remain, on average, subdominant
compared to the $m=1$ mode. Therefore the runs are well described by
the $m=0$ and $m=1$ modes, shown in \Fig{fig:m0m1}b.
The axisymmetric energy is dominant at slow rotation, $\Co \le 3$,
while in the range $3 \leq \Co \leq 72$ the first nonaxisymmetric mode is dominant,
but its strength decreases for the low resolution runs for $\Co > 20$, and eventually
there is a return to an axisymmetric configuration at the highest values of
$\Co$.
For the high-resolution runs, however, the $m=1$ mode energy keeps increasing until the
highest rotation rates investigated.

\subsection{Magnetic cycles}

The time evolution of the magnetic field
is not cyclic in the sense that there are not necessarily polarity reversals in all the runs.
Yet, we see cyclic variations around the mean magnetic energy level,
albeit with a poorly defined cycle length.
This would match with an observer's viewpoint, as
most often only light curve variability is observable while the surface magnetic evolution
is hidden. 
Therefore, it makes sense to try to determine the time scale of this
variability for all the runs---not only those for which we can identify cyclic polarity reversals
from the butterfly diagram (the runs underlined in \Table{tab:dec}).  
By counting how many times the mean magnetic energy level is crossed,
sometimes referred to as the syntactic method
\citep[Chapter~9.4]{chen1988signal}, we can assign a
characteristic time 
scale of change, $\tau_{\rm cyc}$. 
For some of the runs, the
time-latitude variability would provide another, more straightforward, 
way to determine the cycle length. For consistency, this approach is used to 
determine the cycle periods for all the runs.
A comparison with cycle determination using all magnetic field
components at all latitudes shows good agreement between these two
methods for these kinds of simulations \citep{W17}.
The last column in \Table{tab:dec} shows $\tau_{\rm cyc}$.
We use the syntactic method on the dominant modes
($m=0$ and $m=1$) and indicate those by a subscript.
The syntactic method, however, has a limitation in that
counting the fluctuations around a mean value means that
we always count at least one oscillation.
This makes the $\tau_{\rm cyc}$ values for Runs~D, K2, and M$^a$ questionable,
as they are roughly half of the run time of the simulations.
This time is denoted by $\Delta t$ and is listed in \Table{tab:runs}.
One could instead determine the characteristic time by running the aforementioned runs for a longer time.
Retrieving cycle periods of the same order as the data set lengths, however, is not uncommon
in observational studies \citep[see e.g.][]{Baliunas1995}, so we have decided to retain
these values with the other, more trustworthy ones, in our analysis.

\begin{figure*}[t]
\centering
\includegraphics[width=\textwidth]{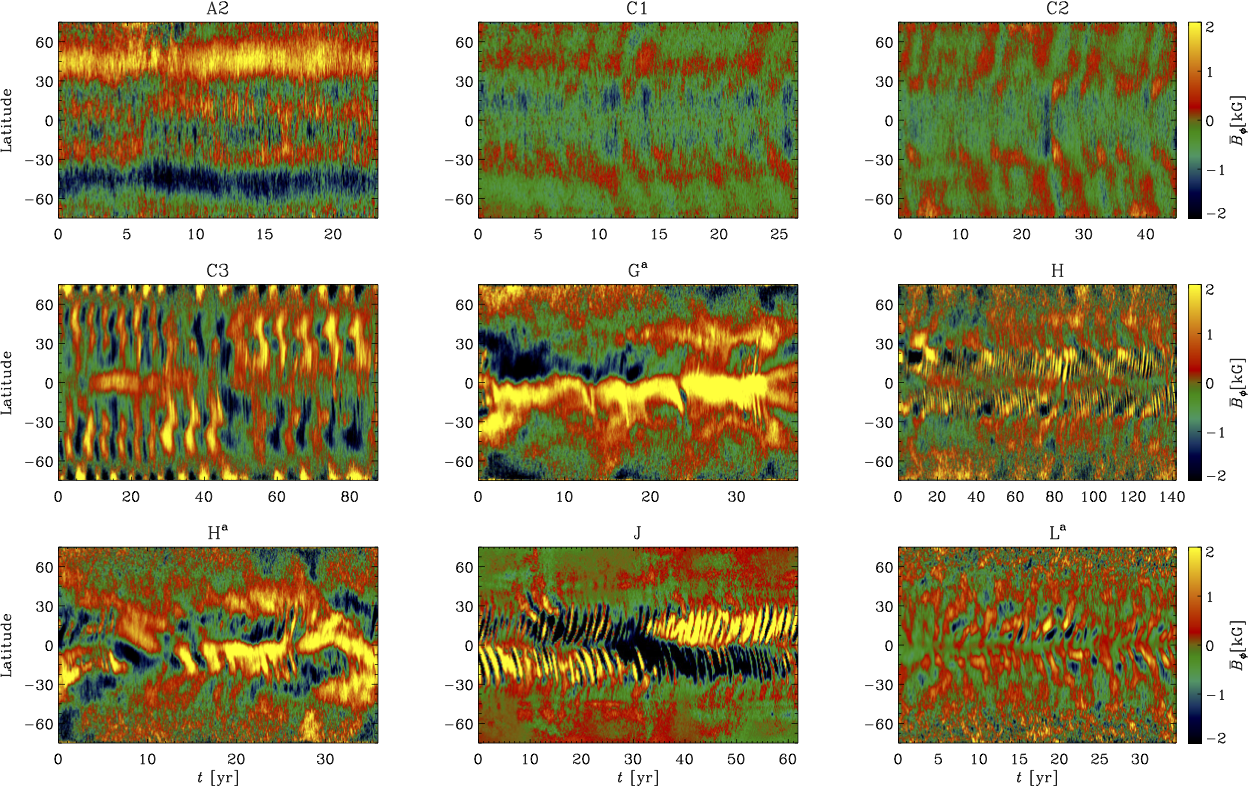}
\caption{Mean toroidal magnetic field $\mean{B}_\phi$
  for nine representative runs
  near the surface $r=0.98R$.}
\label{fig:pbutter}
\end{figure*}

In the Runs~A1, A2, and B,
all with antisolar differential rotation profiles, we do not see much time
dependence in the time--latitude (butterfly) diagrams of the mean
toroidal magnetic fields; see the upper left panel of \Fig{fig:pbutter} for an
example from Run~A2.
Starting from Runs~C1, C2, and C3 onwards to higher rotation rates (other panels of \Fig{fig:pbutter}),
however, more systematic patterns are discerned in the time series and the butterfly diagrams.
Runs~C1 and C2 present two
interesting cases, as it is very rare to obtain cyclic dynamo solutions in the regime
of antisolar rotation profiles \citep[e.g.][]{KKKBOP15,W17}, which these runs clearly possess.
Furthermore, it is clear that simulations with a 2$\pi$ azimuthal extent are
capable of producing oscillating dynamo solutions at lower rotation rates
than the corresponding $\pi/2$ wedges \citep[see comparison in][]{W17}.
In the rapid rotation regime, the time variability is always linked to the 
non-axisymmetric component, especially in the high-resolution runs.

\begin{figure}[t!]
\centering
\includegraphics[width=.87\columnwidth]{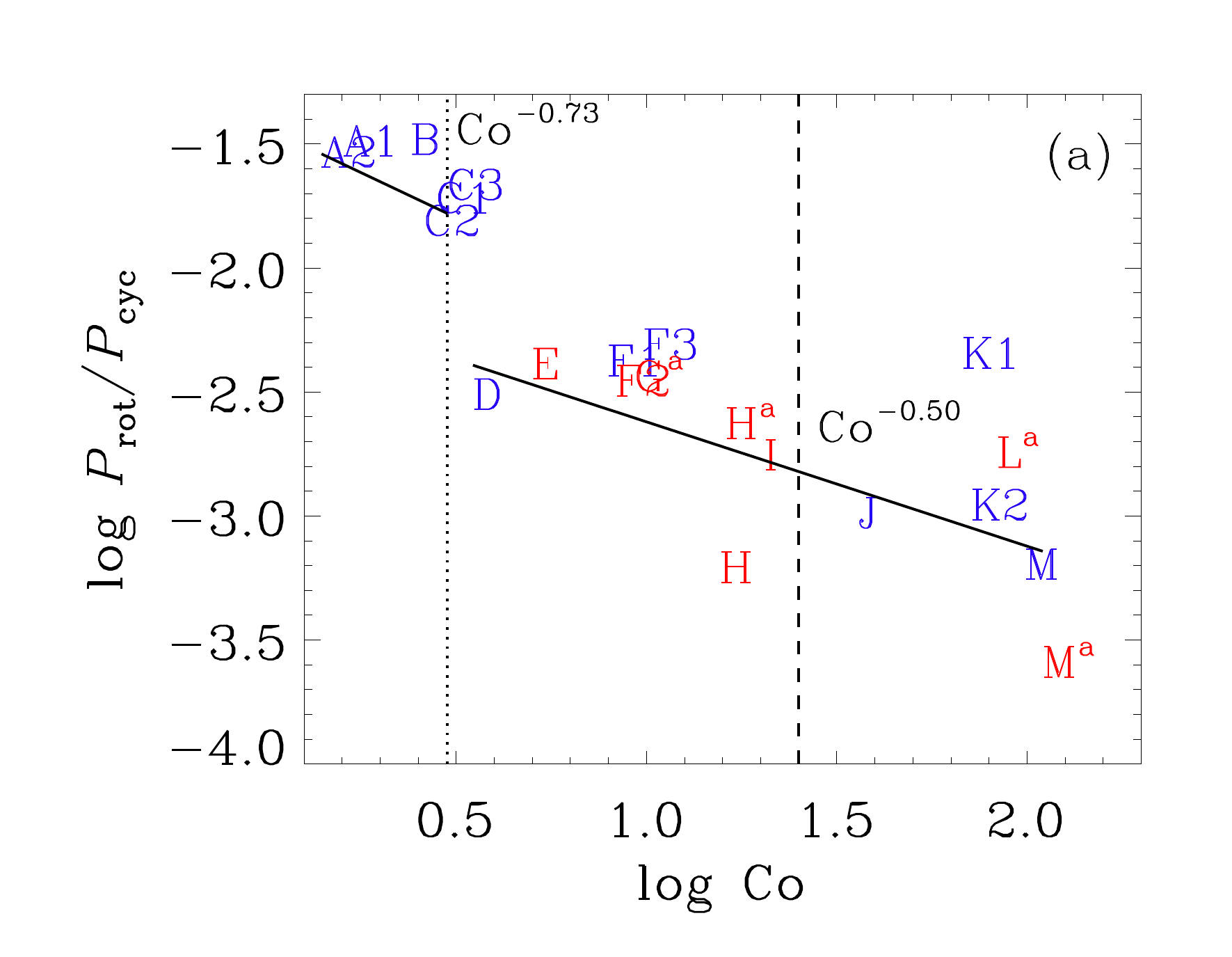}
\includegraphics[width=.87\columnwidth]{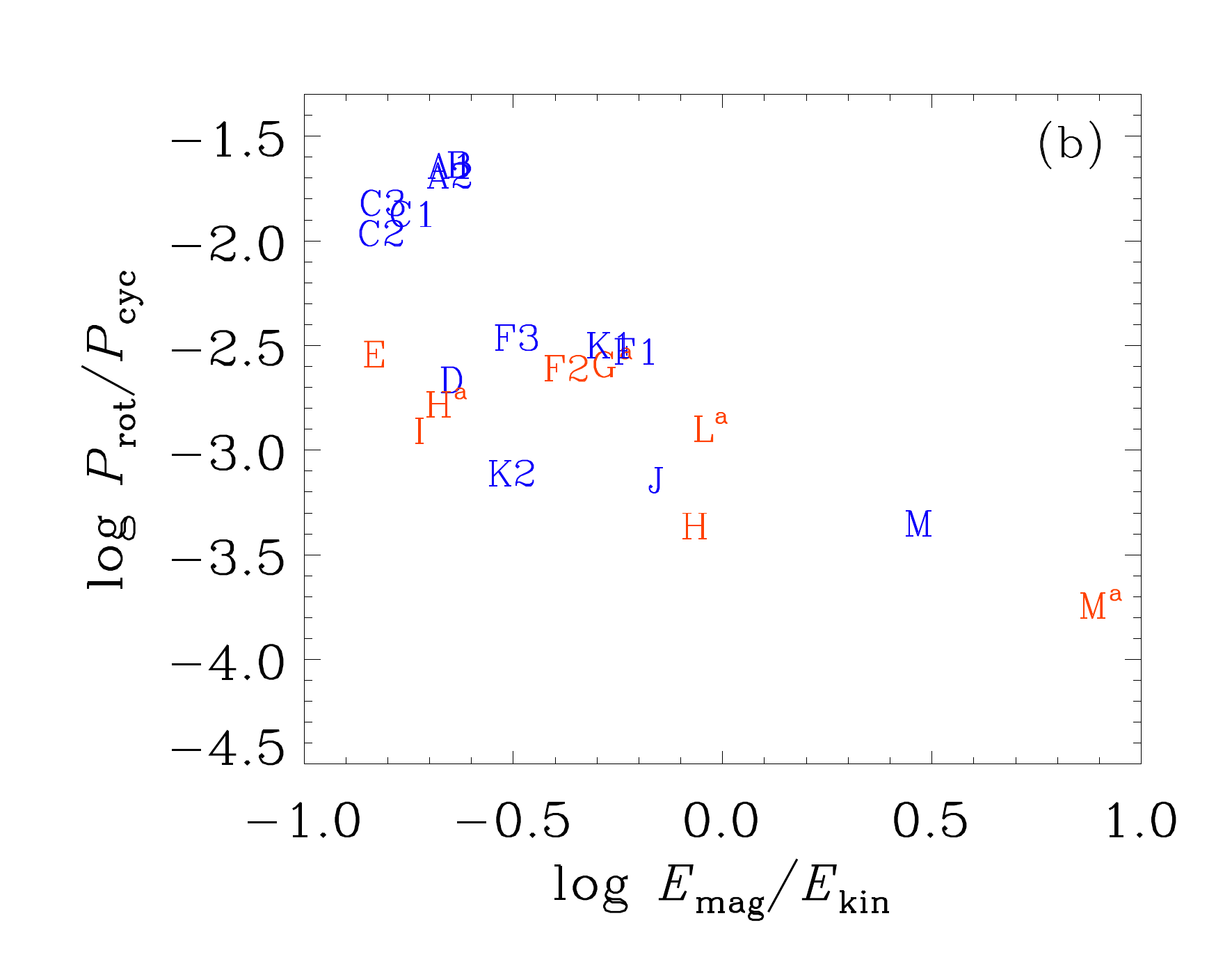}
\includegraphics[width=.87\columnwidth]{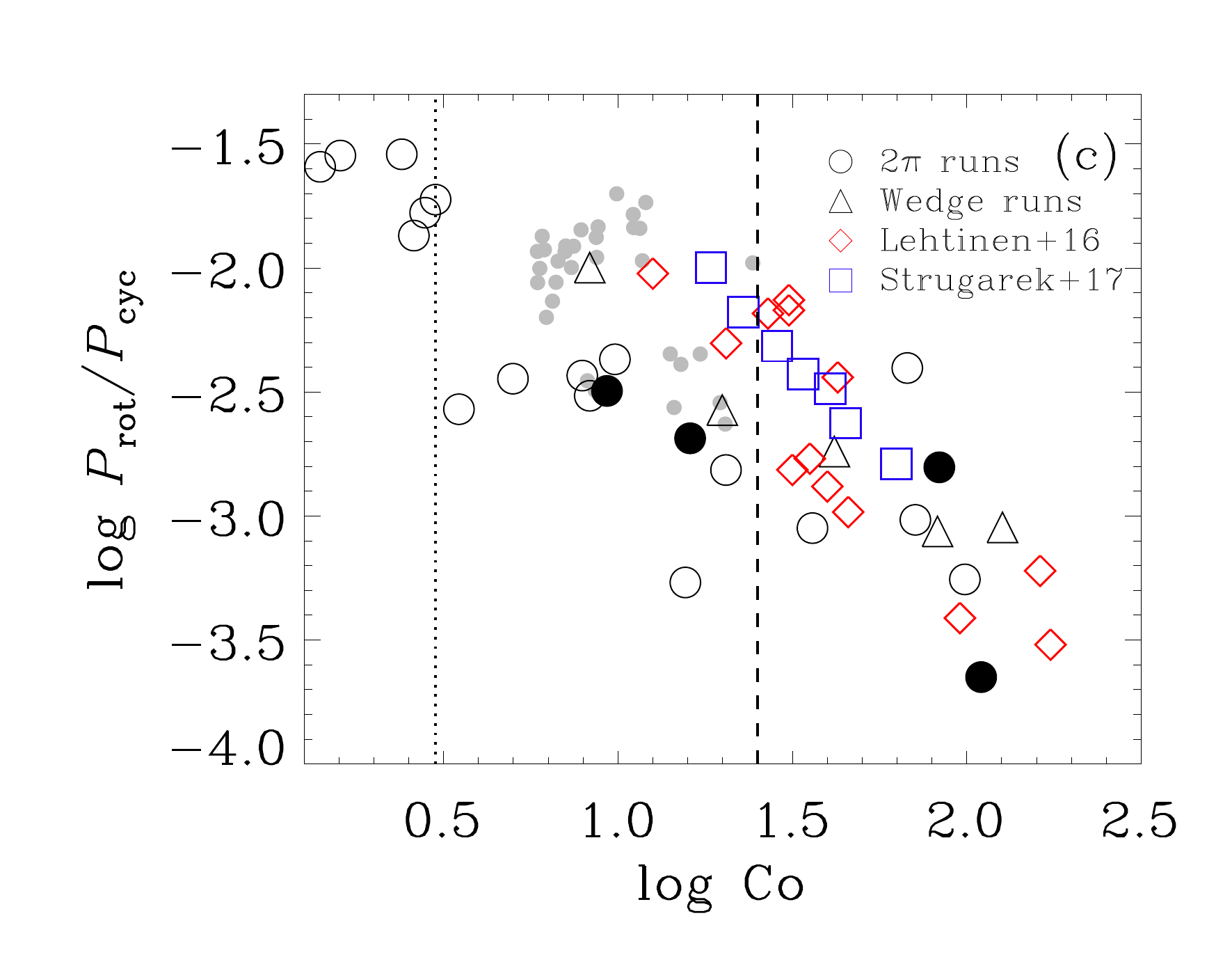}
\caption{Ratio of the rotation period to the cycle period 
  as a function of Coriolis number (a).
  The two black lines indicate the fit to the axisymmetric and 
  rapid rotation runs, respectively.
  The vertical lines denote the nonaxisymmetric transition found in
  our simulations (dotted) ($ \Co \ge
  3$) and from the observational study of \cite{Lehtinen16} (dashed),
  respectively.
  Runs are plotted after their labels.
  The color indicates the mode chosen for calculating $\tau_{\rm cyc}$: blue for $m=0$, 
  red for $m=1$.
  $P_{\rm rot}/P_{\rm cyc}$ as function of activity, represented by
  $E_{\rm mag}/E_{\rm kin}$ is shown in panel b.
  Panel c: comparison between the results presented in this paper, 
  \cite{Lehtinen16}, and \cite{SBCBN17}.
  Black circles and triangles denote, respectively, high resolution and $\pi/2$ wedges in our set.
  The gray dots are M dwarfs and F and G stars from \cite{Brandenburg2017}.}
\label{fig:ProtPcyc}
\end{figure}

After estimating the characteristic time, we can determine 
the activity cycle period as $P_{\rm cyc}= \tau_{\rm cyc}$ and
see how it varies with rotation, and compare with observational
results \citep{SB99,Lehtinen16}.
We show the results in \Fig{fig:ProtPcyc}a, where we plot the
ratio between rotation and activity period against the Coriolis number. 
We see that the transition line $\Co=3$ divides the runs into two
populations: one where the antisolar axisymmetric runs cluster and
another where the solar-like nonaxisymmetric runs cluster.
The former population is located in the upper left corner of the plot
showing a negative slope.
At this location, \cite{1984ApJ...287..769N} found, however, a population of
stars with a positive slope. \cite{BST98} denoted this
the inactive (I) branch -- to distinguish it from another active (A) one.
At even higher rotation rate, \cite{SB99} found yet another ``superactive'' (S) branch.
It has a negative slope,
which coincides with our solar-like nonaxisymmetric population (shown in red).
The simulation data yields $\Co^{-0.50}$ which
agrees quite well with the slope $\Co^{-0.43}$ determined by \cite{SB99} for the
S branch.
However, we cannot clearly identify an A branch nor a transition
between the A and S
branches, which are clearly present in \cite{SB99}.
The dashed vertical line denotes the observational transition of stars without active longitudes
to ones with them in a sample of solar-like rapid rotators \citep{Lehtinen16}.
We note that in our simulations, active longitudes occur for considerably lower
Coriolis numbers ($\Co > 3$, corresponding to the leftmost dotted line).

The best available measure of the magnetic activity from our simulations
is the ratio of magnetic to kinetic energy, which can be directly thought of as a
measure of the efficiency of the dynamo; see
\Fig{fig:EkinEmag}. Figure~\ref{fig:ProtPcyc}b
shows the
rotation--activity period ratio as a function of this quantity. 
In this plot, our runs again cluster near the I branch
and a well separated ``A--S branch.''
In contrast to Figure~\ref{fig:ProtPcyc}a, 
the correlation on the I branch now appears positive, but there are not
enough points to reliably conclude whether either of the
correlations seen on this branch are
significant.
The S branch still remains inseparable, but the population of runs
falling onto this branch shows a distinct negative slope.

In Figure~\ref{fig:ProtPcyc}c we show a comparison between observational results and
the models of \cite{SBCBN17} using again the Coriolis number on the $x$ axis. In this representation,
although the I branch still clearly exists, 
none of the modelled points coincide with the observed I branch.
Instead, the slowly rotating models cluster at lower Coriolis numbers than the observed
stars on the inactive branch, although their cycle ratios would rather well match with 
the ones of the observed population.
The Sun is not reproduced in any of those runs.

The moderate and rapid rotation runs are consistent with the S branch
behavior.
\cite{SBCBN17} and the $\pi/2$ wedges
of this study have a slope most closely matching the observed points of \cite{Lehtinen16}.
The runs covering the full longitudinal extent have significantly
shallower slope than the data points for the observed stars. The fact that the \cite{SBCBN17}
results coincide so well with the ones from our 
$\pi/2$ wedges, where the large-scale nonaxisymmetric modes are absent, suggests
that also the former models tend to become axisymmetric.
It needs to be seen to what extent this can be explained by
those runs not being sufficiently supercritical; see again Appendix~A
of \cite{KKOWB16} for a comparison of different setups.
This is clearly seen in our low-resolution models,
in which the magnetic field becomes axisymmetric at rapid
rotation, while in their high-resolution counterparts it remains nonaxisymmetric.
We note that our first run with solar-like differential rotation, C3,
is also the first showing nonaxisymmetric magnetic field.
This is in disagreement with observations, as the Sun has a mostly
axisymmetric field.
Therefore, we conclude that the model of \cite{SBCBN17}
with lower convective velocities, and thereby less supercritical convection,
can better reproduce the behavior in the proximity of the solar rotation rate.
At rapid rotation regime, however, their convective velocities are too low for
the models to capture the transition at all, while ours are too large and push it
to too low Coriolis numbers.

\begin{figure}[t]
\centering
\includegraphics[width=.9\columnwidth]{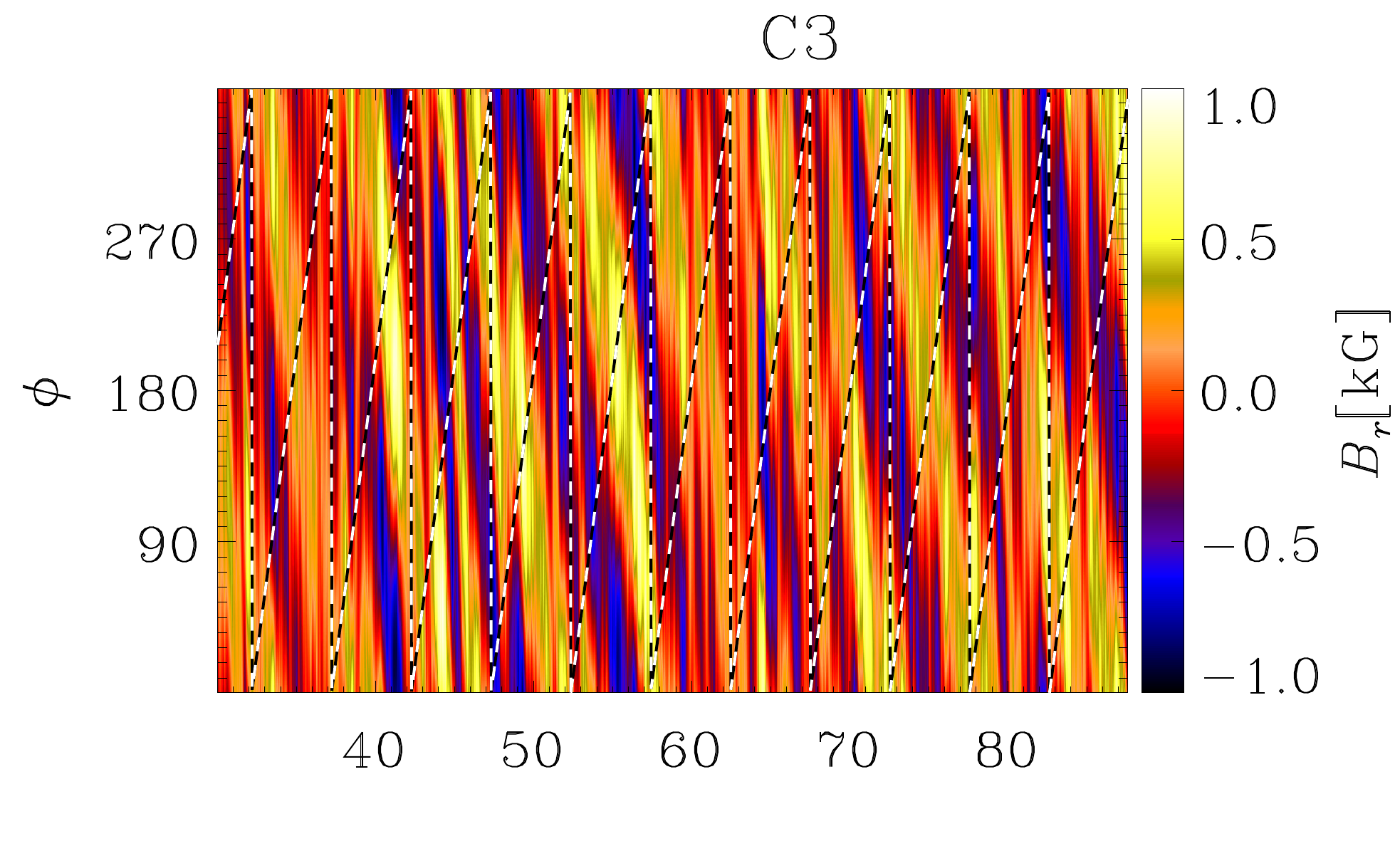}
\includegraphics[width=.9\columnwidth]{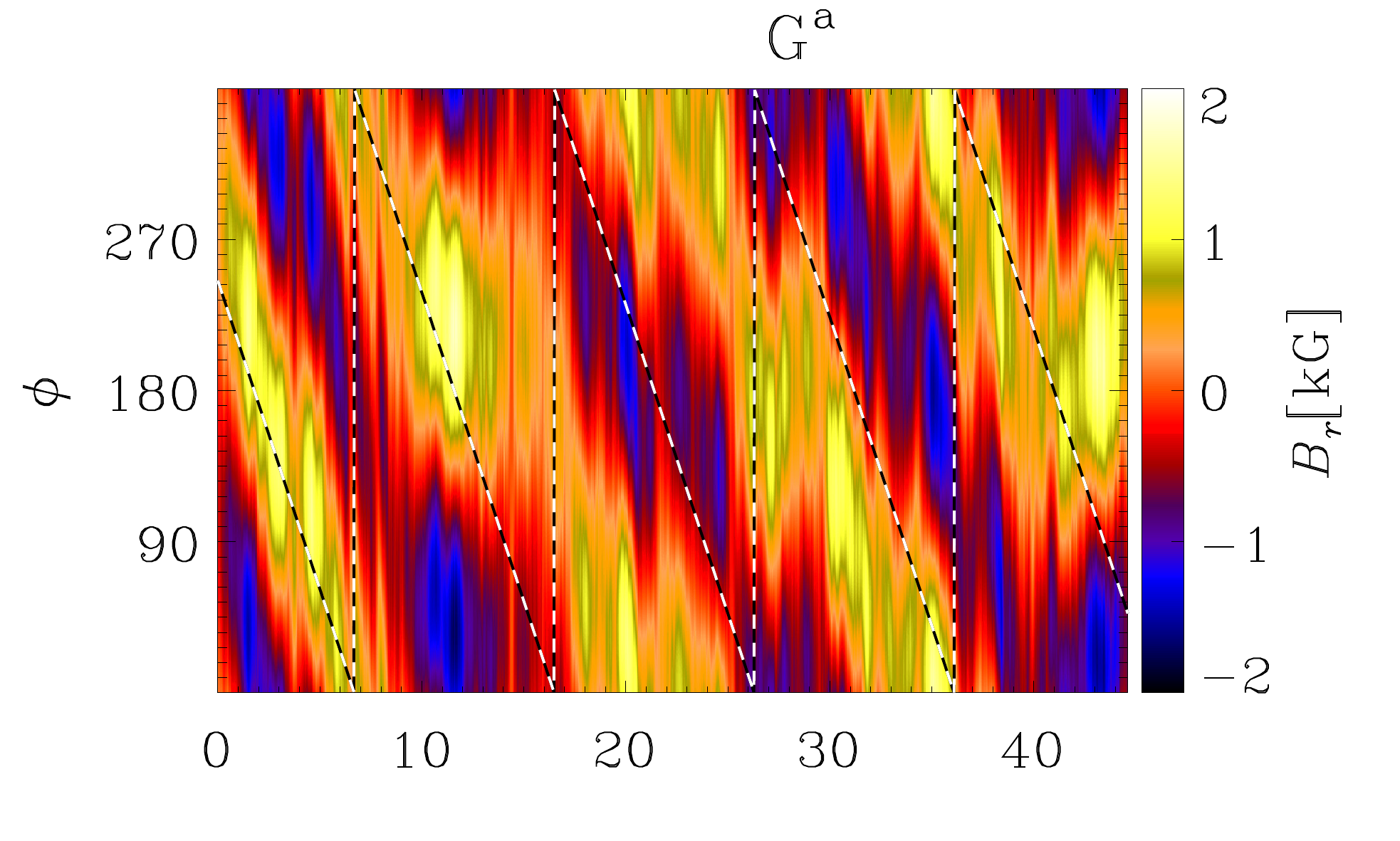}
\includegraphics[width=.9\columnwidth]{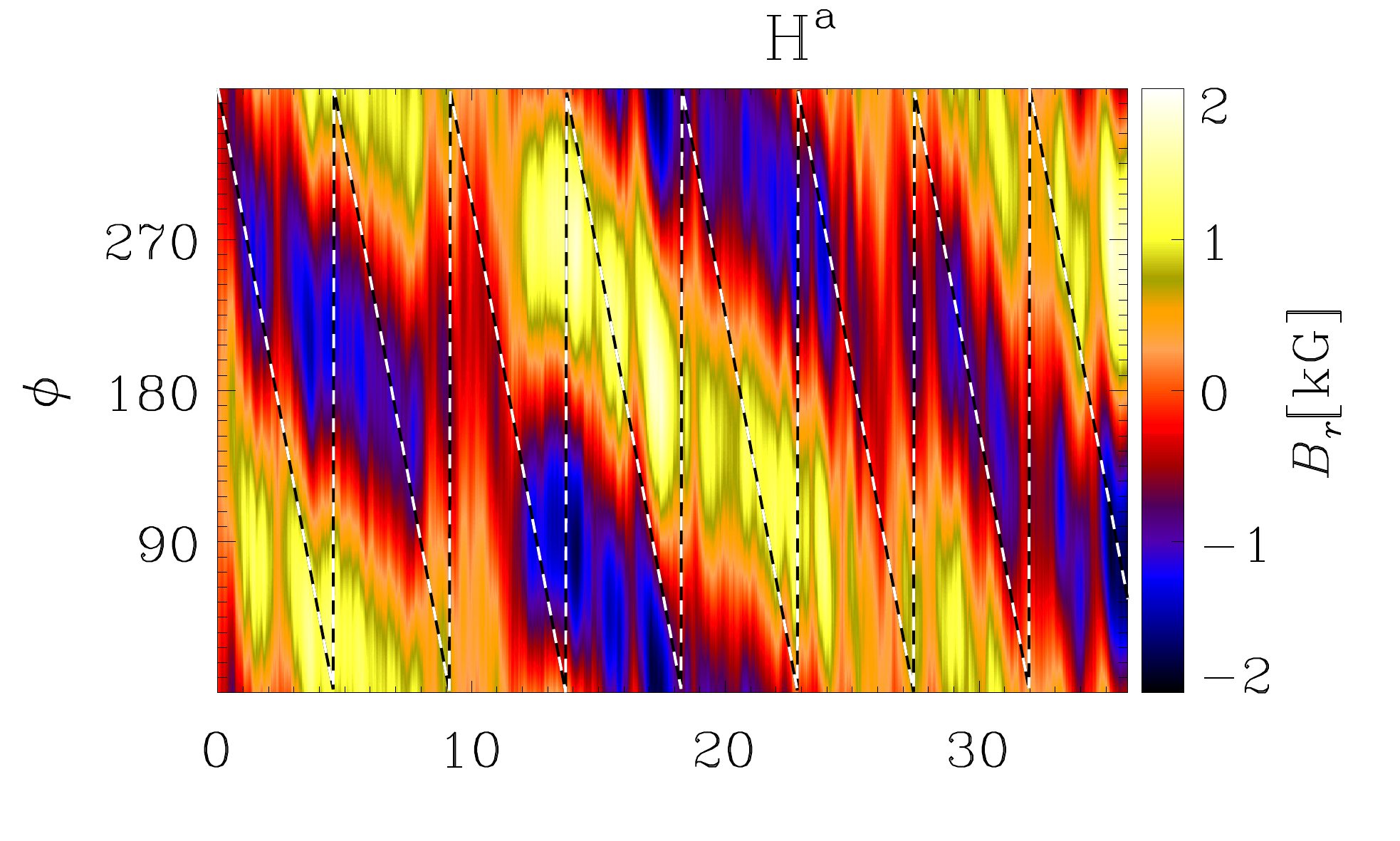}
\includegraphics[width=.9\columnwidth]{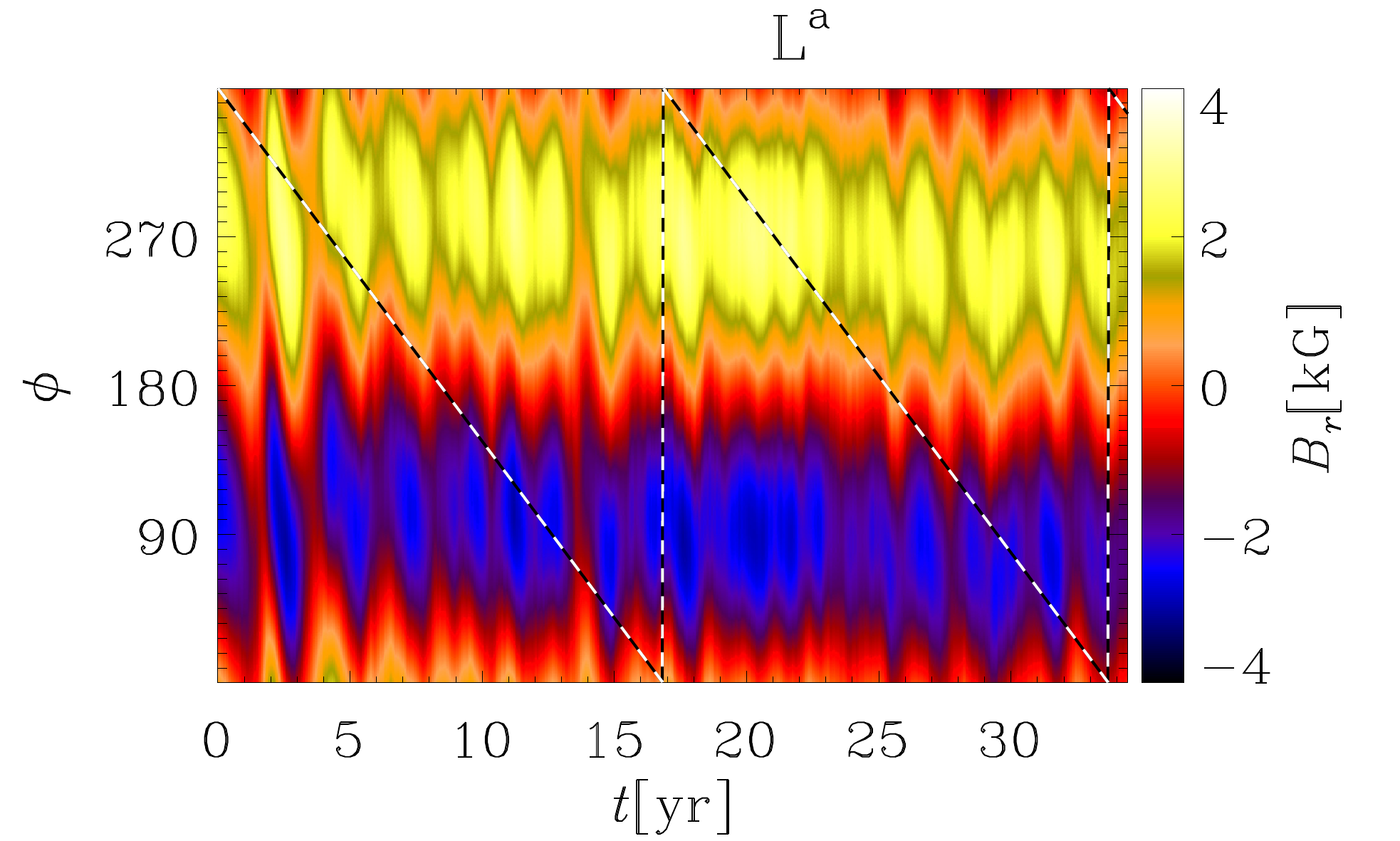}
\caption{Reconstruction of the $m=1$ mode of the magnetic field at 
 the surface of the star for Runs~C3, G$^a$, H$^a$, and L$^a$
 at $\theta=+60^\circ$.
 The black and white line is the path due only to differential rotation.}
\label{fig:adw}
\end{figure}

\subsection{Azimuthal dynamo waves}
\label{sec:ADW}

In stars which rotate more rapidly than the Sun, 
spots tend to emerge at high latitudes and 
unevenly distributed in longitude.
These preferred locations for starspot appearance are called 
active longitudes \citep[][]{Jetsu1996, BT98}.
A phenomenon that has recently been related to active longitudes
from models \citep{CBKK16} and also observations \citep[see e.g.][]{LMOPHHJS13}
is what is now called azimuthal dynamo wave (ADW). This term refers to active
longitude systems which migrate in the orbital reference frame of the star.
A useful comparison is the latitudinal dynamo wave visible in the Sun:
this dynamo wave shows a dependence in latitude, visible as the
appearance of sunspots
at lower latitudes as the solar cycle progresses, but the spots do not 
appear with a preferential location in longitude. 
Instead of its latitude depending on time, in the ADW, the longitude of the
nonaxisymmetric spot-generating mechanism changes periodically in time,
thereby migrating in the rotational frame of reference. Such migration was
already predicted from early linear dynamo models \citep[e.g.][]{KR80}, and
the special case of non-migratory nonaxisymmetric structure could also
be interpreted as a standing ADW. The crucial difference between
latitudinal and azimuthal dynamo waves is that the polarity reversal is always
associated with the former, while not necessarily with the latter.
The migration direction has been observed to be preferentially prograde
\citep[see, e.g.,][]{BT98,LMOPHHJS13,Lehtinen16}, but also a standing wave for $\sigma$ Gem
and a retrograde wave for EI Eri have been reported \citep{BT98}.

We inspect all our runs with a significant $m=1$ mode for the existence
of ADW.
The results for the reconstruction of the first nonaxisymmetric mode of the
radial magnetic field as functions of time and longitude for Runs~C3, G$^a$,
H$^a$, and L$^a$ are shown in \Fig{fig:adw} for $60\degr$ northern latitude.
In all the runs presented here, the $m=1$ mode is rigidly rotating and
has a different pattern speed than the gas.
To verify that the magnetic field is detached from the flow, we overplot the
expected advection due to differential rotation with black-white lines
at the same latitude.
If the magnetic field was advected by the mean flow, its maxima
and minima would fall on this line.
In the range $3\leq \Co \leq 68$, 
the magnetic field follows a pattern different from 
the differential rotation at the surface of the star
at all latitudes.

The parameters related to the ADW are listed in columns 11--14 of \Table{tab:dec}.
The period of the ADW, $P_{\rm ADW}$, 
is calculated using the first derivative 
with respect to time
of the maximum of the phase of the $m=1$ mode,
averaged over time and latitude.
We compare it with the bulk rotation, $P_{\rm ADW}/P_{\rm 0}$, and the
differential rotation, $P_{\rm ADW}/P_{\rm DR}$, 
where $P_0=2\pi/\Omega_0$ and 
$P_{\rm DR}=2\pi/\brac{[\Omega-\Omega_0](r=0.98R)}_{\theta}$, respectively,
and indicate the direction
of the wave, retrograde (R, westward) or prograde (P, eastward), in the column
marked $D$.
A retrograde wave is moving in the opposite direction with respect to the 
bulk rotation. Therefore, its period will be longer than the rotation period.
On the other hand, a wave moving in the prograde direction will have a shorter period.
In most of our cases, we find retrograde ADWs, 
but there are some cases (Runs~J, K1, L$^a$, M$^a$) in which
the behavior is different.
Runs~J and K1 are characterized by rapid rotation and a low value of magnetic energy
and the azimuthal dynamo wave has a smaller amplitude than in the other cases.
In Runs~M$^a$, L$^a$ and J the dynamo wave is drifting very slowly\footnote{A video of the
  surface radial magnetic field evolution of Run~L$^{a}$ can be found from https://www.youtube.com/watch?v=2g4r1uanrj4 }.
During the saturated stage,
these represent standing waves rather than migratory
phenomena (therefore the identifier SW in \Table{tab:dec}). Their almost insignificant
migrations occur in opposite directions with Runs~J and M$^a$ showing
prograde migration, 
and Run~L$^a$ exhibits retrograde migration. 
In the parameter regime included in this study, the retrograde
migration is clearly the dominant one.
The magnetic cycle seems not to be related in any way to the migration period of the
ADW.

\subsection{Time variation and flip-flop phenomenon}

In some cases we find an equatorward migrating oscillatory magnetic
field in the initial stages of the simulation (for example, Runs~G and
H), see \Fig{fig:pbutter}.
Later, however, the dominant
dynamo mode changes to a nonaxisymmetric one soon after the 
large-scale field reaches dynamically significant strengths. 
This behavior has been found in \cite{KMCWB13}, where the $\pi/2$
and 2$\pi$ versions of Run~F1 have been compared.
Thus, we conclude that a reduced $\phi$ extent significantly changes the behavior
of the dynamo by suppressing the large-scale nonaxisymmetric modes ($m=1$, $2$, $3$).
Also, we observe that for cyclic solutions to emerge in $\pi/2$ wedges,
we require a generally higher
Coriolis number than in runs with full azimuthal extents.

\begin{figure}[t]
\centering
\includegraphics[width=\columnwidth]{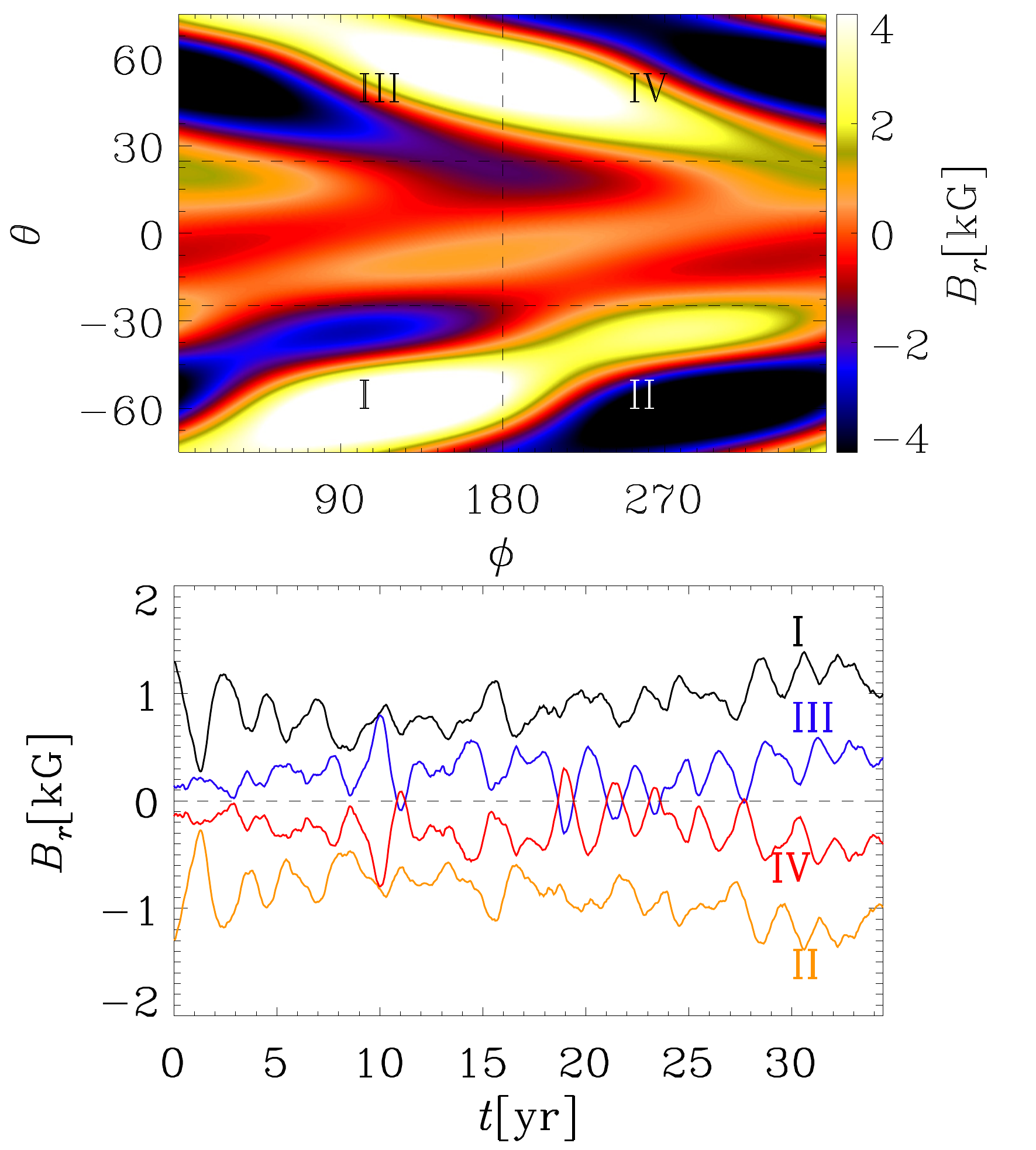}
\caption{Standing dynamo wave of  Run~L$^a$.
The lower panel shows the time variation of the four regions indicated
in the upper plot.}
\label{fig:La}
\end{figure}

Time variations are also seen in the cases of nearly purely nonaxisymmetric
solutions, one such example being the high-resolution Run~L$^{a}$.
The magnetic field in this run forms two active longitudes that remain fixed
on the stellar surface, having opposite polarities in each hemisphere, but
exhibiting a quadrupolar symmetry with respect to the equator. The weak axisymmetric
component also exhibits time variability, as can be seen from the butterfly diagram
plotted in \Fig{fig:pbutter}. Both the axi- and nonaxisymmetric components
develop time variability over a similar time scale of roughly 3 years.
The strength of the active longitudes is modulated on this time scale
in such a way that the ones in the same hemisphere grow simultaneously;
see Figs.~\ref{fig:La} and \ref{fig:adw}, while
the ones on the opposite one decay, followed by a reversed behavior; see \Fig{fig:La}.
However, there are no clear polarity reversals that could be related to this time variation.
In other words, we observe that maximum and minimum on the same hemisphere
never switch in intensity, as is happening in the flip-flop phenomenon
\citep{Hackman2013, BT98}.
It has been postulated that a polarity reversal of the active longitudes would
happen during a flip flop event, observable through ZDI \citep[e.g.][]{Carroll09,KMHI13}, but the effect
of ADWs has never been considered, making these conclusions uncertain.

\begin{figure}[t]
\centering
\includegraphics[width=\columnwidth]{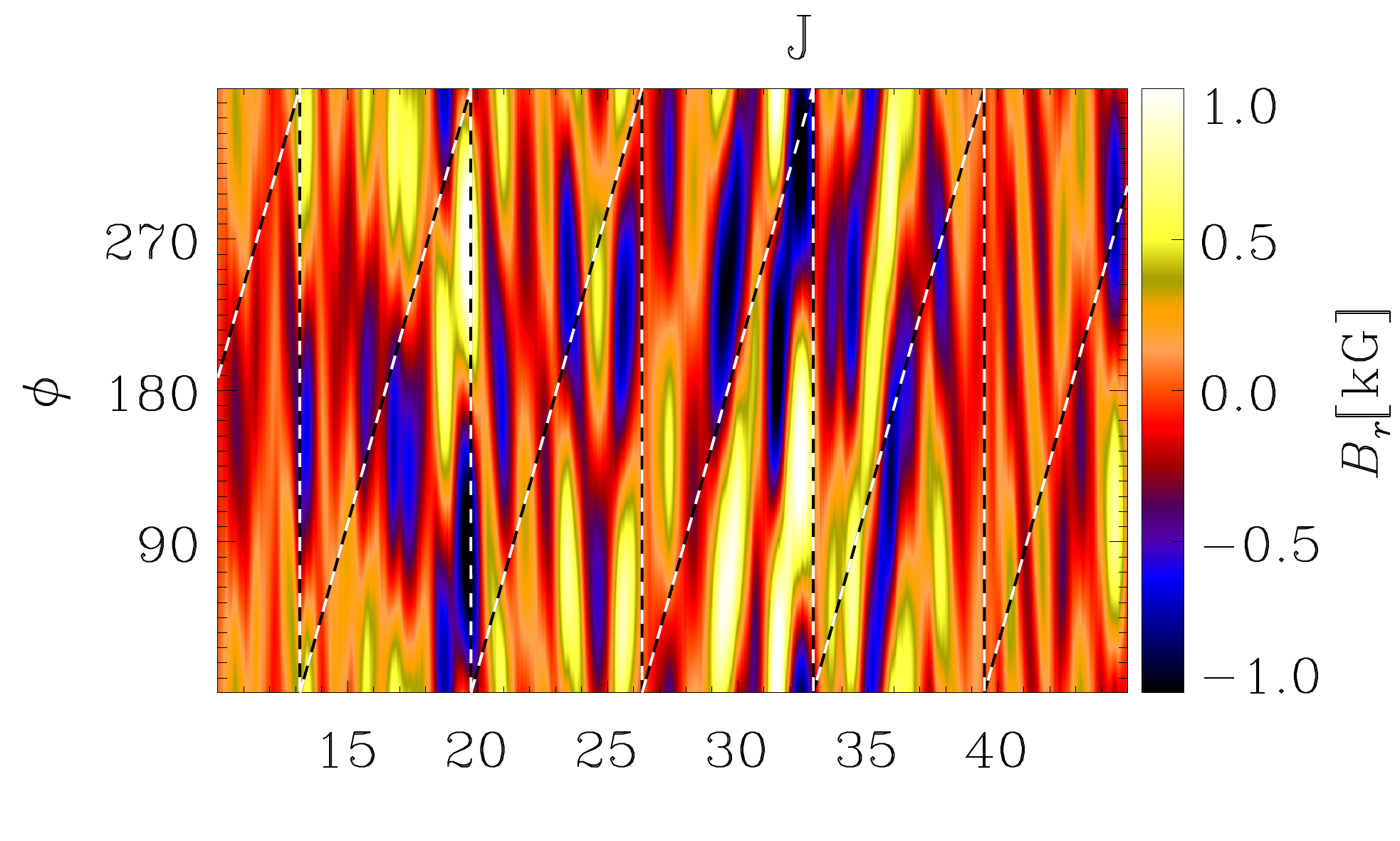}
\includegraphics[width=\columnwidth]{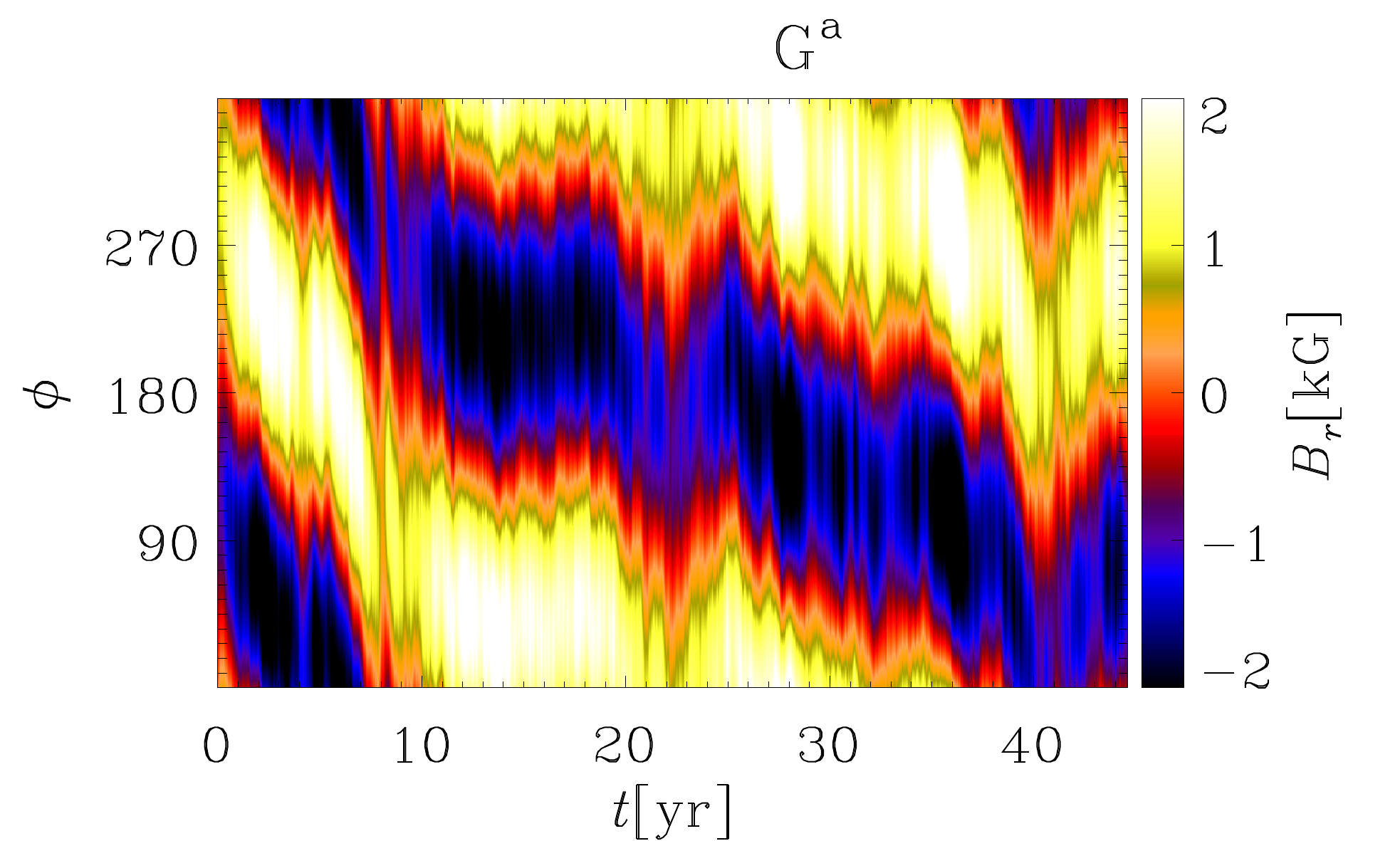}
\caption{
{\it Upper panel}: flip-flop for Run J.
The dashed line is the differential rotation at $\theta=+45^{\circ}$.
{\it Lower panel}: same as \Fig{fig:adw}, 
but at $\theta=+45^{\circ}$, for Run~G$^a$.
The ADW has been de-migrated to better show the active longitudes.}
\label{fig:activelong}
\end{figure}

To see whether flip-flops can occur in systems where there is a competition
between the $m=0$ and $m=1$
modes, we now analyze Run~G$^a$ in detail. 
As discussed in Sect.~\ref{sec:ADW}, this run exhibits an azimuthal
dynamo wave that is migrating in the retrograde direction.
To better see the time evolution of the
active longitudes, this migration has to be removed, as done in \Fig{fig:activelong}, lower panel.
After this systematic motion is removed, however, as in the case of L$^a$, the active 
longitudes are not switching in intensity between maxima and minima, but grow and decay
together on the same hemisphere, while out of phase in the opposite hemisphere.
In Run~J, producing only a very weak dynamo solution with almost a standing azimuthal dynamo
wave, a polarity change can, however, be detected, as is depicted in \Fig{fig:activelong}, upper panel.
The active longitudes are seen to stay nearly fixed in the orbital frame of reference, and after
quasi-regular time points, the polarity of both reverses quite abruptly. In this case the magnetic
field is clearly sub-dominant with respect to the velocity field, but nevertheless the advection
by the differential rotation explains very poorly the time evolution of the active longitudes.
Distinguishing between such a polarity reversal and the mere migration of the 
active longitude poses a challenge to the observations. According to our models, the 
migration speeds are always very distinct from the rotation periods, so any behavior 
caused by such systematic movement would appear smooth to a real flip flop.

\section{Conclusions}
\label{sec:conclusions}

In this paper, we have performed an extensive study of
the effect of rotation rate on convection-driven spherical dynamos, 
covering a range from $1$ to $31$ times the solar value,
$\Omega_{\odot}$, corresponding to $\Co=1.6$ to $127$.
The dependence of stellar dynamos on rotation speed has been
assessed over a range that is much wider than what has been
studied previously.
For example, \cite{SBCBN17} studied the change of cycle frequency while
changing the rotation rate by a factor of two, resulting in a change
in $\Co$ of about a factor of three.
We found that, for $\Omega \gtrsim 1.8\Omega_{\odot}$ ($\Co \gtrsim 3$),
nonaxisymmetric modes are excited and azimuthal dynamo waves are present;
see \Table{tab:results}.
The most commonly excited configuration in our models is the $m=1$ mode
accompanied with an $m=0$ mode comparable (for moderate rotation) or sub-dominant (for rapid rotation) in strength.
The magnetic field near the surface is symmetric (quadrupolar) with respect to the equator in all
cases with an antisolar differential rotation profile.
The axisymmetric part of the magnetic field is more toroidal at moderate rotation,
while preferentially more poloidal configurations are indicated from
the highest rotation rates studied. In the slow rotation regime with
antisolar differential rotation, the solutions are preferentially
axisymmetric and poloidal.

\begin{table}[b!]
\centering
\caption[]{Summary of the transitions from solar-like to antisolar-like
differential rotation and between predominantly nonaxisymmetric and
axisymmetric large-scale fields from observations and our simulations
functions of increasing Coriolis number.}
          \label{tab:results}
\begin{tabular}{lcccc}
\hline
\hline
          & \multicolumn{2}{c}{Observations} & \multicolumn{2}{c}{Simulations} \\
transition & $\tilde\Omega$ & $\Co$ & $\tilde\Omega$ & $\Co$ \\
\hline
antisolar/solar-like DR       & $\approx1$ & $\approx6$ & $1.8$      & $3$ \\
axi/nonaxisymmetric           & $3$--$5$   & $13$--$25$ & $1.8$ & $3$  \\
return to axisymmetry &            &            &            &            \\
(low-res, high $\Co$) &            &            & $15$--$22$ & $37$--$83$ \\
\hline
\end{tabular}
\tablefoot{
Observations refer to \cite{Lehtinen16} for the
nonaxisymmetric to axisymmetric transition and to
\cite{2018ApJ...855L..22B} for the
solar-like to antisolar-like transition using the semi-empirical
$\tau_{\rm c}$ values from \cite{1984ApJ...287..769N}.
}\end{table}

The same pattern over the azimuthal direction can be seen
observationally
in the distribution of active longitudes or the magnetic field geometries of stars
with different rotation rates. \cite{Lehtinen16} found from time series photometry
of active solar-type stars that there is an onset of active longitudes at around
$\Co \approx 25$, corresponding to $\tilde\Omega \approx 4-5$. 
Similarly, surface magnetic field mapping using ZDI has shown
that solar-type stars have a transition between axisymmetric poloidal and
nonaxisymmetric toroidal field geometries at around $\Co \approx 13$
\citep[or $\Ro = P_{\rm rot}/\tau_{\rm c} \approx 1$;][]{Donati2009,
  2016MNRAS.462.4442S}, where $\tau_{\rm c}$ is the convective
turnover time.
This split is not absolute and the rapidly rotating stars can still alternate
between toroidal and poloidal fields \citep{KMHI13}.
Moreover, \cite{Rosen2016} observed that for rapid rotators
the degree of nonaxisymmetry tends to increase
towards more poloidal field geometries. This may indicate a similar behavior
as in the high resolution models which develop nonaxisymmetric poloidal fields
at the highest rotation rates.
We note here that the toroidal and poloidal fields are computed
differently from observations, namely from the total surface field,
than here, namely computing it only for the axisymmetric mean field of
the whole convection zone.

The differences in the rotation rates and Coriolis numbers of the
axisymmetric to nonaxisymmetric transition between observations and
simulations may be
due to several factors. First, the criteria for detecting nonaxisymmetric
structures may not be fully comparable between the different studies. Second,
the observational studies use semi-empirical values of the convective
turnover time $\tau_{\rm c}$ while in this study we have used the definition
$\tau_{\rm c}=2\pi\urms/0.3R$. Lastly, it is worth noting that the
simulations do not occupy the same parameter space as real
stars.
Furthermore, a different value of $\Co$ could just be explained by a
different depth in the star where the dynamo is mainly driven, as
$\urms$ has a strong radial dependency.
The observations do not show any indication that the most rapidly rotating stars
would again have axisymmetric fields, as is the case with the
low resolution runs in this study. 
The difference in behavior between high and low resolution runs, for which
low resolution runs turn back to axisymmetric fields and high resolution runs
remain nonaxisymmetric may simply be a symptom of the inability of the
low resolution runs to capture sufficiently small scales.

In our set of runs, we found mostly retrograde azimuthal dynamo waves,
in contrast with observations of solar-like stars that show a preference 
for prograde direction \citep{Lehtinen16}.
The prograde pattern speeds may be analogous to those seen in the Sun.
Its supergranulation pattern is found to rotate a few percent faster
than the gas at the surface \citep{GDS03}.
Similarly, magnetic tracers including sunspots are seen to rotate
faster than the gas \citep{PT98}.
The occurrence of prograde pattern speeds is theoretically associated
with the near-surface shear layer of the Sun \citep{GK06,Bu07,Br07}.
Thus, a reason for this discrepancy could be the fact that we simulate only the 
stellar convection zone and do not include the near-surface shear
layer, which should lead to a prograde directed wave.

In the interval  $1$--$1.8\,\Omega_{\odot}$, corresponding to
$\Co=1.6$--$2.8$, we find antisolar differential rotation,
in agreement with previous studies, such as \cite{KKB14} and \cite{GYMRW14}. 
We do not see any oscillatory behavior of the magnetic field in the
interval $\Co=1.6$--$2.4$ whereas
close to the transition from antisolar to solar rotation profiles, at $\Co=2.6$--$2.8$
even systems with antisolar rotation profiles produce clear cycles in
their axisymmetric fields.
This seems to be quite a robust finding, as this behavior persists even when the
efficiency of convection is varied.
Cyclic magnetic activity has been seen in giants and subgiants that are believed to have
antisolar differential rotation
profiles \citep{WSW05,KBSVO07,HSKCW16}, which, according to our results, would be possible in
a narrow region near the break-point from antisolar/axisymmetric to solar/nonaxisymmetric behavior
(see \Table{tab:results}). 
In dwarfs, antisolar differential rotation is indicated only
indirectly through the occurrence of enhanced activity at slow rotation for $\tilde\Omega\la 1$
\cite{2018ApJ...855L..22B}.

In the rapid rotation regime, both dominantly axisymmetric and nonaxisymmetric solutions produce
time variability of very different nature, which, however, occurs over similar time scales
and produces similar magnitudes of variations, at least in terms 
of the surface magnetic field strength.
In the axisymmetric case, these relate
to the latitudinal dynamo wave and are accompanied by a polarity change.
In the majority of the nonaxisymmetric cases,
the time variability relates to the changing activity levels of active longitudes on different hemispheres
with no associated polarity change.
In one low-resolution case that produces only a very weak dynamo, we found a solution which also
shows flip-flop type polarity reversals, but this particular parameter regime needs to be
studied with high-resolution runs.
The drift period of the active longitude system in the orbital frame of
reference identified in almost every simulation seem to be de-coupled from the magnetic activity cycle, but 
together with the variations in the active longitude strengths can be thought to give rise
to an azimuthal dynamo wave.
Also observationally the occurrence of cycles is not related to the axisymmetry
of the stellar magnetic fields. Activity cycles are observed on slow and fast
rotating stars alike, regardless if they have active longitudes or not
\citep{Lehtinen16}.

The extensive study on rotation rate allowed us to investigate 
the existence of activity branches \citep[see, e.g.][]{SB99,Distefano2017,Reinhold2017}.
In our $P_{\rm rot} / P_{\rm cyc}$ versus $\Co$ plot, the runs are separated
into two populations: one for the axisymmetric runs at low Coriolis
numbers,
whose slope of $\Co^{-0.73}$ seems to be similar to that found in the $\pi/2$ wedges
of \cite{W17}, 
the other at higher $\Co$ representing the nonaxisymmetric population, 
whose slope of $\Co^{-0.50}$ is close to the superactive branch
reported in \cite{SB99} and whose
behavior resembles that of the transitional branch of
\cite{Distefano2017}.
However, when comparing to observations, our inactive population
does not match the inactive branch seen in observational studies
\citep[e.g.,][]{1984ApJ...287..769N,BST98,Brandenburg2017}.
A possible explanation for this discrepancy could reside in the different 
ways of calculating $\tau_{\rm c}$ in observations and simulations.
Also, we do not find any clear separation between active and superactive branches.
Moreover, we studied the behavior of $P_{\rm rot} / P_{\rm cyc}$ as a 
function of magnetic activity (represented, in our case, by the ratio $E_{\rm mag} / E_{\rm kin}$).
In this case, the axisymmetric population seems to have a positive slope,
as seen in observations.
Anyway, our sparse sample at low rotation and 
the inability to 
reliably compute the chromospheric activity index $\RHK$ from the models,
do not allow us to draw any 
significant conclusion.
We also compare our results with the numerical study of \cite{SBCBN17}.
In contrast with our simulations, their solutions show only axisymmetric behavior.
This, and the fact that in the rotation-activity plot their results lay close to 
our models with reduced $\phi$ extent, make us believe that the resolution used in this study 
was not enough to allow for nonaxisymmetric solutions.
We consider this as a further proof of the importance of using high resolution
when investigating high rotation regimes. 

Our results confirm that
the scale at which the power spectrum of the velocity field peaks
shifts to higher values of $\ell$ with increasing rotation speed,
indicating the presence of smaller convective cells at rapid rotation
\citep{Ch61}.
Our results have also demonstrated that sufficiently high numerical resolution
is important for allowing the $m=1$ nonaxisymmetric structure to develop.
The wedge assumption in the azimuthal direction was not found to be a good
one for rapidly rotating stars. First, it suppresses the nonaxisymmetric
modes that emerge close to the solar rotation rate.
Second, there were only indications of oscillatory solutions
in earlier $\pi/2$ wedges with
antisolar rotation profiles \citep{KKKBOP15,W17}, while in this study
we find clear oscillatory solutions with many polarity reversals in
the runs with full azimuthal extents.
The magnetic
structures appearing preferentially at high-latitude regions
with more rapid rotation also put the latitudinal wedge assumption into
a question. A better modelling strategy for the future are full
spherical grids where the parameters are chosen so that the models
are equally supercritical in terms of the Rayleigh number.

\begin{acknowledgements}
We thank the referee for constructive criticism.
  M.V.\ acknowledges postgraduate fellowship from the SOLSTAR 
  Max Planck Research Group and having been enrolled in the 
  International Max Planck Research School for Solar System Science 
  at the University of G\"ottingen (IMPRS) framework.
 The simulations were performed using the supercomputers hosted by
 GWDG,  the Max Planck supercomputer at RZG in Garching and
  CSC -- IT Center for Science Ltd.\ in Espoo, Finland, who are
  administered by the Finnish Ministry of Education. Special Grand
  Challenge allocation DYNAMO13 is acknowledged. 
J.W.\ acknowledges funding by the Max-Planck/Princeton Center for
Plasma Physics and 
 from the People Programme (Marie Curie
Actions) of the European Union's Seventh Framework Programme
(FP7/2007-2013) under REA grant agreement No.\ 623609.
Financial support from
  the Academy of Finland grant No.\ 
  272157 to the ReSoLVE Centre of Excellence (J.W., M.J.K., P.J.K.), 
  Finnish Cultural Foundation grant No. 00170789 (N.O.), as well as
the NSF Astronomy and Astrophysics Grants Program (grant 1615100),
the Research Council of Norway under the FRINATEK grant 231444,
and the University of Colorado through its support of the
George Ellery Hale visiting faculty appointment
are acknowledged (A.B.).
\end{acknowledgements}

\appendix
\section{Decomposition of the magnetic and velocity field in spherical harmonics}
\label{sec:appendixA}

To investigate the scale dependence of the velocity and magnetic fields, it is
instructive to
decompose the solutions into spherical harmonics.
For this purpose, we will only use the radial components of the magnetic
and velocity fields, $B_r$ and $u_r$, respectively.
Those are related to the respective superpotentials via
\begin{equation}
B_r=L^2 {\cal B},\quad
u_r=L^2 {\cal U},
\end{equation} 
where $L^2(\cdot)=-\sin\!^{-1}\partial_\theta
(\sin\theta\partial_\theta\,\cdot)-\sin\!^{-2}\partial_\phi^2$
is the angular momentum operator and ${\cal B}$ and ${\cal U}$ are the
poloidal superpotentials that can be expanded in terms of spherical
harmonics $Y_\ell^m(\theta,\phi)$ as \citep{KR80}
\begin{equation}
{\cal U}(\theta,\phi)=\sum_{\ell=0}^{\ell_{\max}}\sum_{m=-\ell}^\ell
\tilde{\cal U}_\ell^m Y_\ell^m(\theta,\phi),
\end{equation} 
where $\tilde{\cal U}_\ell^m$ are computed as
\begin{equation}
\tilde{\cal U}_\ell^m=\int_0^{2\pi}\int_{\theta_0}^{\pi-\theta_0}
{\cal U}(\theta,\phi)\,Y_\ell^{m\,*}(\theta,\phi)\,\sin\theta\,d\theta\,d\phi,
\end{equation} 
and likewise for $\tilde{\cal B}_\ell^m$.
Owing to the absence of magnetic monopoles, however, there is no
contribution to the magnetic field for $\ell=0$.
In practice, we work directly with the radial components of velocity
and magnetic field, whose transforms are related to ${\cal B}$ and ${\cal U}$ via
$\hat{B}_{\ell, r}^{m}=\ell(\ell+1)\hat{\cal B}_\ell^m$ and
$\hat{u}_{\ell, r}^{m}=\ell(\ell+1)\hat{\cal U}_\ell^m$, respectively.

While testing the decomposition, we noticed that the large scale field
features were fairly well described by the first few modes ($0\leq \ell \leq 5$).
Therefore, in order to obtain a complete picture,
we decompose the magnetic 
and velocity field in the first eleven spherical harmonics ($0 \leq \ell \leq 10$) 
and consider $0\leq \ell \leq 5$ and $0 \leq m \leq 5$ as the large-scale
fields and the rest as small-scale fields.
Throughout this work we use the decomposition for the
radial velocity and magnetic field on a slice
at a fixed radial position of $r=0.98R$.

\begin{figure*}[t]
 \centering
 \includegraphics[width=0.32\textwidth]{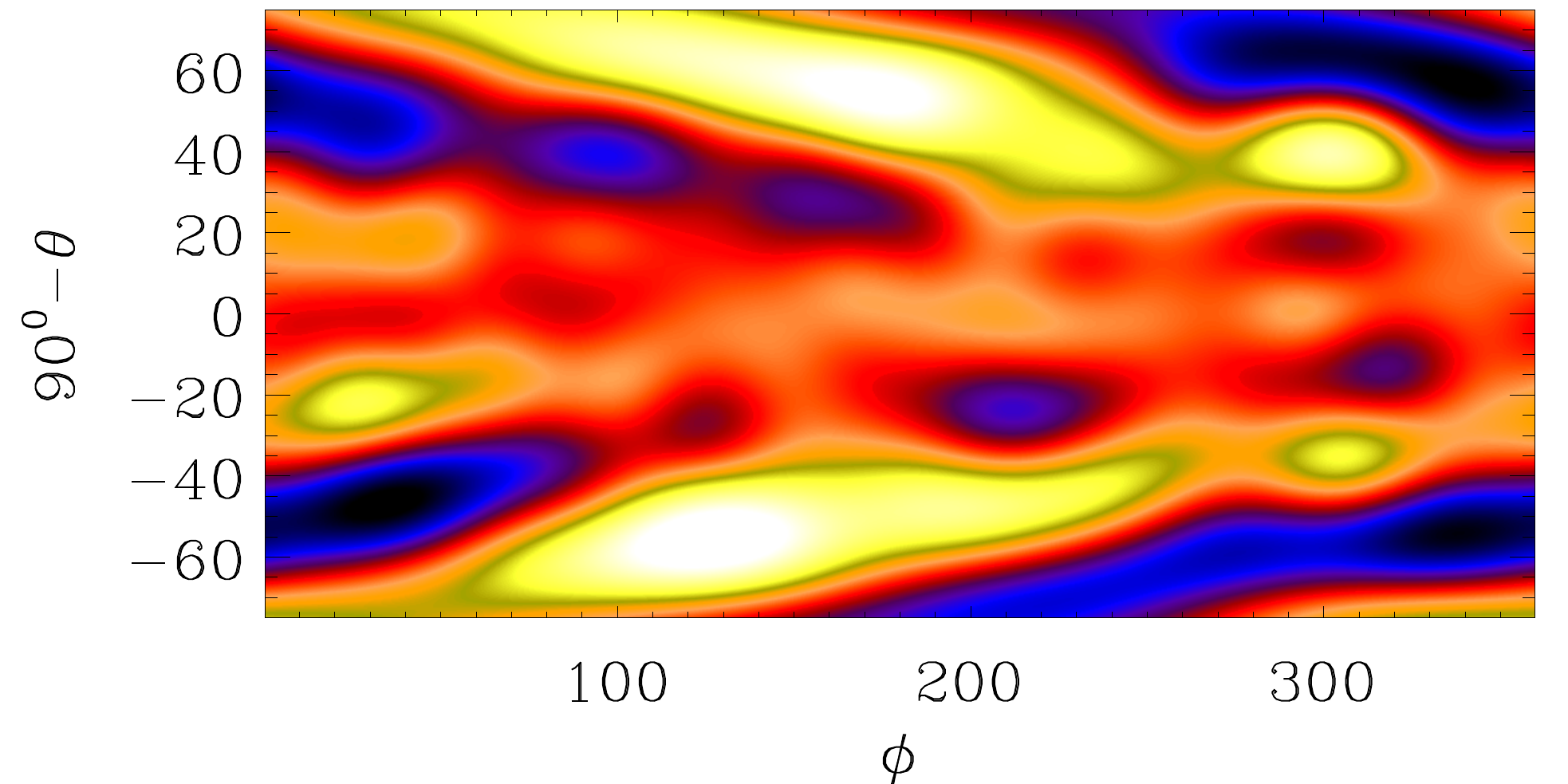}
 \includegraphics[width=0.32\textwidth]{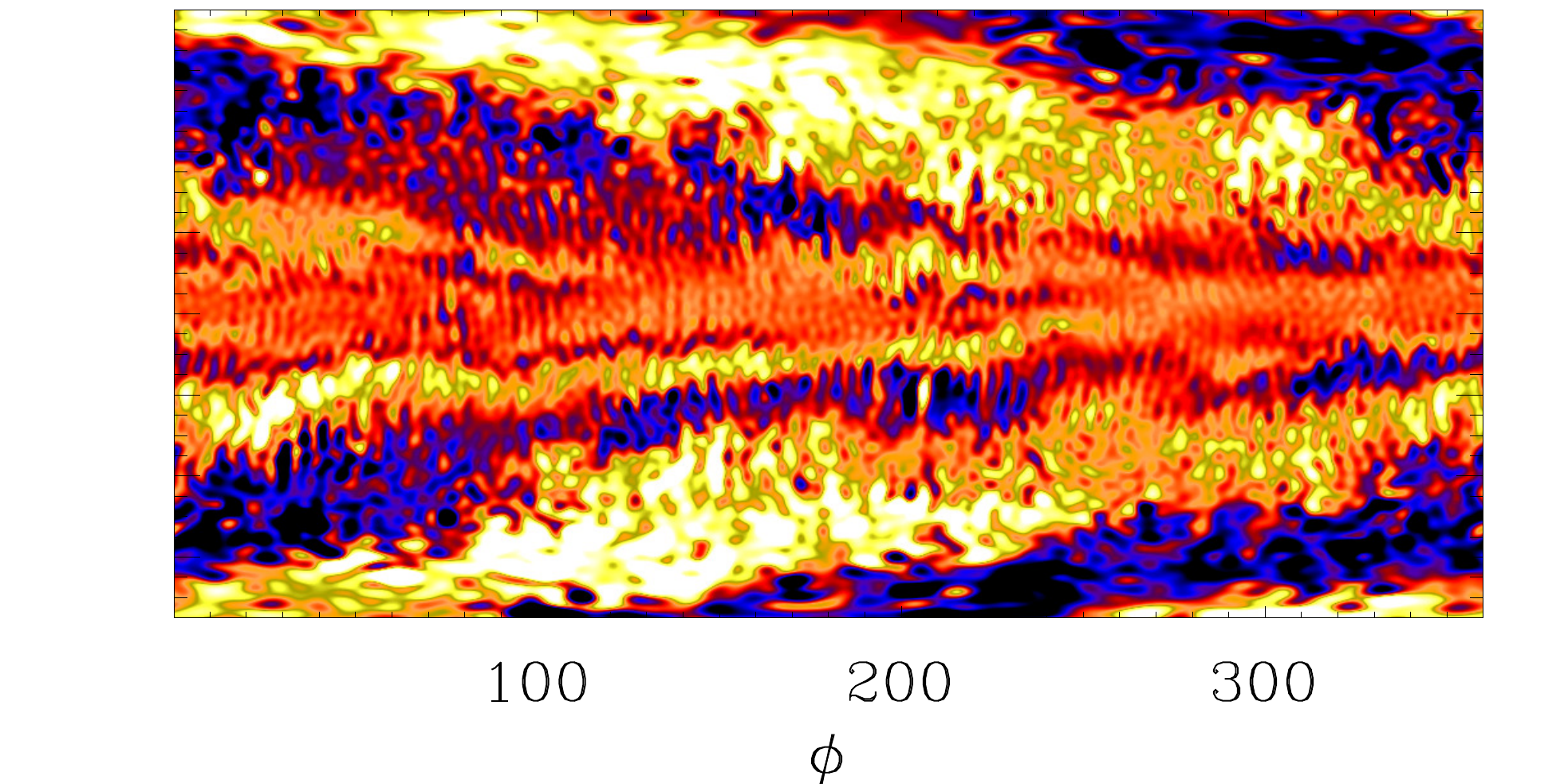}
 \includegraphics[width=0.32\textwidth]{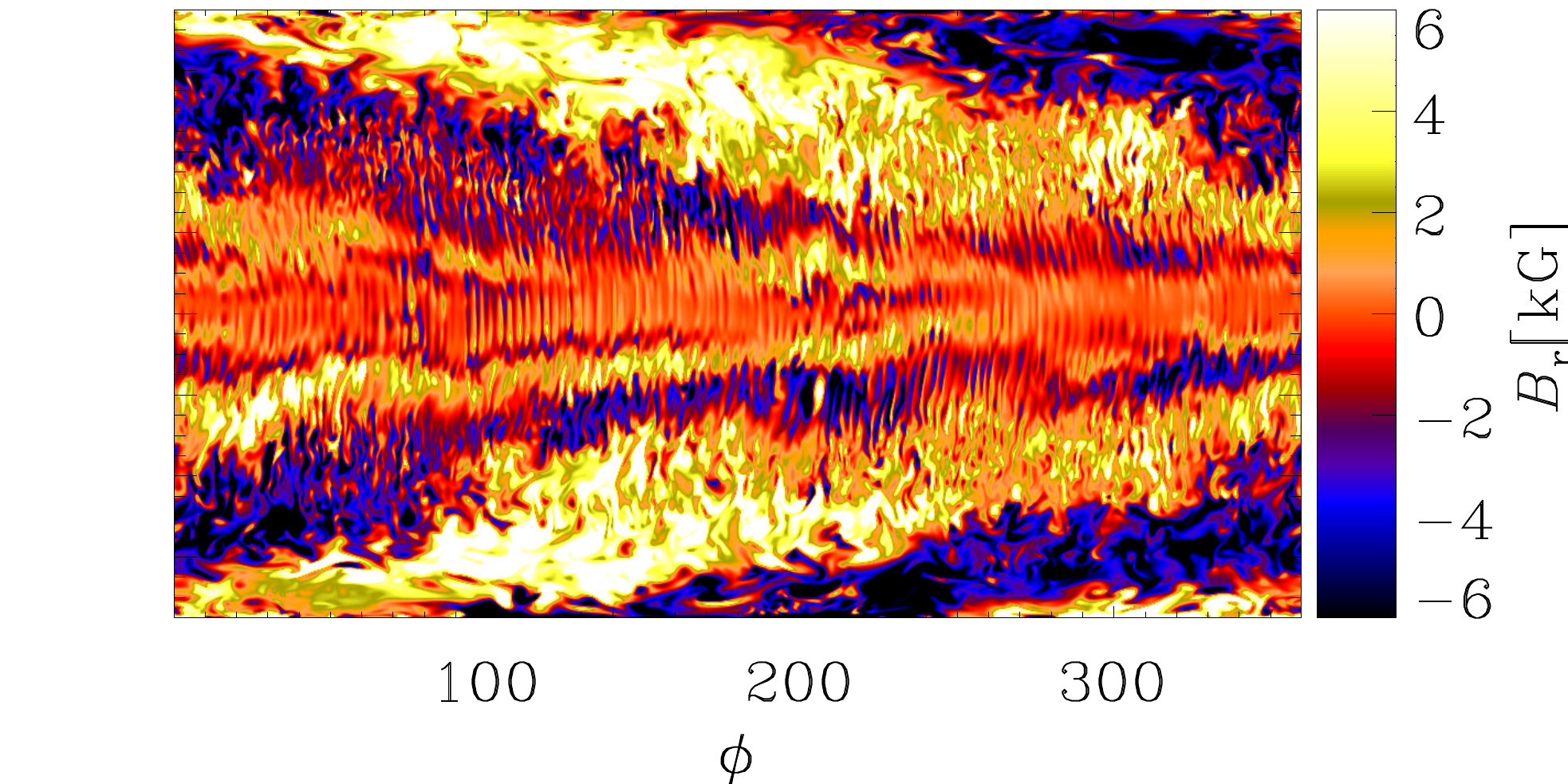}
 \caption{Spherical harmonic reconstruction, using $1 \leq \ell \leq 10$ harmonics (left panel)
 and $1 \leq \ell \leq 100$ harmonics (middle panel),
  and original of the radial magnetic field (right panel) near the surface
  ($r=0.98R$) of Run~L$^a$.}
 \label{fig:reconstruction}
 \end{figure*}

\begin{figure}[t]
 \centering
 \includegraphics[width=0.98\columnwidth]{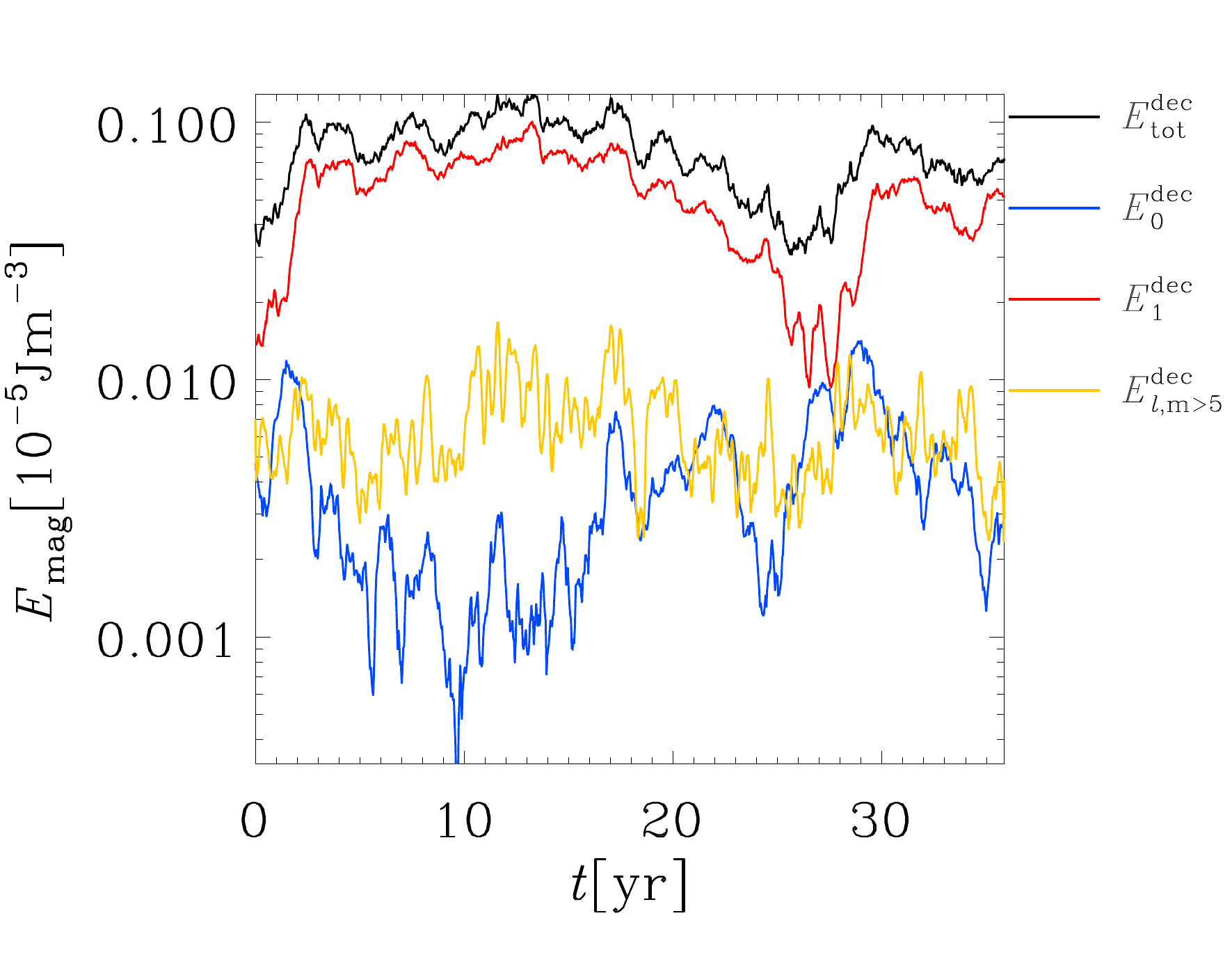}
 \caption{Time series of the total radial magnetic energy from decomposition
   (in black) and each of the $m$ modes
   for Run~H$^a$.
Blue: $m=0$, red: $m=1$, 
yellow: small-scale magnetic field.}
 \label{fig:dectime}
 \end{figure}

We illustrate the quality of the reconstruction in \Fig{fig:reconstruction} showing the radial
magnetic field from Run~L$^a$ using different numbers of spherical
harmonics.
In the left panel, the reconstruction was obtained using $1 \leq \ell
\leq 10$,
while in the central panel, the reconstruction is obtained summing over the first $100$ spherical harmonics.
The right panel shows the original data slice.
It is clear from \Fig{fig:reconstruction} that $l_{max}=10$ allows us 
to capture the main features  of the magnetic field and a reasonable 
amount of the surface total energy (see, \Table{tab:dec}).
We show a typical time series of different $m$-mode energies from the surface radial magnetic field
reconstruction of Run~H$^a$ in \Fig{fig:dectime}.
This run is dominated by the $m=1$ mode, which shows cyclic
variations over time, and also long-term changes, during which the axisymmetric modes become
comparable to the dominant mode for a short period of time.

\bibliographystyle{aa}
\bibliography{paper}

\end{document}